\documentclass[cernpreprint,texlive=2011,txfonts,UKenglish]{latex/atlasdoc} 

\usepackage[subfigure=true,block=none]{latex/atlaspackage}
\usepackage{latex/atlasphysics}

\usepackage{amssymb}
\usepackage[centering,scale=0.75]{geometry}
\usepackage{graphicx}

\usepackage{adjustbox,lipsum}
\usepackage{rotating}
\usepackage{pdflscape}

\usepackage{hyperref}

\usepackage{latex/atlasbiblatex}
\usepackage{mathrsfs}
\usepackage{xspace}
\usepackage{float}
\usepackage{color, colortbl}
\definecolor{Gray}{gray}{0.9}

\newcommand{\ttb}{\ensuremath{{\ttbar}}}

\newcommand{\absyttbar}{\ensuremath{\left|y^{\ttbar}\right|}\xspace}

\newcommand{\absyt}{\ensuremath{\left|y^{t}\right|}\xspace}
\newcommand{\absythad}{\ensuremath{\left|y^{t,{\rm had}}\right|}\xspace}

\newcommand{\mttbar}{\ensuremath{m^{\ttbar}}}

\newcommand{\ptt}{\ensuremath{\pt^t}}
\newcommand{\ptthad}{\ensuremath{\pt^{t,{\rm had}}}}

\newcommand{\ptttbar}{\ensuremath{\pt^{\ttbar}}}
\newcommand{\Etmiss}{\ensuremath{\ET^{\rm miss}}}

\newcommand{\DeltaPhittbar}{\ensuremath{\Delta\phi^{\ttb}}}

\newcommand{\ystar}{{\ensuremath{y^{\star}}}}
\newcommand{\deltaPhittbar}{\ensuremath{\Delta\phi^\ttb}}

\newcommand{\Poutttbar}{\ensuremath{p_{\rm out}^\ttb}}
\newcommand{\absPoutttbar}{\ensuremath{\left|\Poutttbar\right|}}

\newcommand{\HTttbar}{\ensuremath{H_{\rm T}^\ttb}}

\newcommand{\boostttbar}{\ensuremath{y_{\rm boost}^\ttb}}
\newcommand{\chittbar}{\ensuremath{\chi^\ttb}}

\newcommand{\RWtttbar}{\ensuremath{R_{Wt}}}

\newcommand{\ejets}{{$e+$jets}\xspace}
\newcommand{\mujets}{{$\mu+$jets}\xspace}

\newcommand{\btagged}{\ensuremath{b\mbox{-tagged}}\xspace}
\newcommand{\bjet   }{\ensuremath{b\mbox{-jet}}\xspace}

\newcommand{\ie}{{\sl i.e.~}}
\newcommand{\eg}{{\sl e.g.~}}

\newcommand{\lumitot}{\mbox{20.3\,fb$^{-1}$}}

\newcommand{\Powheg}{{\sc Powheg}\xspace}
\newcommand{\PowHeg}{{\sc Powheg}\xspace}

\newcommand{\PowHegBox}{{\sc Powheg-Box}\xspace}

\newcommand{\Pythia}{{\sc Pythia}\xspace}

\newcommand{\PythiaEight}{{\sc Pythia8}\xspace}

\newcommand{\MadGraph}{{\sc MadGraph}\xspace}

\newcommand{\HDamp}{\ensuremath{h_{\rm damp}}}
\newcommand{\HDampMT}{\ensuremath{h_{\rm damp}\!=\!m_{t}}}

\newcommand{\NLO}{{\sc NLO}}
\newcommand{\NNLL}{{\sc NNLL}}
\newcommand{\aNNLO}{a{\sc NNLO}}
\newcommand{\aNNNLO}{a{\sc N$^3$LO}}
\newcommand{\NNLO}{{\sc NNLO}}

\AtlasTitle{Measurements of top-quark pair differential cross-sections in the lepton+jets channel in $pp$ collisions at $\sqrt{s}=8\,\TeV$ using the ATLAS detector}
\author{The ATLAS Collaboration}

\AtlasAbstract{
Measurements of normalized differential cross-sections of top-quark pair production are presented as a~function of
the top-quark, \ttbar{} system and event-level kinematic observables in proton--proton collisions at a~centre-of-mass energy of $\sqrt{s}=8\,{\rm TeV}$. The observables have been chosen to emphasize the \ttbar{}~production process and to be sensitive to effects of initial- and final-state radiation, to the different parton distribution functions, and to non-resonant processes and higher-order corrections. The dataset corresponds to an integrated luminosity of \lumitot{}, recorded in 2012 with the ATLAS detector at the CERN Large Hadron Collider. Events are selected in the lepton+jets channel, requiring exactly one charged lepton and at least four jets with at least two of the jets tagged as originating from a~$b$-quark. The measured spectra are corrected for detector effects and are compared to several Monte Carlo simulations. The results are in fair agreement with the predictions over a wide kinematic range. Nevertheless, most generators predict a harder top-quark transverse momentum distribution at high values than what is observed in the data. Predictions beyond NLO accuracy improve the agreement with data at high top-quark transverse momenta. Using the current settings and parton distribution functions, the rapidity distributions are not well modelled by any generator under consideration. However, the level of agreement is improved when more recent sets of parton distribution functions are used. 
} 
\AtlasRefCode{TOPQ-2015-06}
\PreprintIdNumber{CERN-PH-EP-2015-239}

\author{The ATLAS Collaboration}

\AtlasJournal{Eur. Phys. J. C}
\AtlasJournalRef{Eur. Phys. J. C76 (2016) 538}
\AtlasDOI{10.1140/epjc/s10052-016-4366-4}
\date{\today}

\AtlasCoverCommentsDeadline{Jun 7, 2016}

\AtlasCoverAnalysisTeam{L.~Bellagamba, V.~Boisvert, M.~Capua, R. Di Sipio, M.~Faucci~Giannelli, S.~Honda, J.~Kanzaki, J.~Kvita, J.~Pacalt, M.~Romano, V.~Scarfone, F.Span\`{o}, S.~Suzuki, E.~Tassi, T.~Towstego}

\AtlasCoverEdBoardMember{M. J. Costa, C. Guzman, C. Helsens, R. Schwienhorst (chair)}

\AtlasCoverEgroupEditors{atlas-topq-2015-06-editors@cern.ch}

\AtlasCoverEgroupEdBoard{atlas-topq-2015-06-editorial-board@cern.ch}

\begin{document}

\title{Measurements of top-quark pair differential cross-sections in the lepton+jets channel in $pp$ collisions at $\sqrt{s}=8\,\TeV$ using the ATLAS detector}  
\maketitle
\author{The ATLAS Collaboration}

\tableofcontents
\clearpage

\section{Introduction}\label{sec:Introduction}
The large top-quark pair production cross-section at the LHC allows detailed studies of the characteristics of $\ttbar{}$ production to be performed with respect to different kinematic variables, providing a unique opportunity to test the Standard Model (SM) at the \TeV{} scale. Furthermore, effects beyond the SM can appear as modifications of $\ttbar{}$ differential distributions with respect to the SM predictions~\cite{Frederix:2009} which may not be detectable with an inclusive cross-section measurement. A precise measurement of the $\ttbar{}$ differential cross-section therefore has the potential to enhance the sensitivity to possible effects beyond the SM, as well as to clarify the ability of the theoretical calculations in describing the cross-section.

The ATLAS~\cite{atlasDiff1,atlasDiff2,atlasDiff3} and CMS~\cite{CmsDiff1} experiments have published measurements of the $\ttbar{}$ differential cross-sections at a centre-of-mass energy $\sqrt{s}=7$ \TeV{} in $pp$ collisions, both in the full phase space using parton-level variables and in fiducial phase-space regions using observables constructed from final-state particles (particle level); the CMS experiment also published measurements of the $\ttbar{}$ differential cross-sections with data taken at $\sqrt{s}=8$ \TeV{}~\cite{CmsDiff2}. The results presented here represent the natural extension of the previous ATLAS measurements of the $\ttbar{}$ differential cross-sections to the $\sqrt{s}=8$ \TeV{} dataset, and benefit from higher statistics and reduced detector uncertainties.

In the SM, the top quark decays almost exclusively into a~\Wboson{} boson and a~$b$-quark. The signature of a~\ttbar{} decay is therefore determined by the \Wboson{} boson decay modes. This analysis makes use of the lepton$+$jets \ttbar{} decay mode, where one \Wboson{} boson decays into an electron or a muon and a~neutrino and the other \Wboson{} boson decays into a~pair of quarks, with the two decay modes referred to as the $e$+jets and $\mu$+jets channel, respectively. Events in which the \Wboson{} boson decays to an electron or muon through a~$\tau$ lepton decay are also included.

This paper presents a set of measurements of the \ttbar{} production cross-section as a function of different properties of the reconstructed top quark and of the \ttbar{} system. The results, unfolded both to a fiducial particle-level phase space and to the full phase space, are compared to the predictions of Monte Carlo (MC) generators and to perturbative QCD calculations beyond the next-to-leading-order (NLO) approximation. The goal of unfolding to a fiducial particle-level phase space and of using variables directly related to detector observables is to allow precision tests of QCD, avoiding large model-dependent extrapolation corrections to the parton-level top-quark and to a phase space region outside the detector sensitivity. However, full phase-space measurements represent a valid test of higher-order calculations for which event generation with subsequent parton showering and hadronization is not yet available. A~subset of the observables under consideration has been measured by CMS \cite{CmsDiff1}. 

In addition to the variables measured at $\sqrt{s}=$7 \TeV{} \cite{atlasDiff1,atlasDiff2,atlasDiff3}, a set of new measurements is presented. These variables, similar to those used in dijet measurements at large jet transverse momentum~\cite{ATLAS:dijet8TeV, PhysRevLett.81.2642}, are sensitive to effects of initial- and final-state radiation, to the different parton distribution functions (PDF), and to non-resonant processes including particles beyond the Standard Model \cite{Degrande:2010kt}. Finally, observables constructed as a function of the transverse momenta of the $W$ boson and the $b$-quark originating from the top quark have been found to be sensitive to non-resonant effects (when one or both top-quarks are off-shell) \cite{Denner:2010jp} and non-factorizable higher-order corrections \cite{Bevilacqua:2010qb}.

The paper is organized as follows: Section~\ref{sec:Detector} briefly describes the ATLAS detector, while Section~\ref{sec:DataSimSamples} describes the data and simulation samples used in the measurements. The reconstruction of physics objects and the event selection is explained in Section~\ref{sec:EventReco}. Section~ \ref{sec:PseudoTop} describes the  kinematic reconstruction of the \ttb{}~pairs using the pseudo-top algorithm. Section~\ref{sec:BackgroundDetermination} discusses the background processes affecting these measurements. Event yields for both the signal and background samples, as well as distributions of measured quantities before unfolding, are shown in Section~\ref{sec:YieldsAndPlots}. The measurements of the cross-sections are described in Section~\ref{sec:unfolding}. Statistical and systematic uncertainties are discussed in Section~\ref{sec:Uncertainties}. The results are presented in Section~\ref{sec:Results}, where the comparison with theoretical predictions is also discussed. Finally, a summary is presented in Section~\ref{sec:Conclusion}.

\section{The ATLAS detector}\label{sec:Detector}

ATLAS is a multi-purpose detector \cite{Aad:2008zzm} that provides nearly
full solid angle coverage around the interaction point.
This analysis
exploits all major components of the detector. Charged-particle trajectories
with pseudorapidity\footnote {ATLAS uses a
right-handed coordinate system with its origin at the nominal
interaction point (IP) in the centre of the detector and the $z$-axis
along the beam pipe. The $x$-axis points from the IP to the centre of
the LHC ring, and the $y$-axis points upward. Cylindrical coordinates
($r$,$\phi$) are used in the transverse plane, $\phi$ being the azimuthal angle
around the beam pipe. The pseudorapidity is defined in terms of the
polar angle $\theta$ as $\eta = - \ln \tan(\theta/2)$ and the angular separation between particles is
defined as $\Delta R = \sqrt{(\Delta \phi)^2 + (\Delta \eta)^2}$.}
$|\eta| <2.5$ are reconstructed
in the inner detector, which comprises a silicon pixel detector, a silicon 
microstrip detector and a transition radiation tracker (TRT). 
The inner detector is embedded in a 2 T axial magnetic field.
Sampling calorimeters with several different
designs span the pseudorapidity range up to $|\eta| = 4.9$. High-granularity 
liquid argon (LAr) electromagnetic (EM) calorimeters are available up
to $|\eta| = 3.2$. Hadronic calorimeters based on scintillator-tile
active material cover $|\eta| < 1.7$ while LAr technology is used
for hadronic calorimetry from $|\eta| = 1.5$ to $|\eta| = 4.9$. The
calorimeters are surrounded by a muon spectrometer within a magnetic field  
provided by air-core toroid magnets with a bending integral of about 2.5 Tm 
in the barrel and up to 6 Tm in the endcaps.
Three stations of precision drift tubes and cathode-strip chambers provide 
an accurate measurement of the muon track curvature in the region  $|\eta| < 2.7$. 
Resistive-plate and thin-gap chambers provide muon triggering capability up to $|\eta| = 2.4$.

Data are selected from inclusive $pp$ interactions using a three-level trigger system. A hardware-based trigger (L1) uses custom-made hardware and low-granularity detector data to initially reduce the trigger rate to approximately 75 kHz. The detector readout is then available for two stages of software-based triggers. In the second level (L2), the trigger has access to the full detector granularity, but only retrieves data for regions of the detector identified by L1 as containing interesting objects. Finally, the Event Filter (EF) system makes use of the full detector readout to finalize the event selection. During the 2012 run period, the selected event rate for all triggers following the event filter was approximately 400~Hz.

\section{Data and simulation samples} \label{sec:DataSimSamples}
The differential cross-sections are measured using a dataset collected by the ATLAS detector
during the 2012 LHC $pp$ run at $\rts =8$~TeV, which corresponds to an
integrated luminosity of $20.3~\pm~0.6$~\ifb. The luminosity is measured using
techniques similar to those described in Ref.~\cite{Aad:2013ucp} with a
calibration of the luminosity scale derived from beam-separation scans.
The average number of interactions per bunch crossing in 2012 was 21.  
Data events are considered only if they are acquired under stable beam conditions 
and with all sub-detectors operational. The data sample is collected using single-lepton triggers; for each lepton type the logical OR of two triggers is used in order to increase the efficiency for isolated leptons at low transverse momentum. The triggers with the lower \pt{} thresholds include isolation requirements on the candidate lepton, resulting in inefficiencies at high \pt{} that are recovered by the triggers with higher \pt{} thresholds. For electrons the two transverse momentum thresholds are 24~\GeV{} and 60~\GeV{} while for muons the thresholds are 24~\GeV{} and 36~\GeV{}.

Simulated samples are used to characterize the detector
response and efficiency to reconstruct $\ttbar$ events, estimate systematic uncertainties and predict
the background contributions from various processes.
The response of the detector is simulated~\cite{ATLASsim} using a detailed model
implemented in GEANT4~\cite{Geant4}.
For the evaluation of some systematic uncertainties, generated samples are passed through a fast 
simulation using a parameterization of the performance of the
ATLAS electromagnetic and hadronic calorimeters~\cite{ATL-SOFT-PUB-2014-001}.
Simulated events include the effect of multiple $pp$ collisions from the same 
and previous bunch-crossings (in-time and out-of-time pile-up) and are re-weighted 
to match the same number of collisions as observed in data. All simulated samples are normalized to the integrated luminosity of the data sample; in the normalization procedure the most precise cross-section calculations available are used. 

The nominal signal $\ttbar$ sample is generated using the \PowHegBox~\cite{Frixione:2007vw} generator, based on next-to-leading-order QCD matrix elements.
The CT10 \cite{CT10} parton distribution
functions are employed and the top-quark mass ($m_{t}$) is set to $172.5$~\GeV{}. The \HDamp~ parameter, which effectively regulates the high-\pt{} radiation in \PowHeg, is set to the top-quark mass.
Parton showering and hadronization are simulated with \Pythia~\cite{Sjostrand:2006za} (version 6.427) using the Perugia 2011C set of tuned parameters~\cite{perugia}. 
The effect of the systematic uncertainties related to the PDF for the signal simulation are evaluated using samples generated with {\sc MC@NLO}~\cite{MCATNLO} (version 4.01) using the CT10nlo PDF set, interfaced to {\sc Herwig}~\cite{HERWIG} (version 6.520) 
for parton showering and hadronization, and {\sc Jimmy}~\cite{JIMMY} (version 4.31) for the modelling of multiple 
parton scattering.
For the evaluation of systematic uncertainties due to the parton showering model, 
a \PowHeg{}+\Herwig{}~sample is compared to a \PowHeg{}+\Pythia{}~sample. The \HDamp~ parameter in the \PowHeg{}+\Herwig{}~sample is set to infinity.
The uncertainties due to QCD initial- and final-state radiation (ISR/FSR) modelling are estimated 
with samples generated with \PowHegBox interfaced to \Pythia for which the parameters of the generation ($\Lambda_{\rm QCD}$, $Q^2_{\rm max}$
scale, transverse momentum scale for space-like parton-shower evolution and the \HDamp~ parameter) are 
varied to span the ranges compatible with the results of measurements of $\ttbar$ production in association with jets~\cite{rapidity_gap,atlas_ttbar_njets,atlas_mc_predictions}. 
Finally, two additional $\ttbar$ samples are used only in the comparison against data. The first one is a sample of \PowHeg{} matrix elements generated with the nominal settings interfaced to \PythiaEight{}~\cite{pythia8} (version 8.186 and Main31 user hook) and the AU14 \cite{A14tune} set of tuned parameters. In the second sample, \MadGraph~\cite{Alwall:2011uj} \ttb{} matrix elements with up to three additional partons are interfaced to \Pythia using the matrix-element to parton-shower MLM matching scheme~\cite{MLMatching} and the Perugia 2011C set of tuned parameters \cite{perugia}.

The $\ttbar$ samples are normalized to the 
NNLO+NNLL cross-section of $\sigma_{t\bar{t}}=253^{+13}_{-15}$~pb (scale, PDF and $\alpha_S$), evaluated using the Top++2.0 program~\cite{Czakon:2011xx}, which includes the next-to-next-to-leading-order QCD corrections and resums next-to-next-to-leading logarithmic soft gluon terms~\cite{Cacciari:2011hy,Beneke:2011mq,Baernreuther:2012ws,Czakon:2012zr,Czakon:2012pz,Czakon:2013goa}. The quoted cross-section corresponds to a top-quark mass of 172.5 \GeV{}.
Each $\ttbar$ sample is produced requiring at least one semileptonic 
decay in the \ttb{}~pair.

Single-top-quark processes for the $s$-channel, $t$-channel and $Wt$ associated production constitute the largest
background in this analysis. These processes are simulated with \PowHegBox using the PDF set CT10 and showered with \Pythia (version 6.427) calibrated with the P2011C tune \cite{perugia} and the PDF set CTEQ6L1 \cite{cteq6l1}. All possible production channels containing one lepton in the final state are considered.
All samples are generated requiring the presence of a leptonically
decaying $W$ boson.
The cross-sections multiplied by the branching ratios for the leptonic $W$ decay employed for these processes are 
normalized to NLO+NNLL calculations~\cite{Kidonakis:2011wy, Kidonakis:2010ux, Kidonakis:2010tc}.

Leptonic decays of vector bosons produced in association with high-$\pt$ jets,
referred to as $W$+jets and $Z$+jets, constitute the second largest
background in this analysis. 
Samples of simulated $W/Z$+jets events with up to five additional
partons in the LO 
matrix elements are produced with the {\sc Alpgen} generator (version 2.13)~\cite{ALPGEN} using the PDF set CTEQ6L1 \cite{cteq6l1} and 
interfaced to \Pythia (version 6.427) for parton showering; the overlap between samples is dealt with by using the MLM matching
scheme~\cite{MLMatching}. Heavy-flavour quarks
are included in the matrix-element calculations to produce the $Wb\bar{b}$,
$Wc\bar{c}$, $Wc$, $Zb\bar{b}$ and $Zc\bar{c}$ samples.
 The overlap between the heavy-flavour quarks
produced by the matrix element and by the parton shower is removed.
The $W$+jets samples are normalized 
to the inclusive $W$ boson NNLO cross-section~\cite{Hamberg:1990np,Gavin:2012sy} and corrected by
applying additional scale factors 
derived from data, as described in Section~\ref{sec:BackgroundDetermination}.
 
Diboson production is modelled using {\sc Herwig} and {\sc Jimmy} with the {\sc CTEQ6L1} PDF set \cite{cteq6l1} and the yields are normalized using the NLO cross-sections~\cite{Campbell:2011bn}. All possible production channels containing at least one lepton in the final states are considered.

\section{Object definition and event selection} \label{sec:EventReco}
The lepton+jets \ttbar{} decay mode is characterized by the presence of a~high-\pt{} lepton, missing transverse momentum due to the neutrino, two jets originating from $b$-quarks, and two jets from the hadronic \Wboson{} boson decay.

The following sections describe the detector-level, particle-level and parton-level objects used to characterize
the final-state event topology and to define a fiducial phase-space region for the measurements.

\subsection{Detector-level objects}\label{sec:ObjectDef}

Primary vertices in the event are formed from reconstructed tracks such that they are spatially compatible with the luminous interaction region. The hard-scatter primary vertex is chosen to be the vertex with the highest $\sum \pt^2$ where the sum extends over all associated tracks with $\pt > 0.4\,\mathrm{\GeV{}}$.

Electron candidates are reconstructed by associating tracks in the inner detector with energy deposits in the EM calorimeter. They must satisfy identification criteria based on the shower shape in the EM calorimeter, on the track quality, and on the detection of the transition radiation produced in the TRT detector. The EM clusters are required to be in the pseudorapidity region $|\eta| < 2.47$, excluding the transition region between the barrel and the endcap calorimeters ($1.37 < |\eta| < 1.52$). They must have a transverse energy $\ET>25 \,$ \GeV{}. The associated track must have a longitudinal impact parameter $|z_0|<2$ mm with respect to the primary vertex.
Isolation requirements, on calorimeter and tracking variables, are used to reduce the background from non-prompt electrons. The calorimeter isolation variable is based on the energy sum of cells within a cone of size $\Delta R < 0.2$ around the direction of each electron candidate. This energy sum excludes cells associated with the electron cluster and is corrected for leakage from the electron cluster itself and for energy deposits from pile-up. 
The tracking isolation variable is based on the track \pT\ sum around the electron in a cone of size $\Delta R < 0.3$, excluding the electron track. 
In every \pT\ bin both requirements are chosen to result separately in a 90\% electron selection efficiency for prompt electrons from $Z$ boson decays. 

Muon candidates are defined by matching tracks in the muon spectrometer with tracks in the inner detector. The track $\pt$ is determined through a global fit of the hits which takes into account the energy loss in the calorimeters. The track is required to have $|z_0|<2$ mm and a transverse impact parameter significance, $|d_0/\sigma(d_0)|<3$, consistent with originating in the hard interaction. Muons are required to have $\pt>25\,\mathrm{\GeV{}}$ and be within $|\eta|<2.5$.
To reduce the background from muons originating from heavy-flavour decays inside jets, muons are required to be separated by $\Delta R>0.4$ from the nearest jet, and to be isolated. They are required to satisfy the isolation requirement $I^{\ell}< 0.05$, where the isolation variable is the ratio of the sum of $\pt$ of tracks, excluding the muon, in a cone of variable size $\Delta R = 10\,\mathrm{\GeV{}}/\pt(\mu)$ to the $\pt$ of the muon~\cite{Objects_muon_isolation}. The isolation requirement has an efficiency of about $97\%$ for prompt muons from $Z$ boson decays.

Jets are reconstructed using the anti-$k_{t}$ algorithm~\cite{akt1} implemented in the {\sc FastJet} package \cite{Fastjet} with radius parameter $R = 0.4$. The jet reconstruction starts from topological clusters calibrated and corrected for pile-up effects using the jet area method \cite{Cacciari:2008gn}. A residual correction dependent on the instantaneous luminosity and the number of reconstructed primary vertices in the event~\cite{JVF_2013-083} is then applied.
They are calibrated using an energy- and $\eta$-dependent simulation-based calibration scheme, with {\sl in situ} corrections based on data~\cite{Objects_jet_calibration} and are accepted if $\pT > 25\,\mathrm{\GeV{}}$ and $|\eta| < 2.5$. 
To reduce the contribution from jets associated with pile-up, jets with $\pT < 50\,\mathrm{\GeV{}}$ are required to satisfy $|\mathrm{JVF}| > 0.5$, where JVF is the ratio of the sum of the \pT\ of tracks associated with both the jet and the primary vertex, to the sum of \pT\ of all tracks associated with the jet.
Jets with no associated tracks or with $|\eta| > 2.4$ at the edge of the tracker acceptance are always accepted.

To prevent double-counting of electron energy deposits as jets, the closest jet lying within $\Delta R < 0.2$ from a reconstructed electron is removed.
To remove leptons from heavy-flavour decays, the lepton is discarded if the lepton is found to lie within $\Delta R$ < 0.4 from a selected jet axis.

The purity of the selected \ttbar{}~sample is improved by tagging jets containing $b$-hadrons, exploiting their long decay time and the large mass. Information from
the track impact parameters, secondary vertex location and 
decay topology are combined in a neural-network-based algorithm (MV1) \cite{btag}.
The operating point used corresponds to an overall 70\% $b$-tagging
efficiency in $\ttbar$ events, and to a 
probability to mis-identify light-flavour jets of approximately 1\%. 

The missing transverse momentum $\Etmiss$ is computed from the vector sum of the transverse momenta of 
the reconstructed calibrated physics objects (electrons, photons, hadronically decaying $\tau$ leptons, jets and muons) as well as the transverse energy 
deposited in the calorimeter cells not associated with these objects \cite{atlasEtmisPerf}.
Calorimeter cells not associated with any
physics object are calibrated using tracking information before being included in the $\Etmiss$ calculation.
The contribution from muons is added using their momentum. To avoid double counting of energy, the parameterized muon energy loss in the calorimeters is subtracted in the $\Etmiss$ calculation.
 
\begin{figure*}[htbp]
\centering
\subfigure[]{ \includegraphics[width=0.38\textwidth]{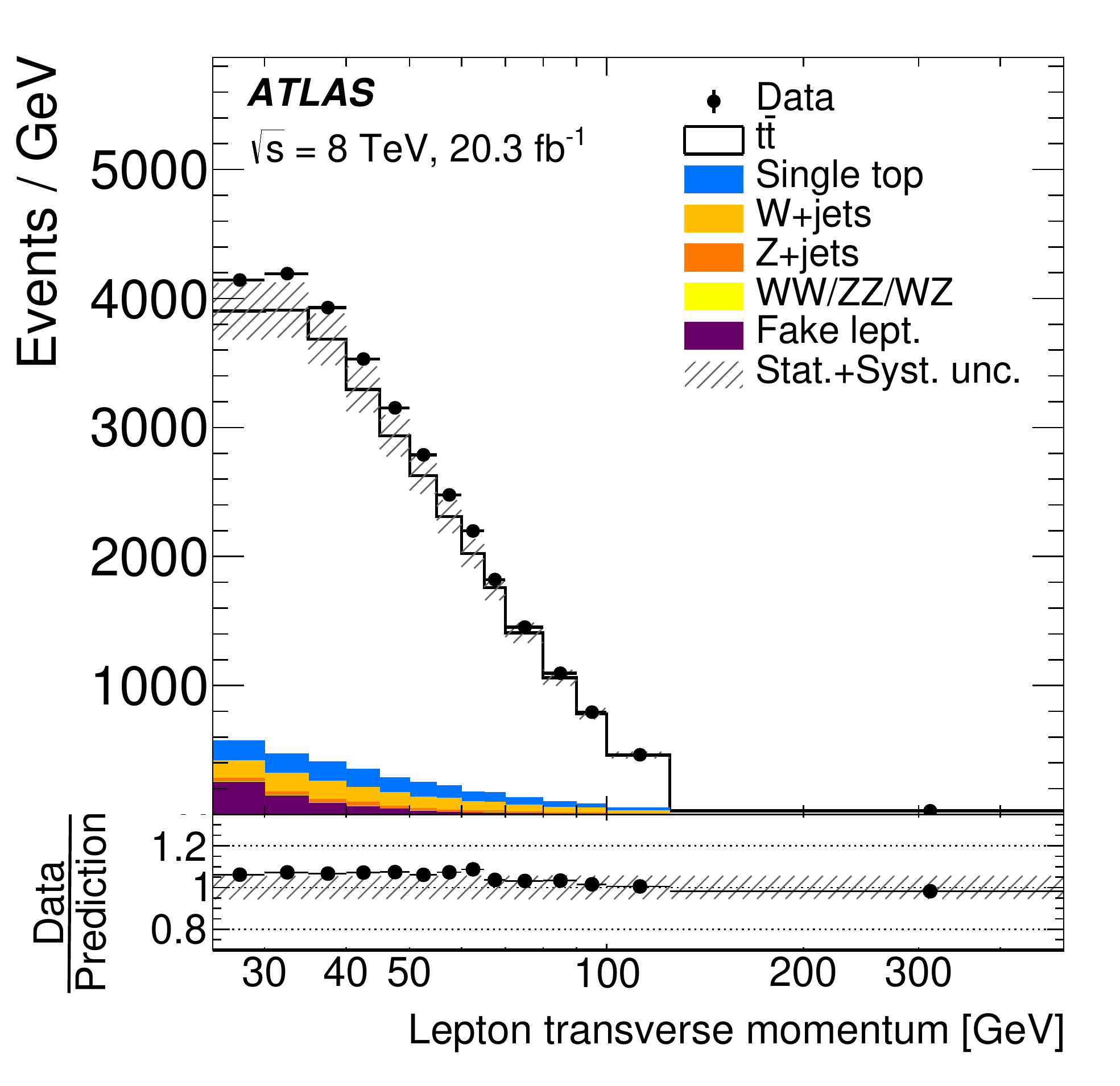}\label{fig:lep_pt_co}}
\subfigure[]{ \includegraphics[width=0.38\textwidth]{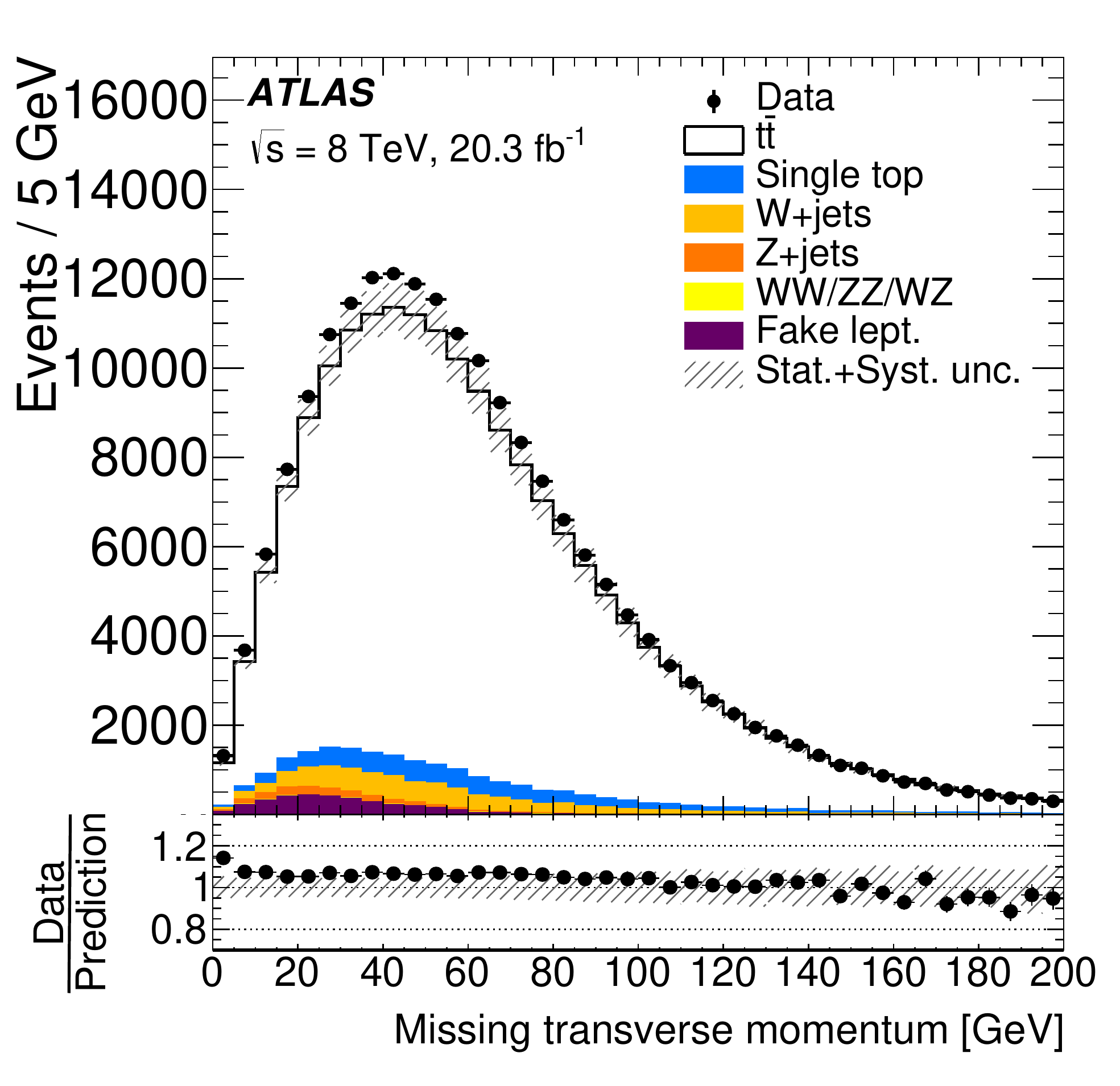}\label{fig:met_co}}
\subfigure[]{ \includegraphics[width=0.38\textwidth]{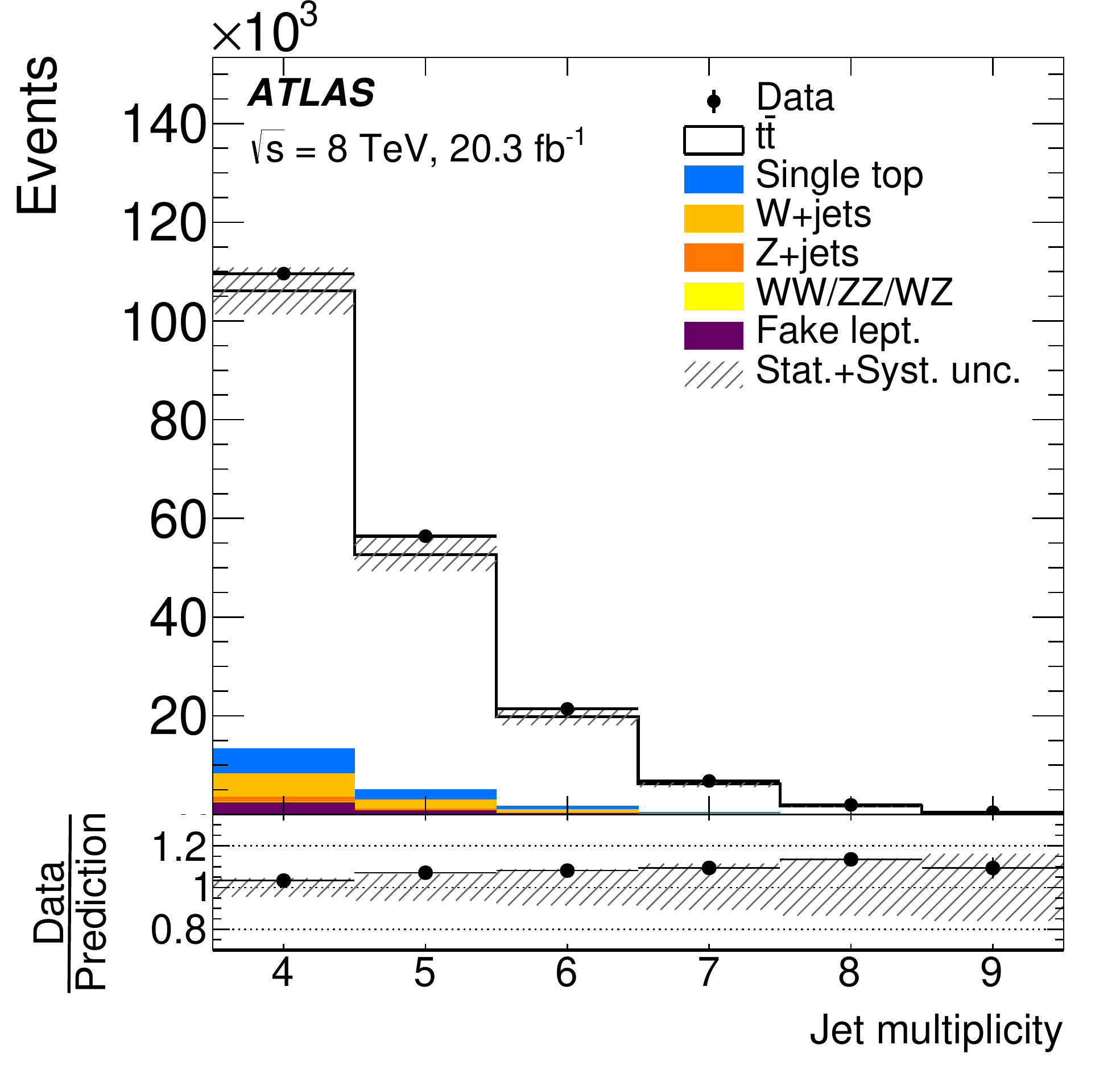}\label{fig:jet_n_co}}
\subfigure[]{ \includegraphics[width=0.38\textwidth]{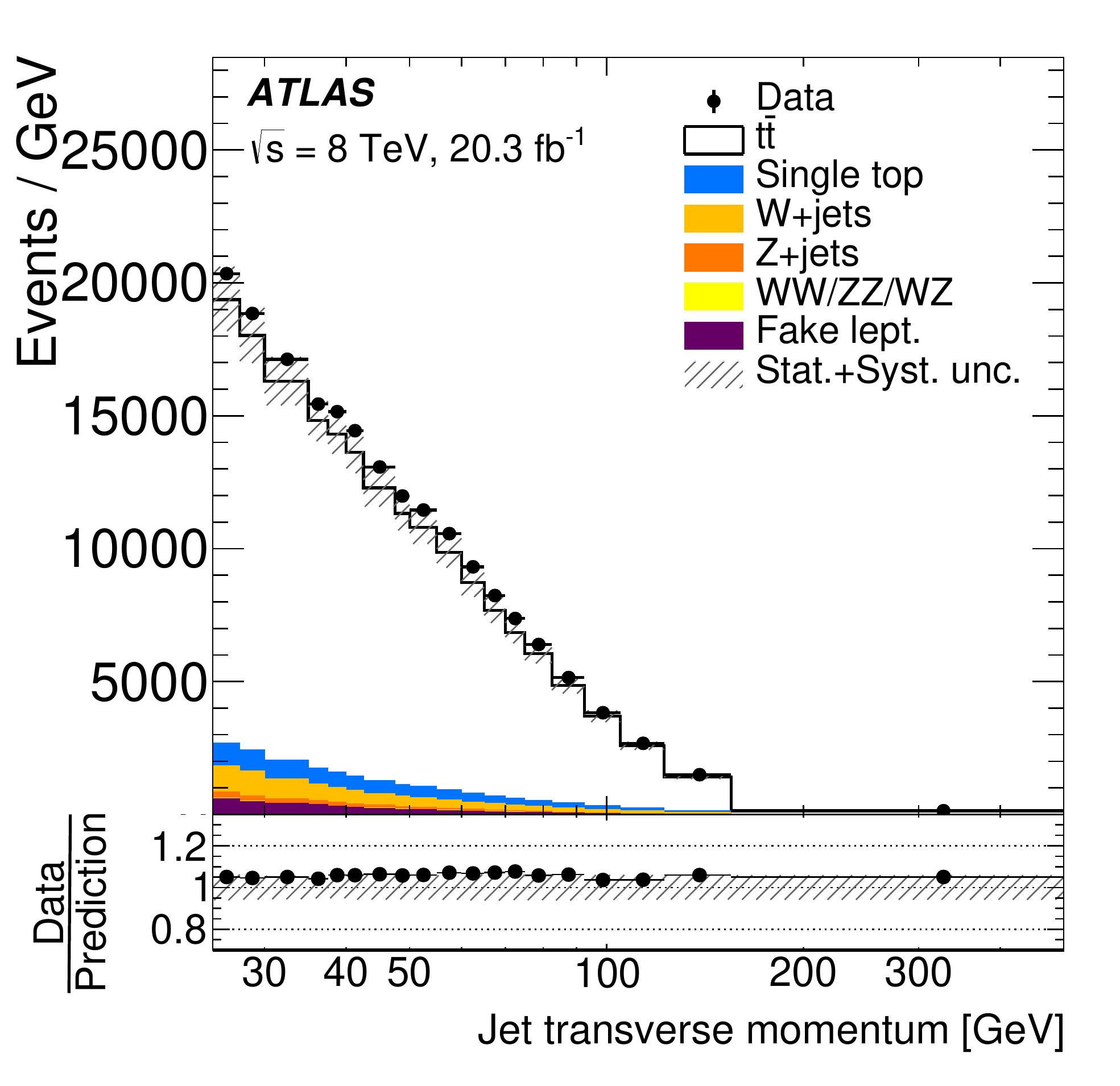}\label{fig:jet_pt_co}}
\subfigure[]{ \includegraphics[width=0.38\textwidth]{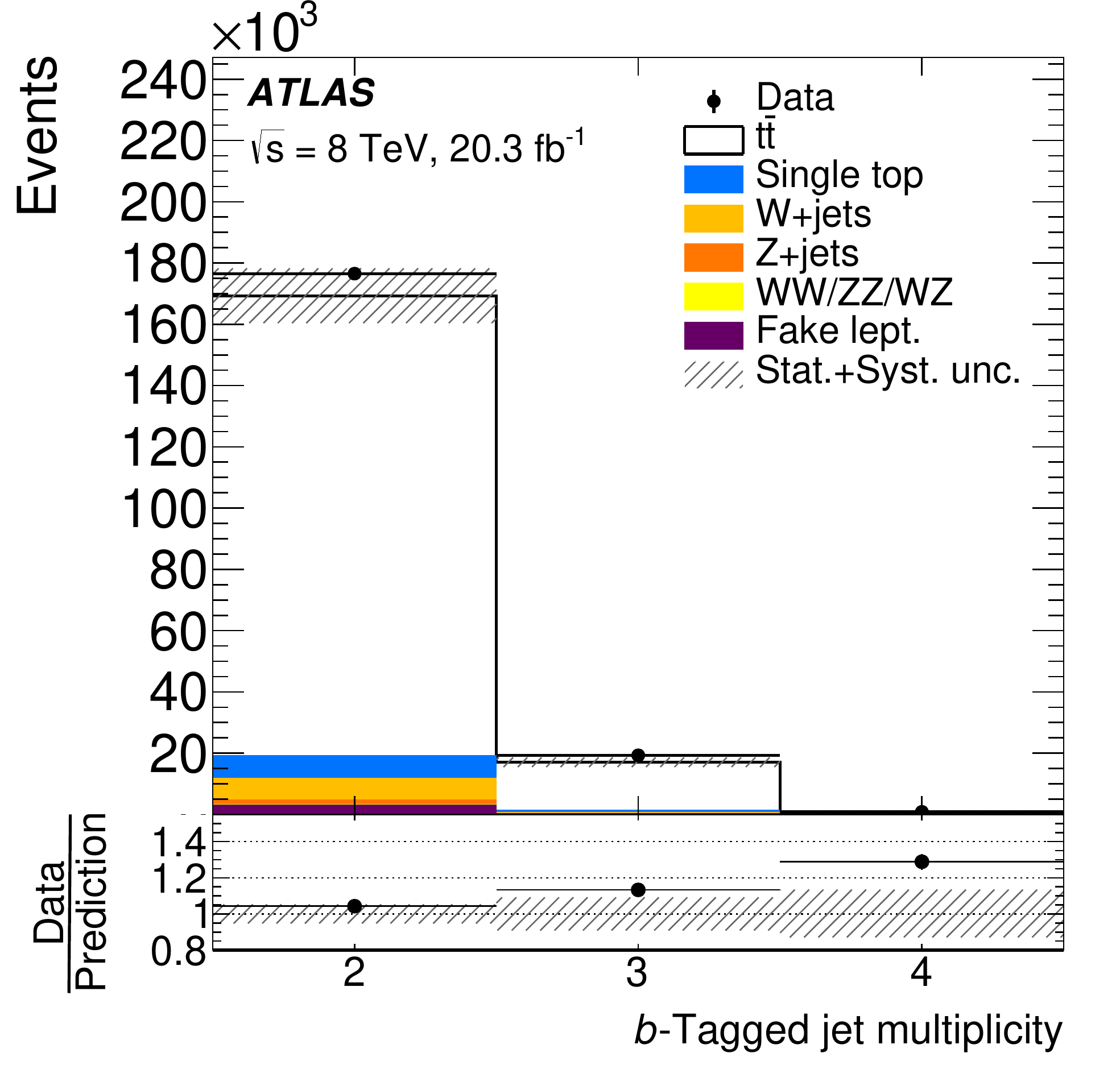}\label{fig:bjet_n_co}}
\subfigure[]{ \includegraphics[width=0.38\textwidth]{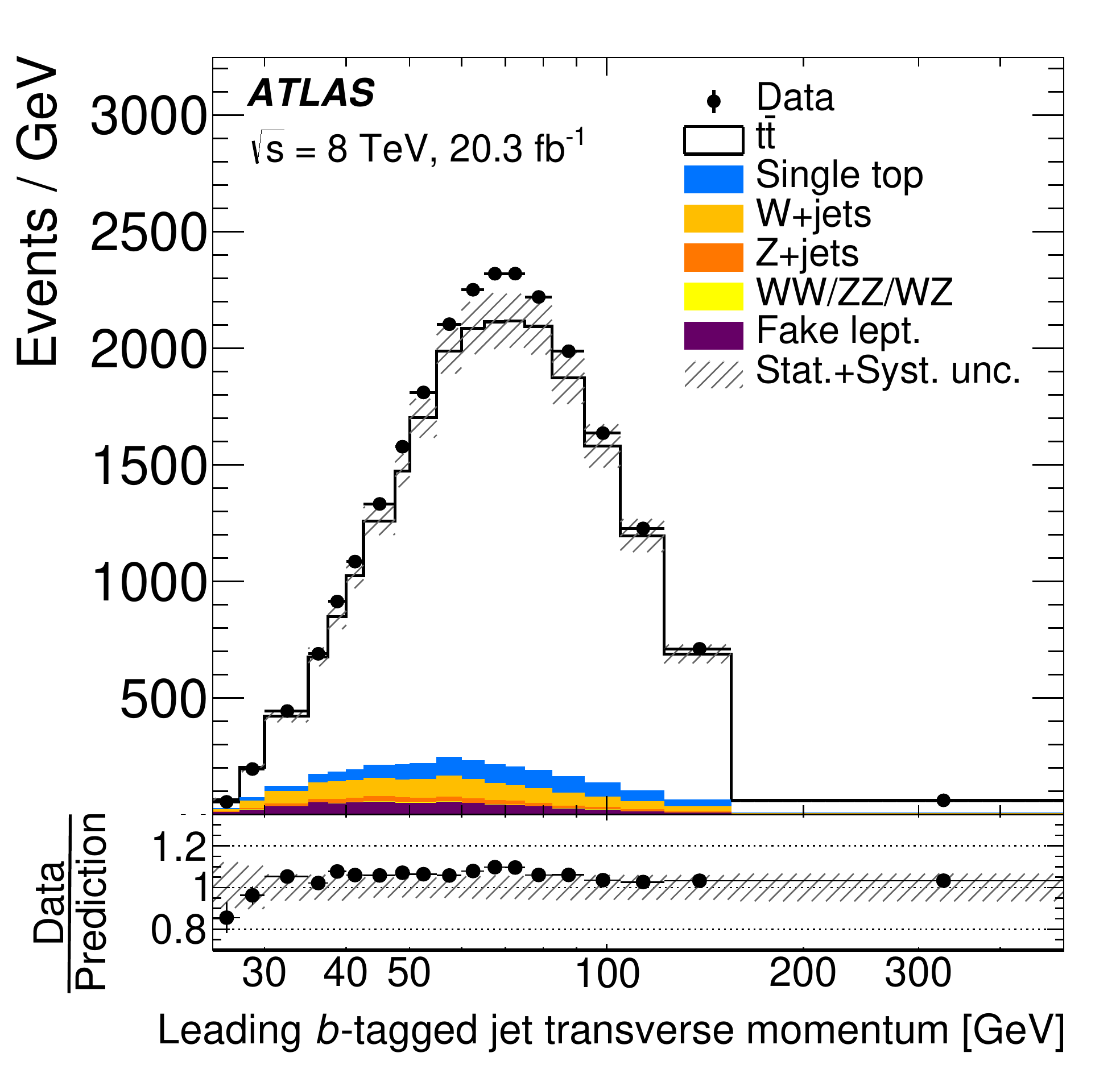}\label{fig:bjet1_pt_co}}
\caption{Kinematic distributions of the combined electron and muon selections at the detector level: \subref{fig:lep_pt_co}~lepton transverse momentum and \subref{fig:met_co}~missing transverse momentum \Etmiss{}, \subref{fig:jet_n_co}~jet multiplicity, \subref{fig:jet_pt_co}~jet transverse momentum, \subref{fig:bjet_n_co}~$b$-tagged jet multiplicity and \subref{fig:bjet1_pt_co}~leading $b$-tagged jet \pt{}. Data distributions are compared to predictions using \Powheg{}+\Pythia{} as the \ttbar{} signal model. The hashed area indicates the combined statistical and systematic uncertainties on the total prediction, excluding systematic uncertainties related to the modelling of the $\ttbar$ system.}
\label{fig:controls_4j2b_detector}
\end{figure*}

\subsection{Event selection at detector level}\label{sec:EventSelection}
The event selection consists of a set of requirements based on the general event quality and on the reconstructed objects, defined above, that characterize the final-state event topology.
Each event must have a reconstructed primary vertex with five or more associated tracks.
The events are required to contain exactly one reconstructed
lepton candidate with $\pt > 25\,\mathrm{\GeV{}}$ geometrically matched to a corresponding object at the trigger level and at least four jets with  $\pt > 25\,\mathrm{\GeV{}}$ and $|\eta| < 2.5$. At least two of the jets have to be tagged as $b$-jets. The event selection is summarized in Table~\ref{tab:selectionList}. The event yields are displayed in Table~\ref{tab:yields} for data, simulated signal, and backgrounds (the background determination is described in Section~\ref{sec:BackgroundDetermination}). Figure~\ref{fig:controls_4j2b_detector} shows, for some key distributions, the comparison between data and predictions normalized to the data integrated luminosity. The selection produces a quite clean \ttbar sample, the total background being at the 10\% level. The difference between data and predicted event yield is $\sim 7\%$, in fair agreement with the theoretical uncertainty on the \ttbar total cross-section used to normalize the signal MC simulation (see Section~\ref{sec:DataSimSamples}).

\begin{table}[!ht]
\begin{center}
\begin{tabular}{ l | l}
\toprule
Cut & Event selection \\
\hline
\hline
Single lepton 			& {\ } Electrons (isolated): \pt{} > 60 (24) \GeV{} \\
trigger					& {\ } Muons (isolated): \pt{} > 36 (24) \GeV{} \\
\hline
Primary vertex 		& {\ }$\ge5$ tracks with \pt{} $ >0.4\,$\GeV{} \\
\hline
Exactly one 		& {\ }Muons: \pt{} $> 25\,$\GeV{}, $|\eta| <  2.5$ \\
isolated lepton 	& {\ }Electrons: \pt{} $>$ $25\,$\GeV{} \\
 					& {\ }$|\eta|< 2.47$, excluding $1.37 < |\eta| < 1.52$ \\
\hline
Jets  			& {\ }$\ge$4 jets \pt{} $> 25\,$\GeV{}, $|\eta| < 2.5$ \\
				& {\ }$\geq 2$ $b$-tagged jets at $\epsilon_b=70\%$ \\
\hline\hline
\end{tabular}
\end{center}
\caption{Summary of all requirements included in the event selection.}
\label{tab:selectionList}
\end{table}

\begin{table}[!ht]
\begin{center}
 \sisetup{
table-number-alignment=left,
separate-uncertainty=true,
table-figures-integer = 8,
table-figures-decimal = 0
}
 \begin{tabular}{
 				l
                S[separate-uncertainty=true,table-figures-uncertainty=1]
                S[separate-uncertainty=true,table-figures-uncertainty=1]
                S
               }
      \toprule
		    & {\ejets} & {\mujets} \\
    \hline\hline
    \vspace{.10cm}
 $t\bar{t}$ ${}^{\phantom{1}^{\phantom{1}}}$ 		& 74000 \pm 4700 & 92000 \pm 5900 &\\
 \vspace{.10cm}
 Single top											& 3600 \pm 200	 &  4400 \pm 300 &\\
  \vspace{.10cm}
 $W+$jets											& 3000 \pm 300  &   4400 \pm 400 & \\
  \vspace{.10cm}
$Z$+jets  											& 1100 \pm 600  &    570 \pm 300 &\\
  \vspace{.10cm}
$WW$/$WZ$/$ZZ$   									&   73 \pm 40  &      67 \pm 35 &\\
  \vspace{.10cm}
 Non-prompt and fake lept.							& 2000 \pm 900  &   1400 \pm 600 &\\
\hline
\vspace{.10cm}
 Prediction${}^{\phantom{1}^{\phantom{1}}}$			& 84000 \pm 4900 & 103000 \pm 6000 &\\
  \vspace{.10cm}
 Data 		 										& 89413			& 108131  &\\ 
    \hline \hline
  \end{tabular}

\end{center}
\caption{Event yields in the \ejets{} and \mujets{} channels after the selection. The signal model, denoted $t\bar{t}$ in the table, is generated using \PowHeg{}+\Pythia{}. The quoted uncertainties represent the sum in quadrature of the statistical and systematic uncertainties on each subsample. Neither modelling uncertainties nor uncertainties on the inclusive \ttb{}~cross-section are included in the systematic uncertainties.}
\label{tab:yields}
\end{table}

\clearpage

\subsection{Particle-level objects and fiducial phase-space definition}\label{sec:TruthObjectDef}
Particle-level objects are defined for simulated events in analogy to the detector-level objects described above. Only stable final-state particles, \ie particles that are not decayed further by the generator, and unstable particles\footnote{Particles with a mean lifetime $\tau$ > 300 ps} that are to be decayed later by the detector simulation, are considered.

The fiducial phase space for the measurements presented in this paper is defined using a series of requirements applied to particle-level objects close to those used in the selection of the detector-level objects. The procedure explained in this section is applied to the \ttbar{} signal only, since the background subtraction is performed before unfolding the data.

Electrons and muons must not originate, either directly or through a $\tau$ decay, from a hadron in the MC particle record. 
This ensures that the lepton is from an electroweak decay without requiring a direct match to a $W$ boson.
The four-momenta of leptons are modified by adding the four-momenta of all photons within $\Delta R=0.1$ that do not originate from hadron decays to take into account final-state QED radiation.
Such leptons are then required to have $\pT > 25\,\mathrm{\GeV{}}$ and $|\eta| < 2.5$. Electrons in the transition region (1.37 < $\eta$ < 1.52 ) are rejected at the detector level but accepted in the fiducial selection. This difference is accounted for by the efficiency correction described in Section~\ref{sec:unfolding:fiducial}.

The particle-level missing transverse momentum is calculated from the four-vector sum of the neutrinos, discarding neutrinos from hadron decays, either directly or through a $\tau$ decay.
Particle-level jets are clustered using the anti-$k_{t}$ algorithm with radius parameter $R = 0.4$, starting from all stable particles, except for selected leptons ($e$, $\mu$, $\nu$) and the photons radiated from the leptons. Particle-level jets are required to have $\pT > 25\,\mathrm{\GeV{}}$ and $|\eta| < 2.5$.
Hadrons containing a $b$-quark with $\pT > 5\,\mathrm{\GeV{}}$ are associated with jets through a ghost matching technique as described in Ref.~\cite{Cacciari:2008gn}.
Particle $b$-tagged jets have $\pT > 25\,\mathrm{\GeV{}}$ and $|\eta| < 2.5$. 
The events are required to contain exactly one reconstructed
lepton candidate with $\pt > 25\,\mathrm{\GeV{}}$ and at least four jets with  $\pt > 25\,\mathrm{\GeV{}}$ and $|\eta| < 2.5$. At least two of the jets have to be $b$-tagged. 
Dilepton events where only one lepton passes the fiducial selection are by definition included in the fiducial measurement.

\subsection{Parton-level objects and full phase-space definition}
Parton-level objects are defined for simulated events. Only top quarks decaying directly to a $W$ boson and a $b$-quark in the simulation are considered\footnote{These particles are labelled by a status code 155 in \Herwig, 3 in \Pythia and 22 in \PythiaEight respectively.}. The full phase space for the measurements presented in this paper is defined by the set of \ttb{}~pairs in which one top quark decays semileptonically (including $\tau$ leptons) and the other decays hadronically. Events in which both top quarks decay semileptonically define the dilepton background, and are thus removed from the signal simulation.

\section{Kinematic reconstruction} \label{sec:PseudoTop}

The pseudo-top algorithm~\cite{atlasDiff3} reconstructs the kinematics of the top quarks and their complete decay chain from final-state objects, namely the charged lepton (electron or muon), missing transverse momentum, and four jets, two of which are $b$-tagged. By running the same algorithm on detector- and  particle-level objects, the degree of dependency on the details of the simulation is strongly reduced compared to correcting to parton-level top quarks.

In the following, when more convenient, the leptonically (hadronically) decaying $W$ boson is referred to as the leptonic (hadronic) $W$ boson, and the semileptonically (hadronically) decaying top quark is referred to as the leptonic (hadronic) top quark.

The algorithm starts with the reconstruction of the neutrino four-momentum. The $z$-component of the neutrino momentum is calculated using the $W$ boson mass constraint imposed on the invariant mass of the system of the charged lepton and the neutrino. If the resulting quadratic equation has two real solutions, the one with smallest absolute value of $|p_z|$ is chosen. If the determinant is negative, only the real part is considered. 
The leptonic $W$ boson is reconstructed from the charged lepton and the neutrino and the leptonic top quark is reconstructed from the leptonic $W$ and the \btagged jet closest in $\Delta R$ to the charged lepton. The hadronic $W$ boson is reconstructed from the two non-$b$-tagged jets whose invariant mass is closest to the mass of the $W$ boson. This choice yields the best performance of the algorithm in terms of the correlation between detector, particle and parton levels. Finally, the hadronic top quark is reconstructed from the hadronic $W$ boson and the other \bjet. In events with more than two $b$-tagged jets, only the two with the highest transverse momentum are considered.

\section{Background determination} \label{sec:BackgroundDetermination}

The single-top-quark background is the largest background contribution, amounting to approximately 4\% of the total event yield and 40\% of the total background estimate. 

The shape of the distributions of the kinematical variables of this background is evaluated with a Monte Carlo simulation, and the event yields are normalized to the most recent calculations of their cross-sections, as described in Section \ref{sec:DataSimSamples}. The overlap between the $Wt$ and \ttb{}~samples is handled using the diagram removal scheme \cite{SingleTopWt}. 

The $W$+jets background represents the second largest background. After the event selection, approximately $3$--$4$\% of the total event yield and 35\% of the total background estimate is due to $W$+jets events. 
The estimation of this background is performed using a combination of
MC simulation and data-driven techniques.
The {\sc Alpgen+Pythia} $W$+jets samples, normalized to
the inclusive $W$ boson NNLO cross-section, are used as a starting point while
the absolute normalization and the heavy-flavour fractions of this process, which are affected by large theoretical uncertainties, are determined from data.

The corrections for generator mis-modelling in the
fractions of $W$ boson production associated with jets of different flavour components ($W+b\bar{b}$, $W+c\bar{c}$, $W+c$) 
are estimated in a sample with the same
lepton and $\met$ selections as the signal selection, but with only
two jets and no
$b$-tagging requirements. The $b$-jet multiplicity, in conjunction
with knowledge of the $b$-tagging and mis-tag efficiency, is used
to extract the heavy-flavour fraction. This information
is extrapolated to the signal region using MC simulation, assuming constant relative rates for the signal and control regions. 

The overall $W$+jets normalization is then obtained by exploiting the expected charge asymmetry in the production of $W^+$ and $W^-$ bosons in $pp$ collisions. This asymmetry is predicted by theory~\cite{Halzen:2013bqa} and evaluated using MC simulation, while other processes in the $\ttbar$ sample are symmetric in charge except for a small contamination from single-top and $WZ$ events, which is subtracted using MC simulation. The total number of $W$+jets events in the sample can thus be estimated with the following equation:

\begin{equation}
N_{W^+} + N_{W^-} = \left(\frac{r_{\rm MC} + 1}{r_{\rm MC} - 1}\right)(D_{\rm+} - D_{\rm-}),
\label{eq:Wchargeasymm}
\end{equation}

where $r_{\rm MC}$ is the ratio of the number of events with positive leptons to the number of events with negative leptons in the MC simulation, and $D_{\rm+}$ and $D_{\rm-}$ are the number of events with positive and negative leptons in the data, respectively.

Multi-jet production processes have a large cross-section and mimic the lepton+jets signature due to jets misidentified as prompt leptons (fake leptons) or semileptonic decays of heavy-flavour hadrons (non-prompt real leptons). This background is estimated directly from data by using the matrix-method technique~\cite{atlasXsec3}. The number of background events in the signal region is evaluated by applying efficiency factors to the number of events passing the tight (signal) and loose selection. The fake leptons efficiency is measured using data in control regions dominated by the multi-jet background with the real-lepton contribution subtracted using MC simulation. The real leptons efficiency is extracted from a tag-and-probe technique using leptons from $Z$ boson decays.
Fake leptons events contribute to the total event yield at approximately the $1$--$2$\% level.

$Z$+jets and diboson events are simulated with MC generators, and the event yields are normalized to the most recent theoretical calculation of their cross-sections. The total contribution of these processes is less than 1\% of the total event yield or approximatively 10\% of the total background.

Top-quark pair events with both top quarks and anti-top quarks decaying semileptonically (including decays to $\tau$) can sometimes pass the 
event selection, contributing approximately 5\% to the total event yield. The fraction of dileptonic $\ttbar$ events
in each $\pt$ bin is estimated with the same MC sample used for the
signal modelling.
In the fiducial phase-space definition, semileptonic top-quark decays to $\tau$ leptons in lepton+jets $\ttbar$ events are
considered as signal only if the $\tau$ lepton decays leptonically.

\section{Observables}\label{sec:YieldsAndPlots}

A set of measurements of the \ttbar~ production cross-sections is presented as a function of kinematic observables. In the following, the indices {\sl had} and {\sl lep} refer to the hadronically and semileptonically decaying top quarks, respectively. The indices 1 and 2 refer respectively to the leading and sub-leading top quark, ordered by transverse momentum.  

First, a set of baseline observables is presented: transverse momentum (\ptthad) and absolute value of the rapidity (\absythad) of the hadronically decaying top quark (which was chosen over the leptonic top quark due to better resolution), and the transverse momentum (\ptttbar), absolute value of the rapidity (\absyttbar) and invariant mass (\mttbar) of the \ttbar~system. These observables, shown in Figure~\ref{fig:controls_4j2b_tt}, have been previously measured by the ATLAS experiment using the 7 \TeV{} dataset~\cite{atlasDiff2,atlasDiff3} except for \absythad which has not been measured in the full phase-space. The level of agreement between data and prediction is within the quoted uncertainties for \absythad, \mttbar{} and \ptttbar. A trend is observed in the \ptthad~distribution, which is not well modelled at high values. A fair agreement between data and simulation is observed for large absolute values of the \ttb{}~rapidity. 

Furthermore, angular variables sensitive to a $\pt$ imbalance in the transverse plane, \ie to the emission of radiation associated with the production of the top-quark pair, are employed to emphasize the central production region \cite{PhysRevLett.81.2642}. The angle between the two top quarks has been found to be sensitive to non-resonant contributions due to hypothetical new particles exchanged in the $t$-channel \cite{ATLAS:dijet8TeV}. The rapidities of the two top quarks in the laboratory frame are denoted by $y^{t,{\rm 1}}$ and $y^{t,{\rm 2}}$, while their rapidities in the $\ttbar{}$ centre-of-mass frame are $\ystar = \frac{1}{2}\left(y^{t,{\rm 1}}-y^{t,{\rm 2}} \right)$ and $-\ystar$. The longitudinal motion of the \ttbar~ system in the laboratory frame is described by the rapidity boost $\boostttbar=\frac{1}{2} \left [  y^{t,{\rm 1}} + y^{t,{\rm 2}} \right ]$ and $\chittbar=e^{2|\ystar|}$, which is closely related to the production angle. In particular, many signals due to processes not included in the Standard Model are predicted to peak at low values of \chittbar{}~\cite{ATLAS:dijet8TeV}. Finally, observables depending on the transverse momentum of the decay products of the top quark have been found to be sensitive to higher-order corrections \cite{Denner:2010jp,Bevilacqua:2010qb}.

The following additional observables are measured:
\begin{itemize}

\item The absolute value of the azimuthal angle between the two top quarks (\deltaPhittbar);

\item the absolute value of the out-of-plane momentum (\absPoutttbar), \ie the projection of top-quark three-momentum onto the direction perpendicular to a plane defined by the other top quark and the beam axis ($z$) in the laboratory frame \cite{PhysRevLett.81.2642}:

\begin{equation}
\absPoutttbar  = \left | \vec{p}^{~t, {\rm had}} \cdot \frac{\vec{p}^{~t,{\rm lep}} \times \hat{z}}{|\vec{p}^{~t,{\rm lep}}\times \hat{z}|} \right | \,;\\
\end{equation}

\item the longitudinal boost of the \ttb{}~system in the laboratory frame (\boostttbar) \cite{ATLAS:dijet8TeV};

\item the production angle between the two top quarks (\chittbar) \cite{ATLAS:dijet8TeV};

\item the scalar sum of the transverse momenta of the two top quarks (\HTttbar) \cite{Denner:2010jp,Bevilacqua:2010qb}

\item and the ratio of the transverse momenta of the hadronic $W$ boson and the top quark from which it originates (\RWtttbar) \cite{Denner:2010jp,Bevilacqua:2010qb}

\begin{equation}
\RWtttbar = \pt^{W,{\rm had}} / \pt^{t,{\rm had}}\,.
\end{equation}

\end{itemize}

These observables are shown in Figure~\ref{fig:controls_4j2b_difference} at detector level. All these variables show only modest agreement with data. In particular, at high values of \HTttbar, fewer events are observed with respect to the prediction. The longitudinal boost \boostttbar~is predicted to be less central than the data. Finally, \RWtttbar~is predicted to be lower than observed in the range $1.5$--$3.0$.

\begin{figure*}[htbp]
\centering
\subfigure[]{ \includegraphics[width=0.38\textwidth]{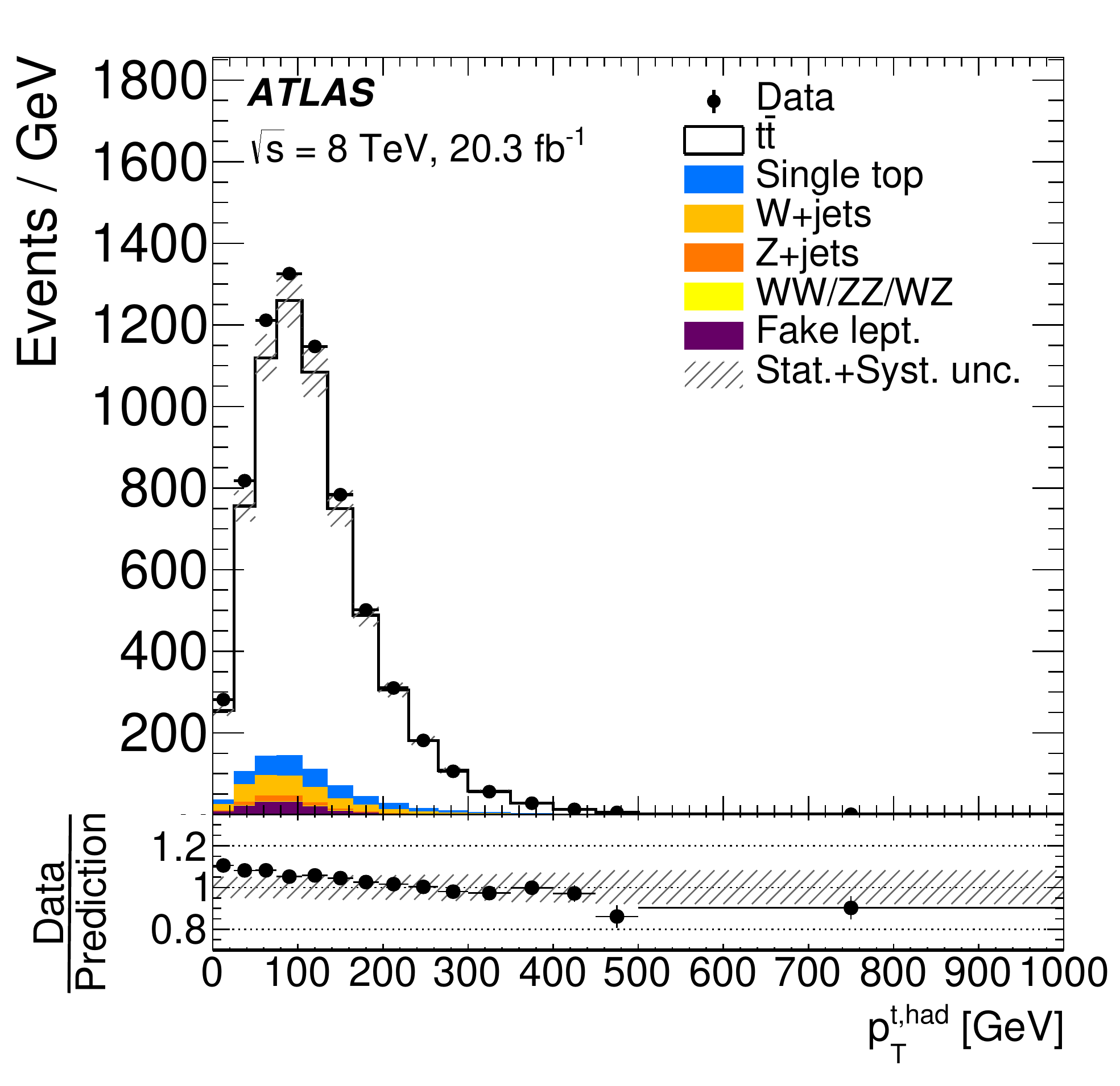}\label{fig:topH_pt_co}}
\subfigure[]{ \includegraphics[width=0.38\textwidth]{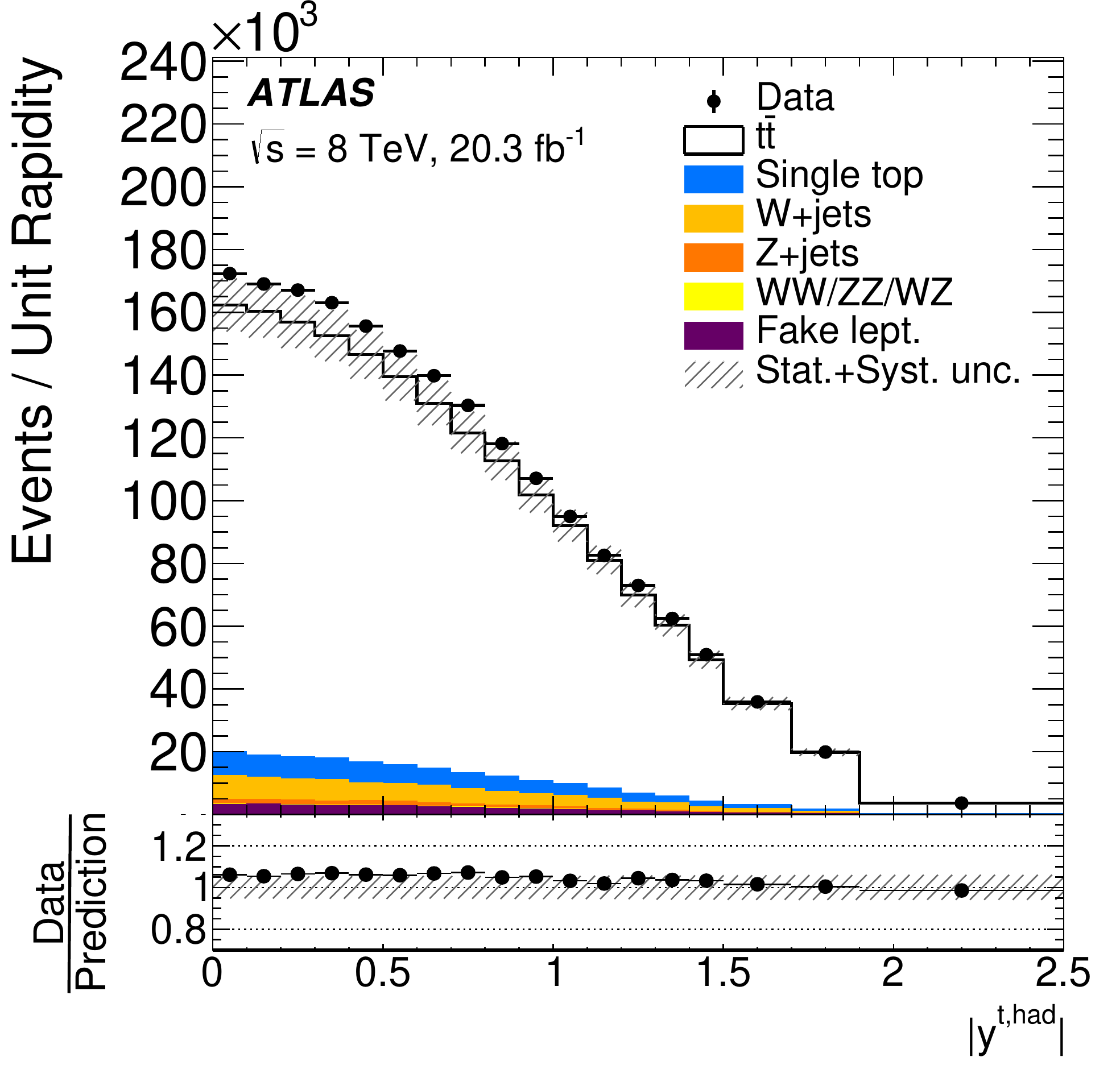}\label{fig:topH_absrap_co}}
\subfigure[]{ \includegraphics[width=0.38\textwidth]{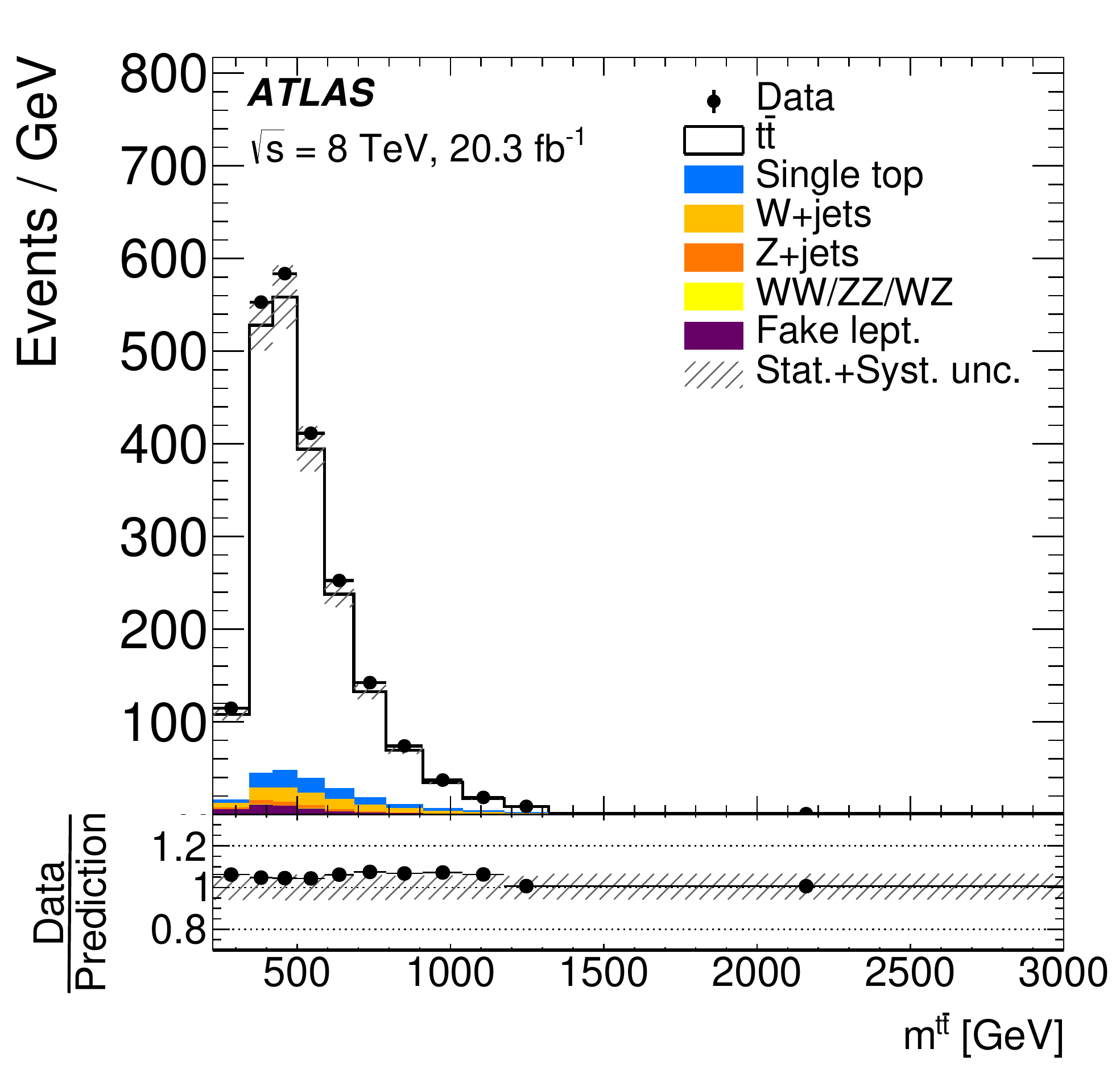}\label{fig:tt_m_co}}
\subfigure[]{ \includegraphics[width=0.38\textwidth]{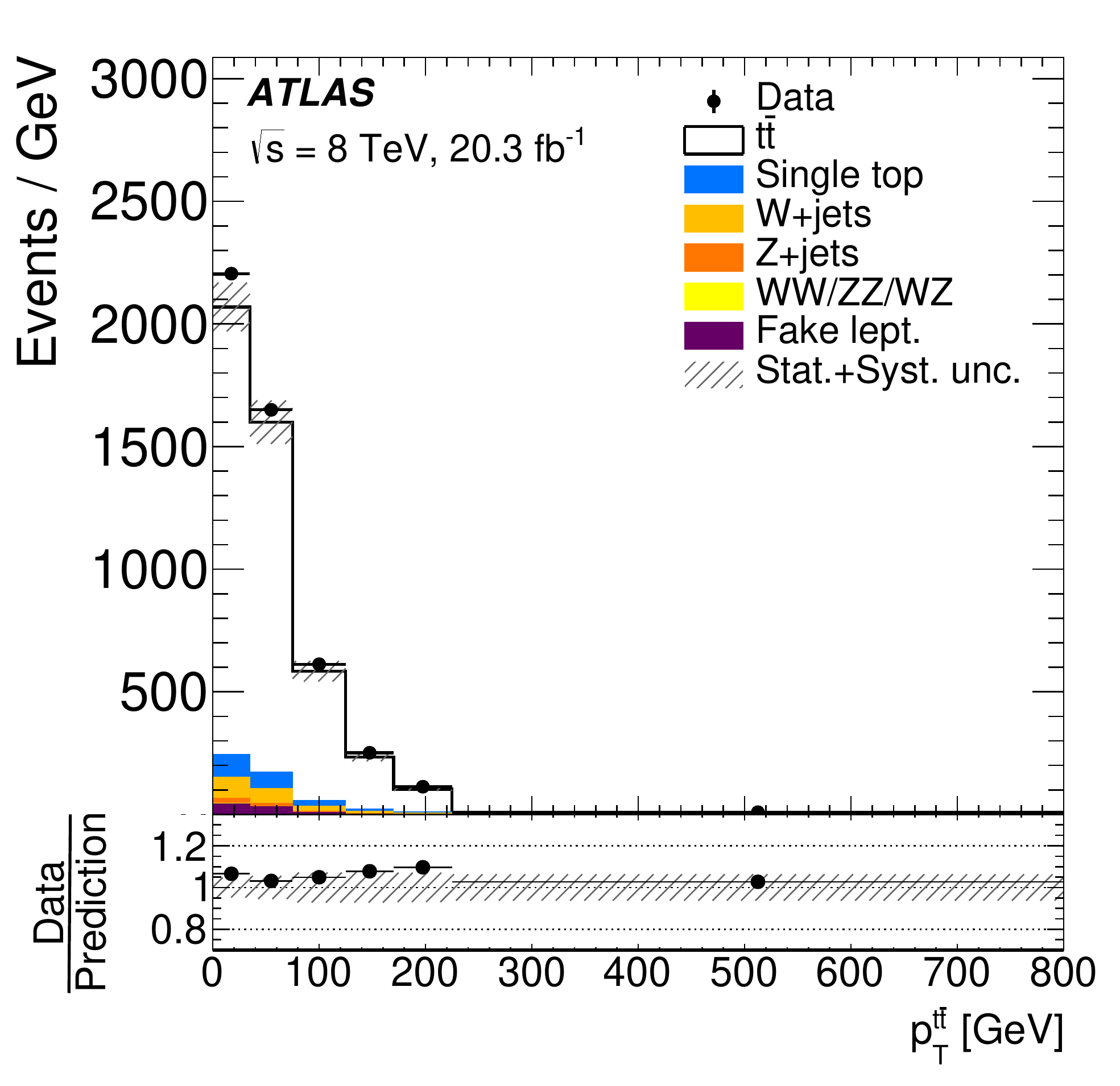}\label{fig:tt_pt_co}}
\subfigure[]{ \includegraphics[width=0.38\textwidth]{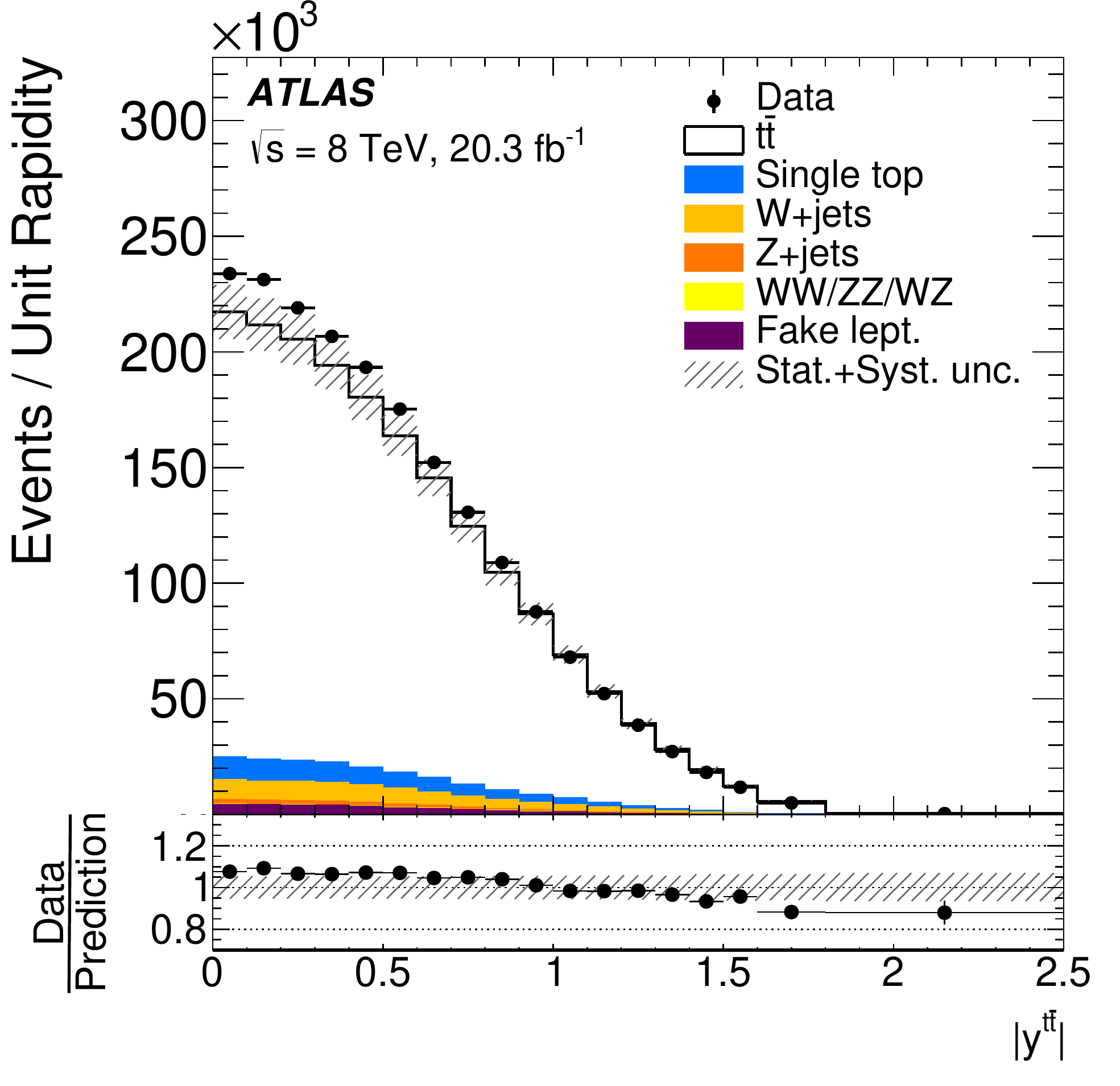}\label{fig:tt_absrap_co}}
\caption{Distributions of observables of the combined electron and muon selections at detector level: \subref{fig:topH_pt_co}~hadronic top-quark transverse momentum \ptthad{} and \subref{fig:topH_absrap_co}~absolute value of the rapidity \absythad{}, \subref{fig:tt_m_co}~\ttb{}~ invariant mass \mttbar{}, \subref{fig:tt_pt_co}~transverse momentum \ptttbar{} and \subref{fig:tt_absrap_co}~absolute value of the rapidity \absyttbar{}. Data distributions are compared to predictions, using \Powheg{}+\Pythia{} as the \ttbar{} signal model. The hashed area indicates the combined statistical and systematic uncertainties (described in Section~\ref{sec:Uncertainties}) on the total prediction, excluding systematic uncertainties related to the modelling of the $\ttbar$ system.}
\label{fig:controls_4j2b_tt}
\end{figure*}

\begin{figure*}[htbp]
\centering
\subfigure[]{ \includegraphics[width=0.38\textwidth]{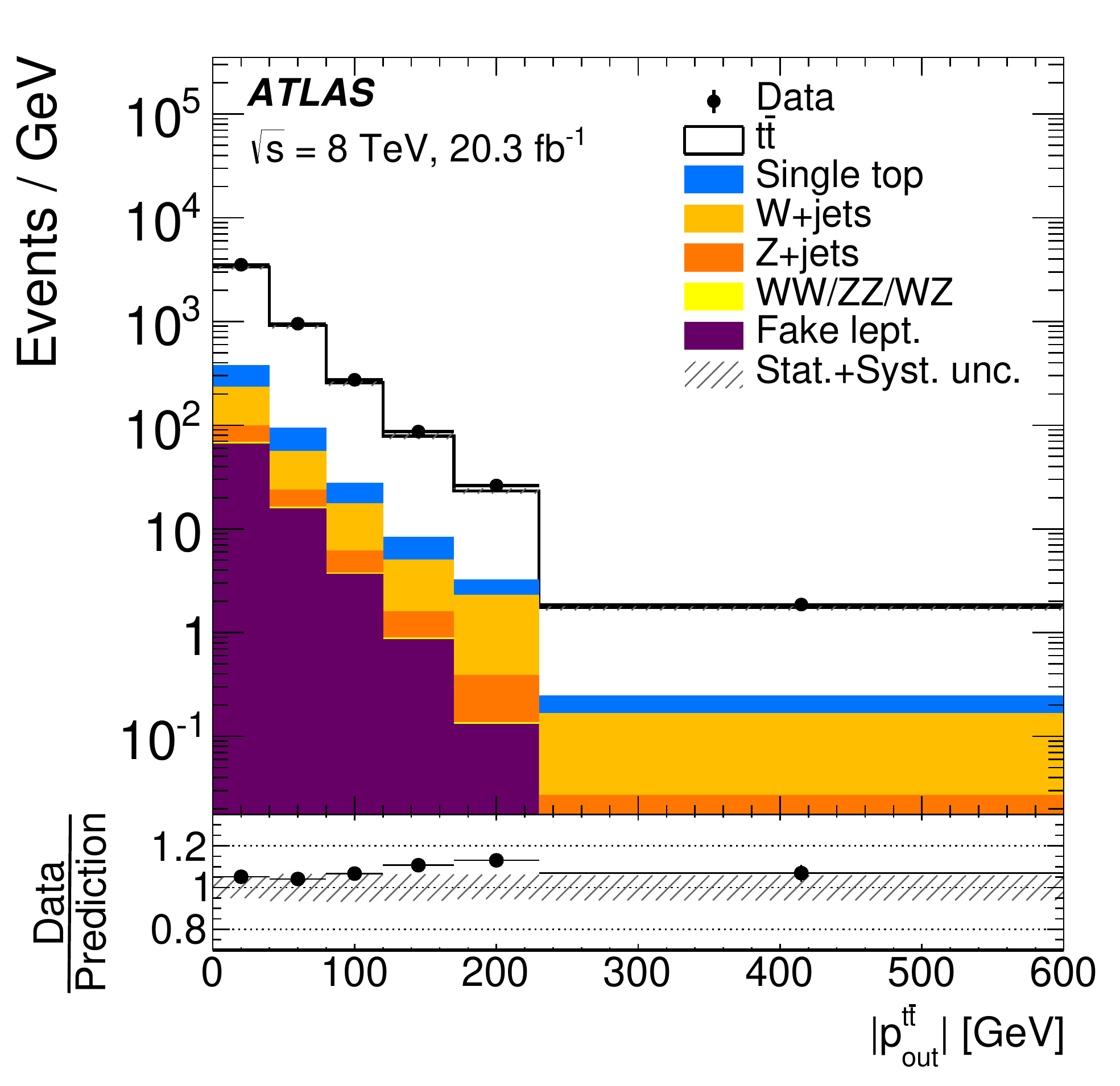}\label{fig:tt_difference_Pout}}
\subfigure[]{ \includegraphics[width=0.38\textwidth]{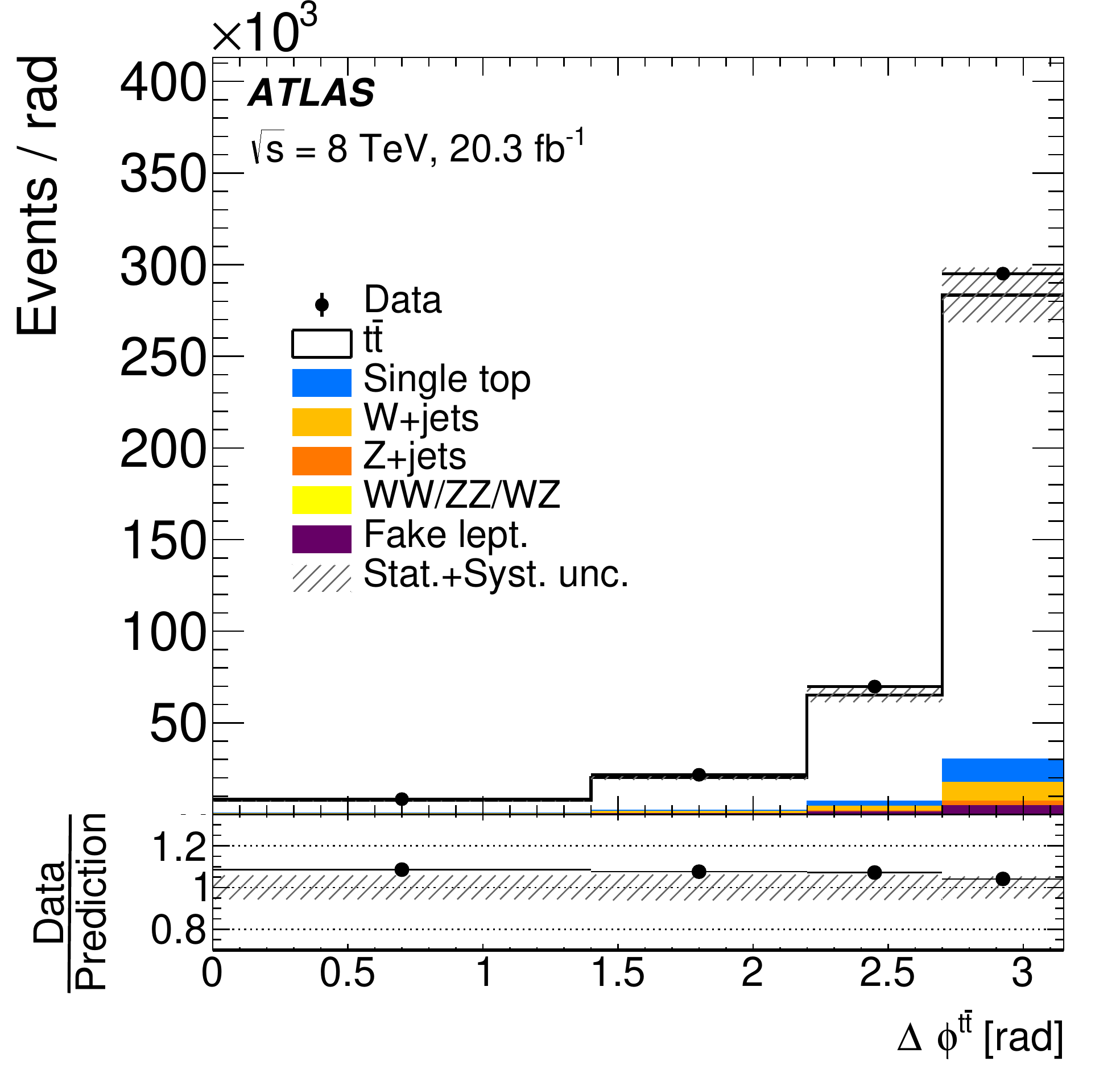}\label{fig:tt_difference_dPhi_ttbar}}
\subfigure[]{ \includegraphics[width=0.38\textwidth]{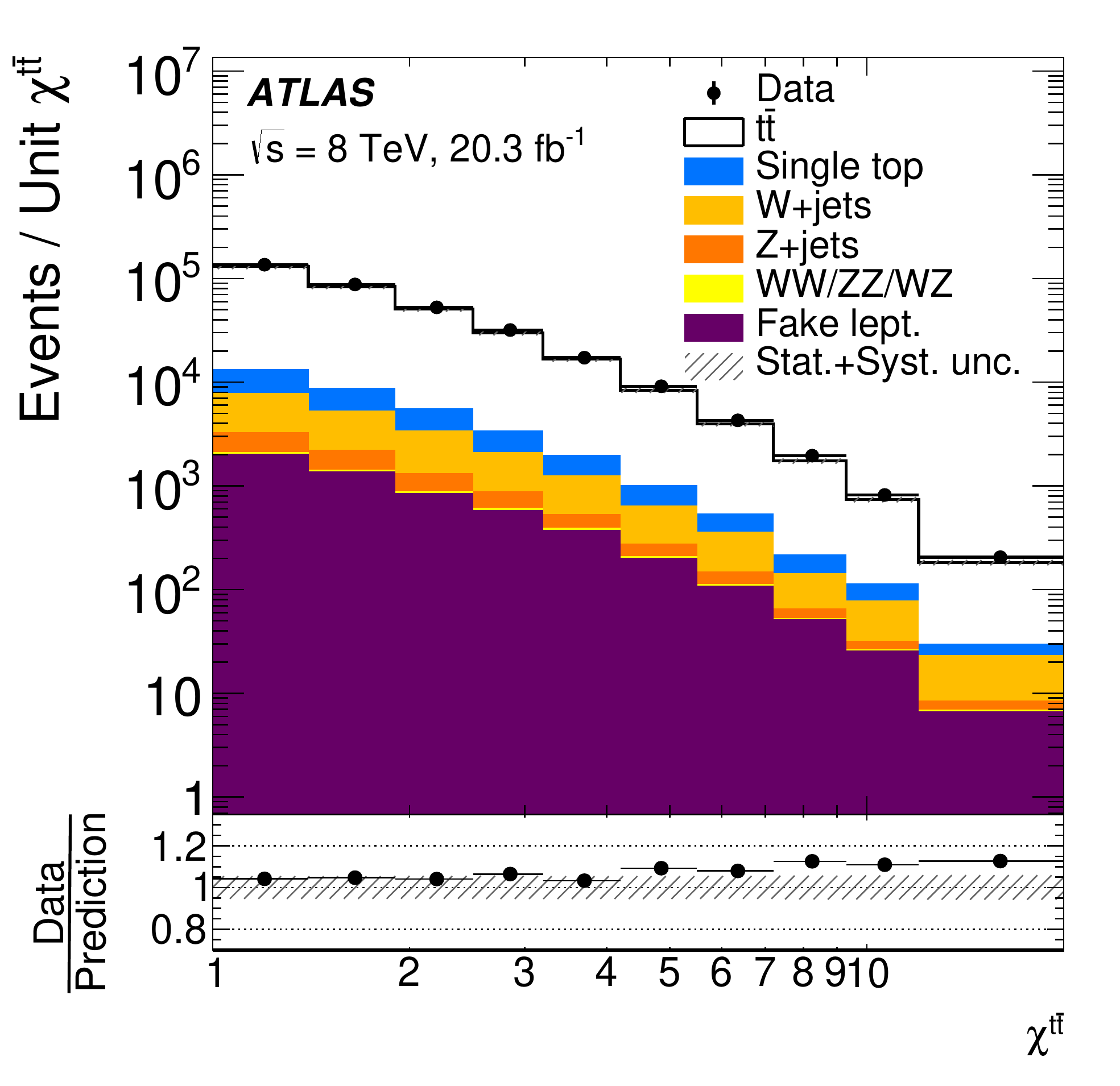}\label{fig:tt_difference_Chi_ttbar}}
\subfigure[]{ \includegraphics[width=0.38\textwidth]{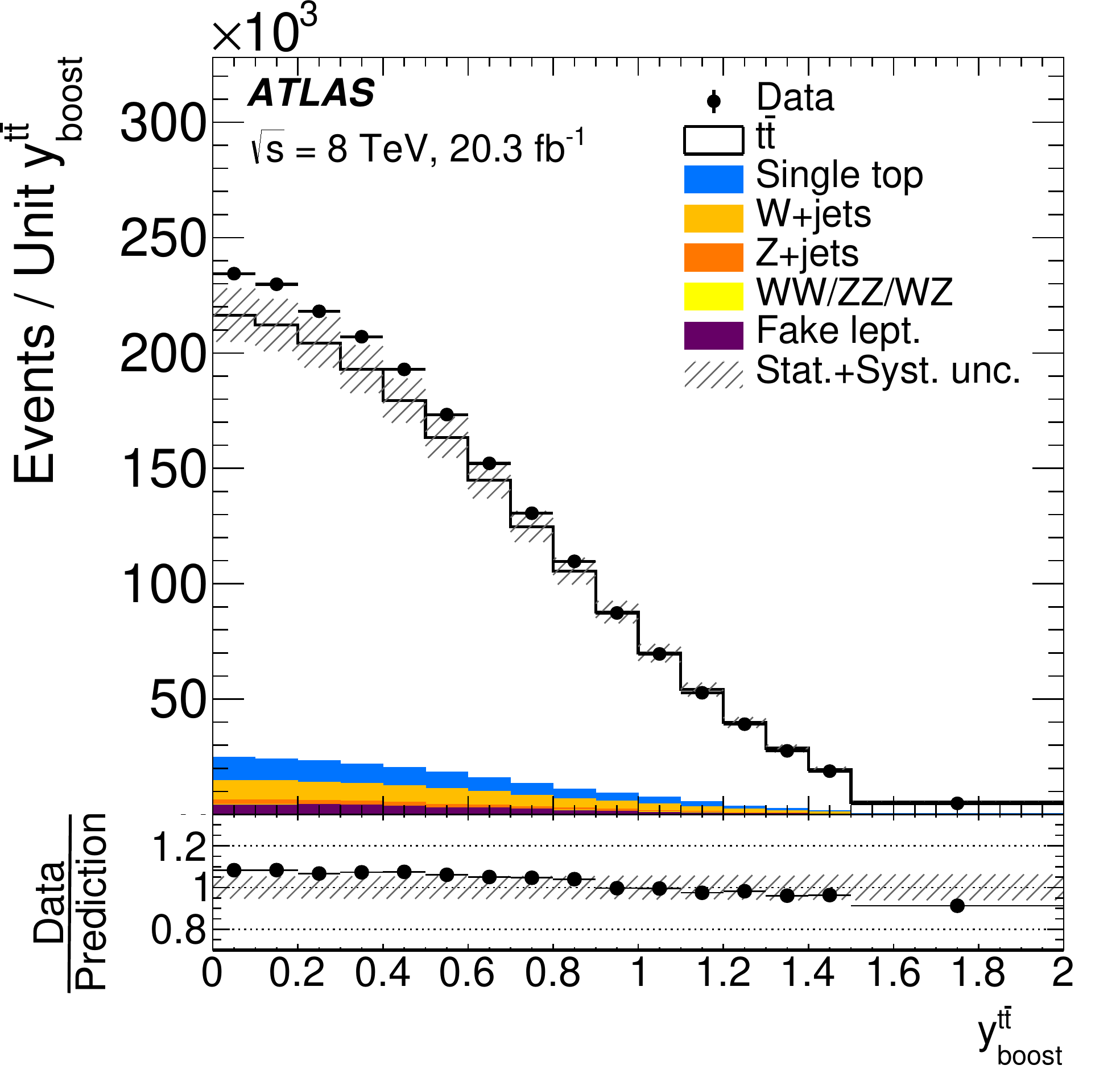}\label{fig:tt_difference_Yboost}}
\subfigure[]{ \includegraphics[width=0.38\textwidth]{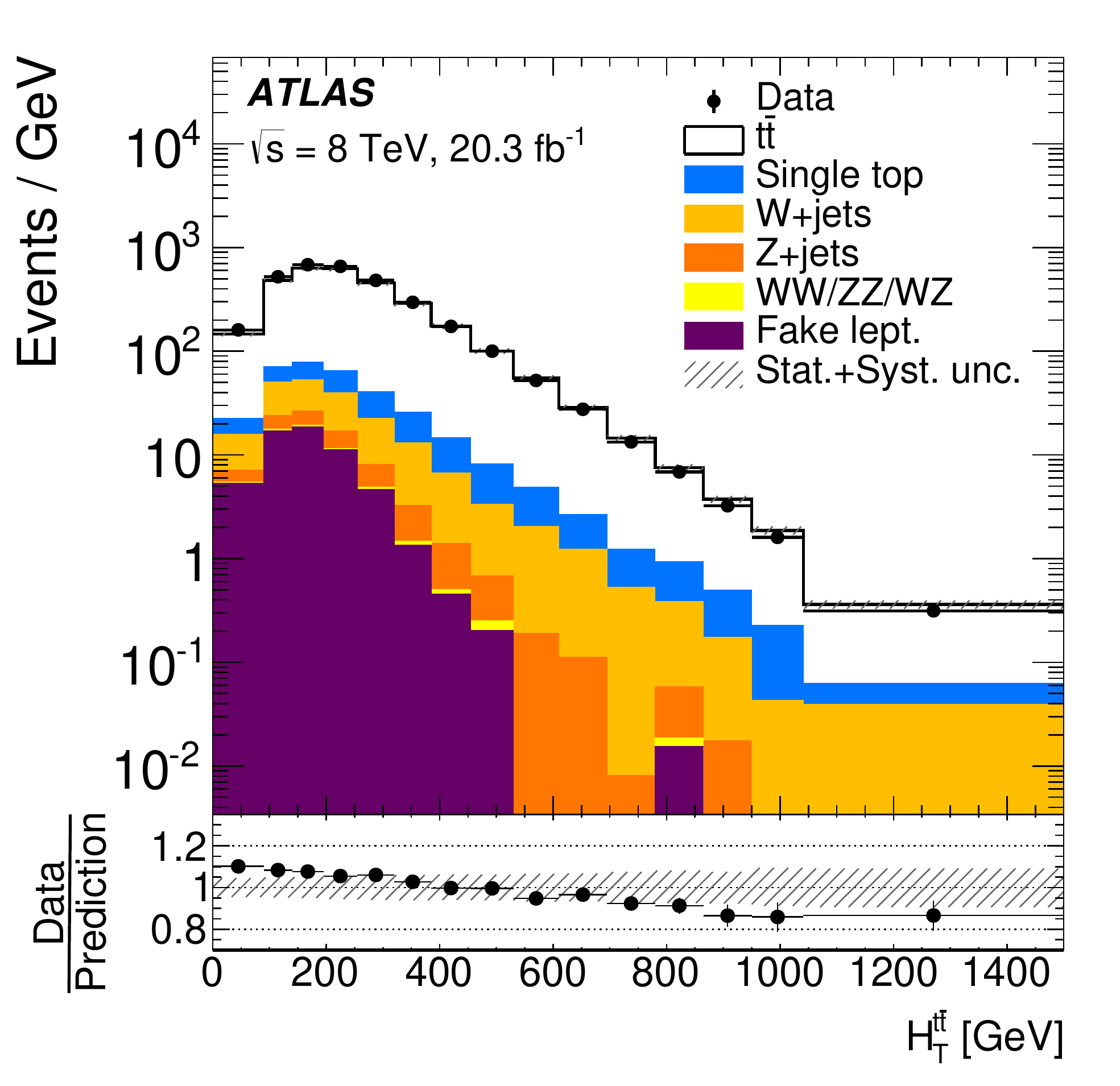}\label{fig:tt_difference_HT_ttbar}}
\subfigure[]{ \includegraphics[width=0.38\textwidth]{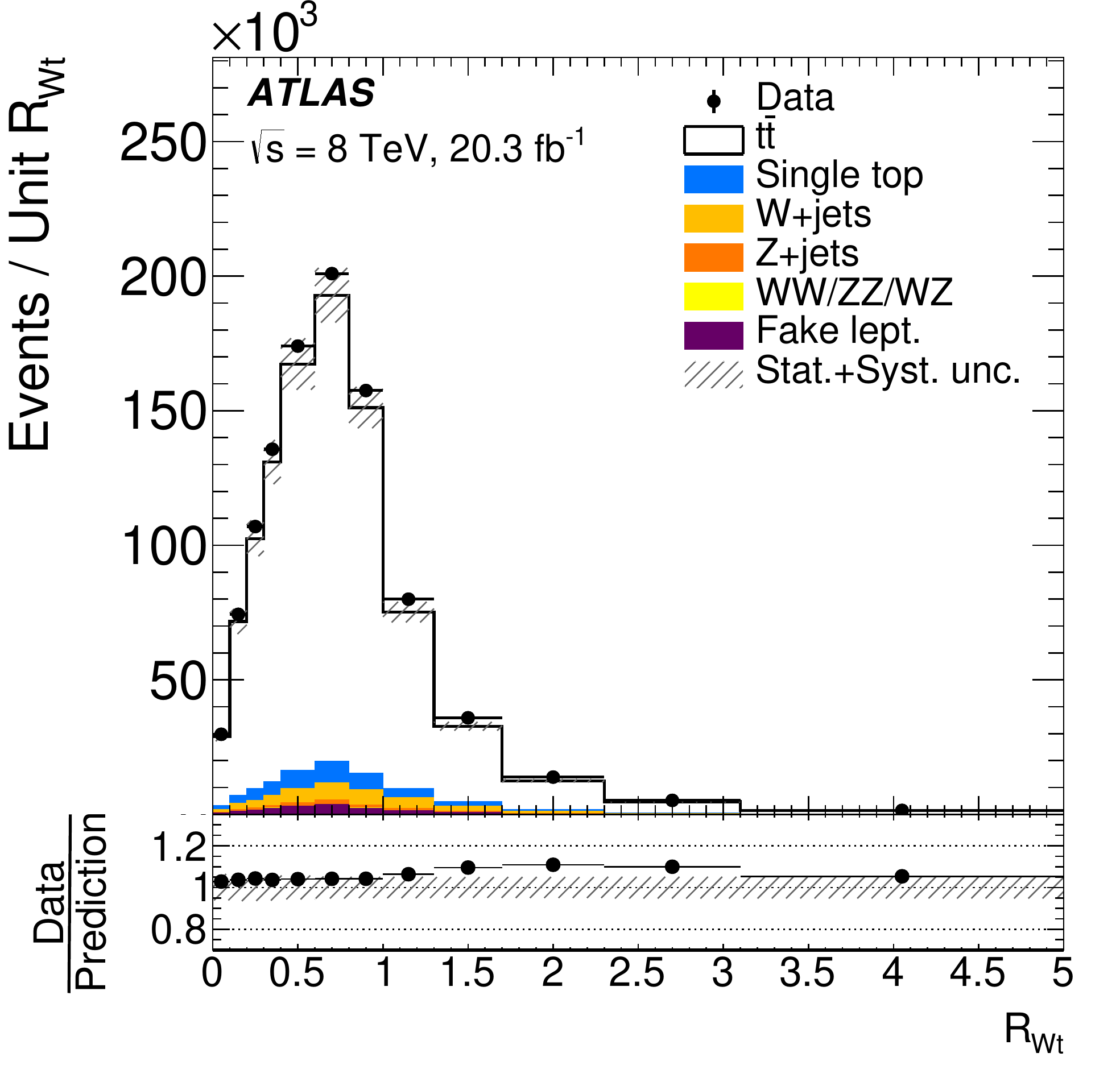}\label{fig:tt_difference_R_Wt_had}}
\caption{Distributions of observables of the combined electron and muon selections at the detector level: 
\subref{fig:tt_difference_Pout}~absolute value of the out-of-plane momentum \Poutttbar, 
\subref{fig:tt_difference_dPhi_ttbar}~azimuthal angle between the two top quarks \deltaPhittbar,
\subref{fig:tt_difference_Chi_ttbar}~production angle \chittbar,
\subref{fig:tt_difference_Yboost}~longitudinal boost \boostttbar,
\subref{fig:tt_difference_HT_ttbar}~scalar sum of hadronic and leptonic top-quarks transverse momenta and
\subref{fig:tt_difference_R_Wt_had} ratio of the hadronic $W$ boson and the hadronic top-quark transverse momenta. 
Data distributions are compared to predictions, using \Powheg{}+\Pythia{} as the \ttbar{} signal model. The hashed area indicates the combined statistical and systematic uncertainties (described in Section~\ref{sec:Uncertainties}) on the total prediction, excluding systematic uncertainties related to the modelling of the $\ttbar$ system.}
\label{fig:controls_4j2b_difference}
\end{figure*}

\clearpage

\section{Unfolding procedure}\label{sec:unfolding}

The underlying differential cross-section distributions are obtained from the detector-level events using an unfolding technique that corrects for detector effects. The iterative Bayesian method~\cite{unfold:bayes} as implemented in RooUnfold~\cite{RooUnfold} is used. 
The individual \ejets{} and \mujets{} channels give consistent results and are therefore combined by summing the event yields before the unfolding procedure.

\subsection{Fiducial phase space}\label{sec:unfolding:fiducial}
 
The unfolding starts from the detector-level event distribution ($N_{\rm reco}$), from which the  backgrounds ($N_{\rm bg}$) are subtracted first. 
Next, the acceptance correction $f_{\rm acc}$ corrects for 
events that are generated outside the fiducial phase-space but pass the detector-level selection.

In order to separate resolution and combinatorial effects, distributions evaluated using a Monte Carlo simulation are corrected to the level where detector- and particle-level objects forming the pseudo-top quarks are angularly well matched.
The matching correction $f_{\rm match}$ accounts for the corresponding efficiency.
The matching is performed using geometrical criteria based on the distance  $\Delta R$. Each particle $e$ ($\mu$) is matched to the closest detector-level $e$ ($\mu$) within $\Delta R < 0.02$. Particle-level jets are geometrically matched to the closest detector-level jet within $\Delta R < 0.4$.
If a detector-level jet is not matched to a particle-level jet, it is assumed to be either from pile-up or matching inefficiency and is ignored.
If two jets are reconstructed as being $\Delta R< 0.4$ from a single particle-level jet, the detector-level jet with smaller $\Delta R$ is matched to the particle-level jet and the other detector-level jet is unmatched.

The unfolding step uses a migration matrix ($\mathcal{M}$) derived from simulated \ttbar{} events which maps the binned generated particle-level events to the binned detector-level events.
The probability for particle-level events to remain in the same bin is therefore represented by the elements on the diagonal, and the off-diagonal elements describe the fraction of particle-level events that migrate into other bins. Therefore, the elements of each row add up to unity as shown in Figure \ref{fig:particle:migra:topH_pt}. The binning is chosen such that the fraction of events in the diagonal bins is always greater than 50\%. The unfolding is performed using four iterations to balance the goodness of fit and the statistical uncertainty. The effect of varying the number of iterations by one was tested and proved to be negligible. Finally, the efficiency correction $f_{\rm eff}$ corrects for events which pass the particle-level selection but are not reconstructed at the detector level.

All corrections are evaluated with simulation and are presented in~Figure~\ref{fig:corrs:fiducial:topH} for the case of the $\pt$ of the top quark decaying hadronically. This variable is particularly representative since the kinematics of the decay products of the top quark change substantially in the observed range. The decrease of the efficiency at high values is primarily due to the increasingly large fraction of non-isolated leptons and close or merged jets in events with high top-quark $\pt$; in order to improve the selection efficiency in this boosted kinematic region, jets with larger $R$ radius, with respect to the one used in this study, are required~\cite{CON-2014-057}. A similar effect is observed in the tail of the \ttbar{} transverse momentum and rapidity, small \deltaPhittbar~angle and high \HTttbar~ distributions. The matching corrections reach the highest values, of the order of $f_{\rm match} = 0.6$--$0.7$, at low \ttbar{} transverse momentum and large \ttbar{} rapidity. Generally, the acceptance corrections are constant and close to unity, indicating very good correlation between the detector- and the particle-level reconstruction. This is also apparent from the high level of diagonality of the migration matrices, with correlations between particle and detector levels of $85$--$95$\%.

The unfolding procedure for an observable $X$ at particle level is summarized by the expression
\begin{equation}
\frac{{\rm d}\sigma^{\rm fid}}{{\rm d}X^i} \equiv \frac{1}{\mathcal{L} \cdot \Delta X^i} \cdot  f_{\rm eff}^i \cdot \sum_j \mathcal{M}_{ij}^{-1} \cdot f_{\rm match}^j \cdot  f_{\rm acc}^j \cdot \left(N_{\rm reco}^j - N_{\rm bg}^j\right)\hbox{,}
\end{equation}
where the index $j$ iterates over bins of $X$ at detector level while the $i$ index labels bins at particle level; $\Delta X^i$ is the bin width while $\mathcal{L}$ is the integrated luminosity and the Bayesian unfolding is symbolized by $\mathcal{M}_{ij}^{-1}$.

The integrated cross-section is obtained by integrating the unfolded cross-section over the kinematic bins, and its value is used to compute the normalized differential cross-section $1/\sigma^{\rm fid}\cdot{\rm d}\sigma^{\rm fid} / {\rm d}X^i$.

\begin{figure*}[htbp]
\centering
\subfigure[]{  \includegraphics[width=0.417\textwidth]{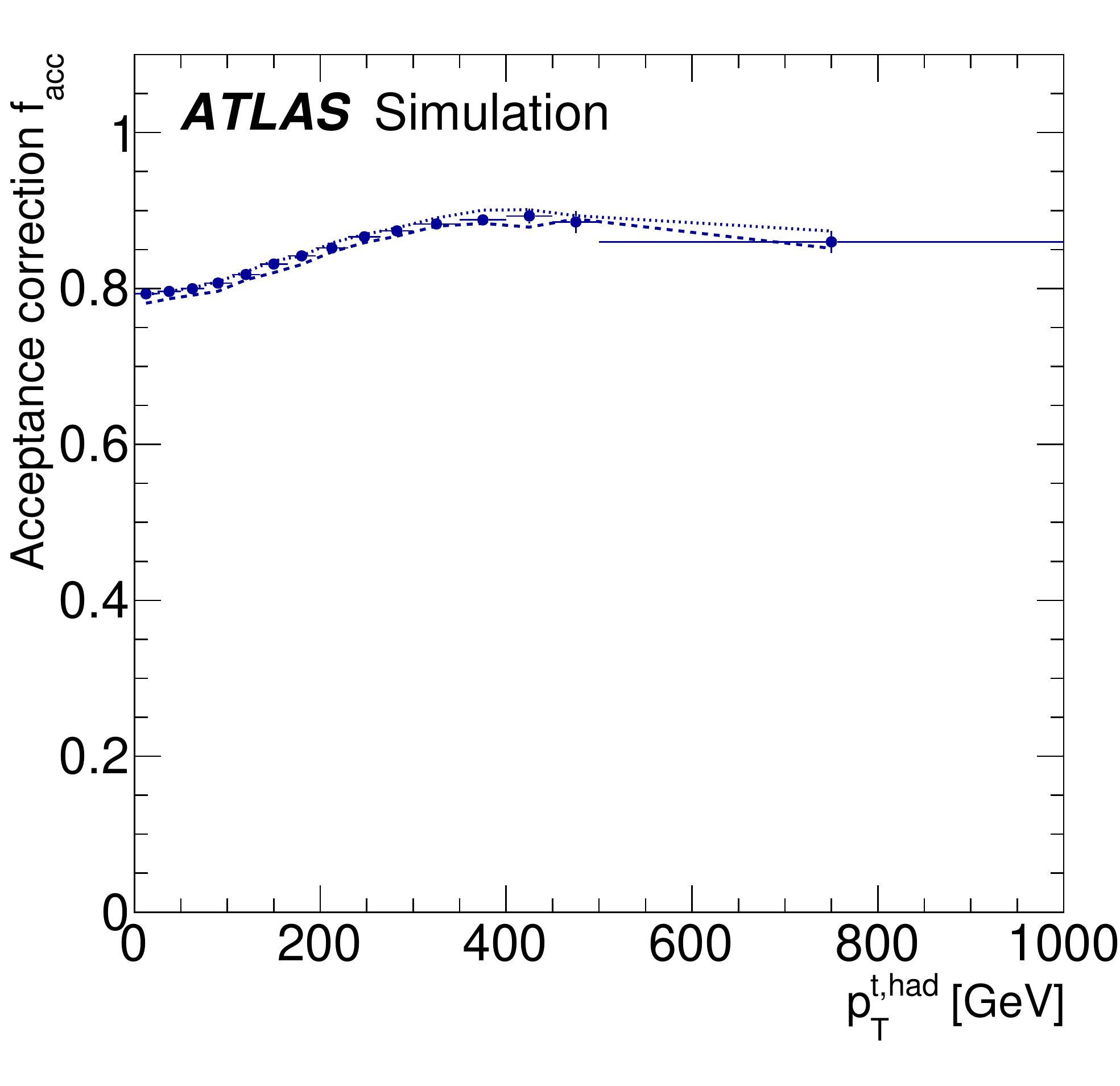}\label{fig:particle:acc:topH_pt} }
\subfigure[]{  \includegraphics[width=0.417\textwidth]{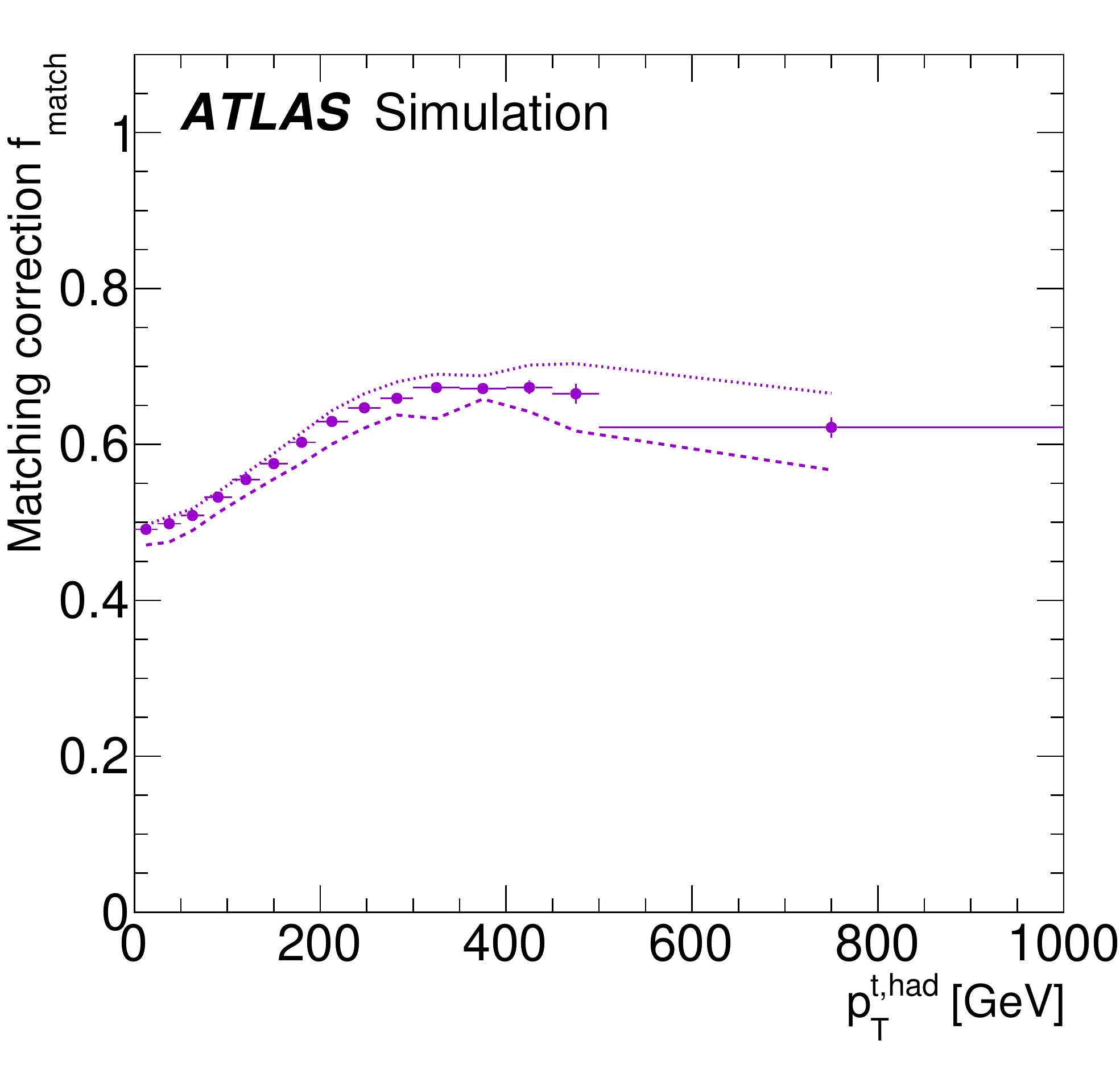}\label{fig:particle:match:topH_pt} }
\subfigure[]{  \includegraphics[width=0.417\textwidth]{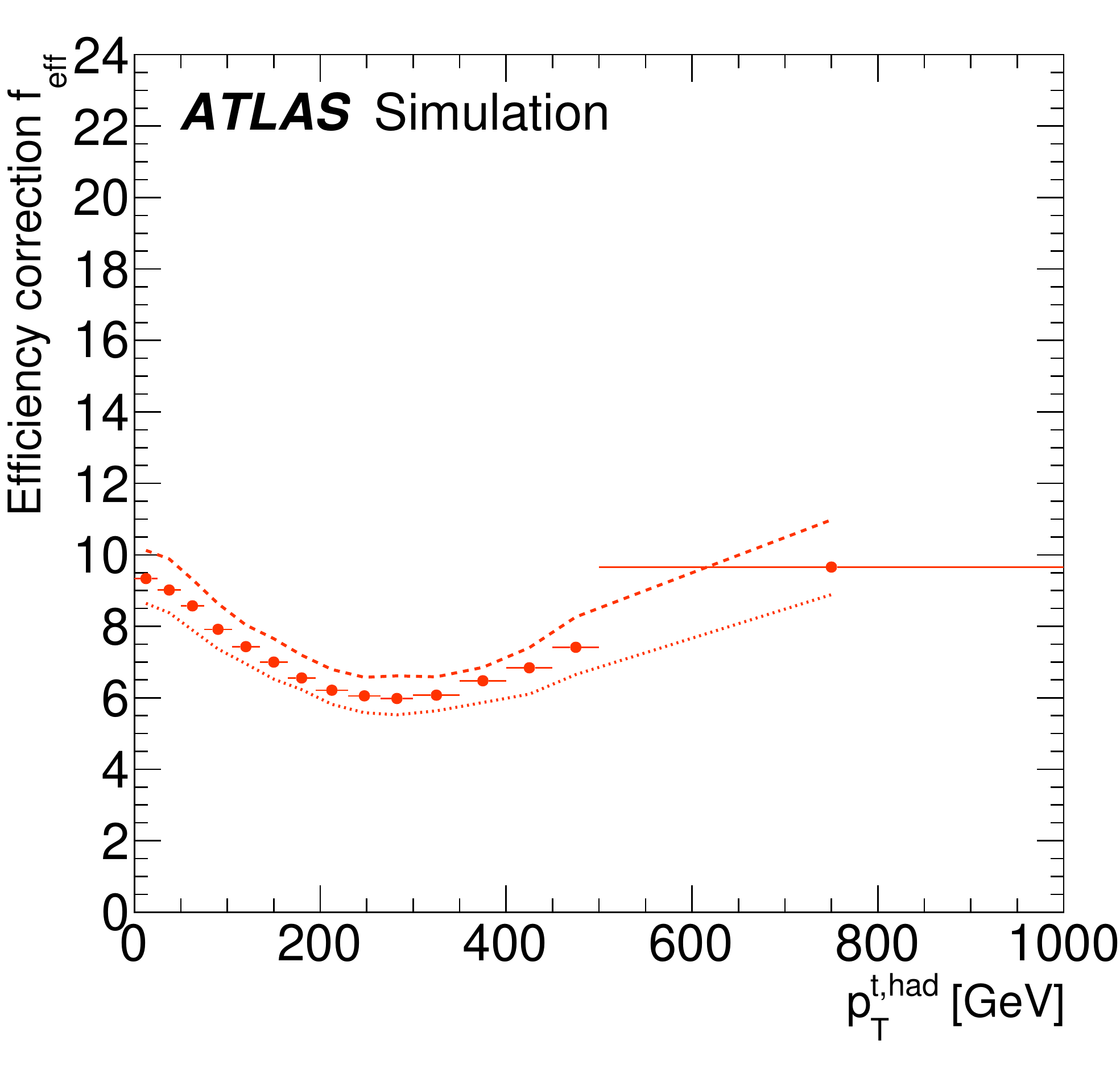}\label{fig:particle:eff:topH_pt} }
\subfigure[]{  \includegraphics[width=0.45\textwidth]{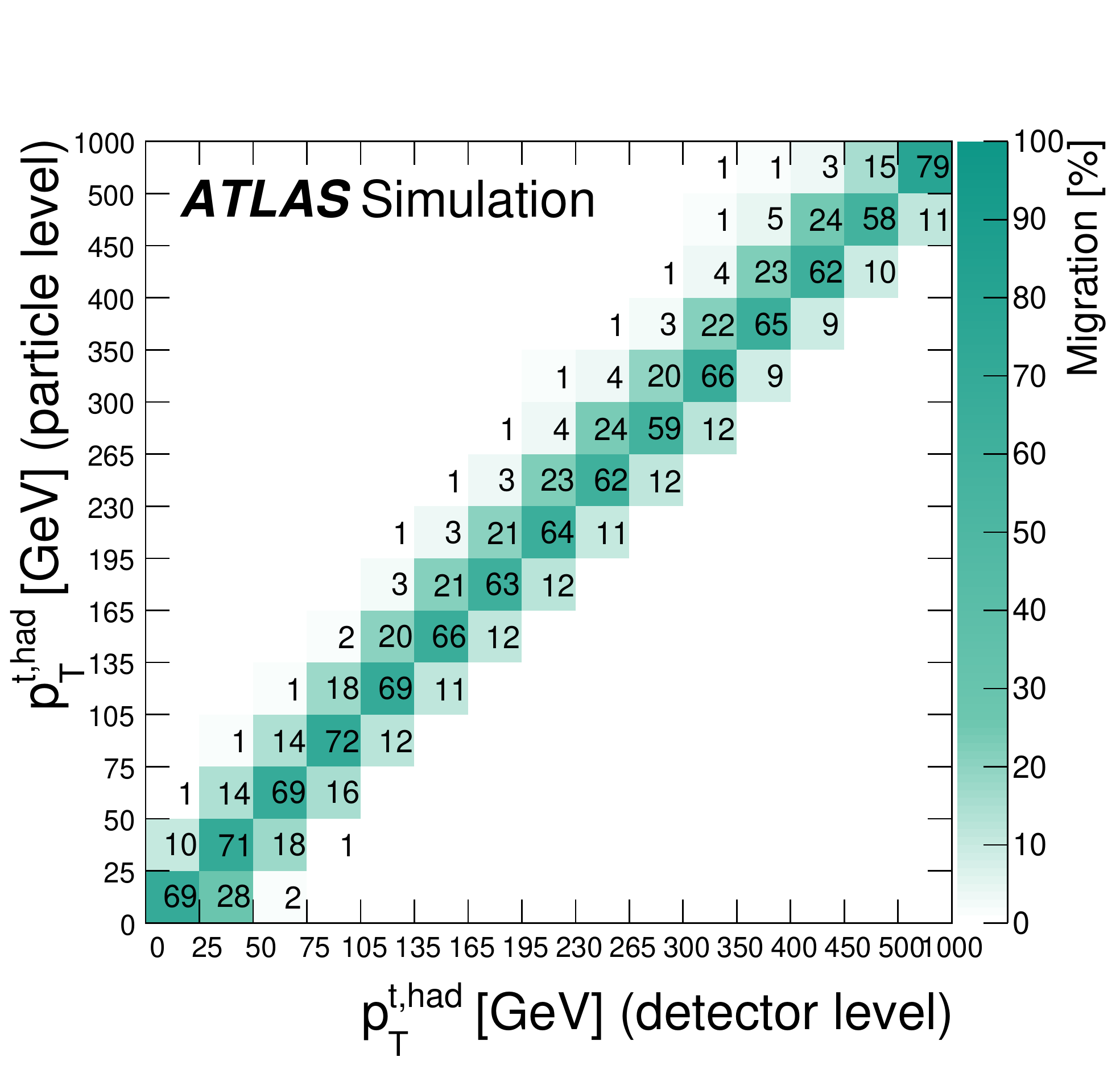}\label{fig:particle:migra:topH_pt} }

\caption{The \subref{fig:particle:acc:topH_pt} acceptance, \subref{fig:particle:match:topH_pt} matching and \subref{fig:particle:eff:topH_pt} efficiency corrections, and the \subref{fig:particle:migra:topH_pt} detector-to-particle level migration matrix for the hadronic top-quark transverse momentum evaluated with the \Powheg{}+\Pythia{} simulation sample with \HDampMT{} and using CT10nlo PDF. In Figures \subref{fig:particle:acc:topH_pt}, \subref{fig:particle:match:topH_pt} and \subref{fig:particle:eff:topH_pt} the dashed lines illustrate the corrections evaluated on alternative ISR/FSR-varied samples. In Figure \subref{fig:particle:migra:topH_pt}, the empty bins contain either no events or the number of events is less than 0.5\%.}
\label{fig:corrs:fiducial:topH}
\end{figure*}

\subsection{Full phase space}\label{sec:unfolding:full}

The measurements are extrapolated to the full phase space of the \ttb{}~ system using a procedure similar to the one described in Section~\ref{sec:unfolding:fiducial}. The only difference is in the value used for the binning. The binning used by the CMS experiment in Ref.~\cite{CmsDiff1} is used for the observables measured by both experiments to facilitate future combinations. This binning is found to be compatible with the resolution of each observable. The fiducial phase-space binning is used for all the other observables. 
In order to unambiguously define leptonic and hadronic top quarks, the contribution of \ttbar{} pairs decaying dileptonically is removed by applying a correction factor $\hat{f}_{\rm ljets}$ which represents the fraction of \ttbar{} single-lepton events in the nominal sample. The $\tau$ leptons from the leptonically decaying $W$ bosons are considered as signal regardless of the $\tau$ decay mode. The cross-section measurements are defined with respect to the top quarks before the decay (parton level) and after QCD radiation. Observables related to top quarks are extrapolated to the full phase-space starting from top quarks decaying hadronically at the detector level. 

The acceptance correction $\hat{f}_{\rm acc}$ corrects for detector-level events which are reconstructed outside the parton-level bin range for a given variable.
The migration matrix ($\hat{\mathcal{M}}$) is derived from simulated \ttbar{} events decaying in the single-lepton channel and the efficiency correction $\hat{f}_{\rm eff}$ corrects for events which did not pass the detector-level selection.

The unfolding procedure is summarized by the expression

\begin{equation}
\frac{{\rm d}\sigma^{\rm full}}{{\rm d}X^i} \equiv \frac{1}{\mathcal{L} \cdot \mathcal{B} \cdot \Delta X^i} \cdot  \hat{f}_{\rm eff}^i \cdot \sum_j \hat{\mathcal{M}}_{ij}^{-1} \cdot \hat{f}_{\rm acc}^j \cdot \hat{f}_{\rm ljets}^i \cdot\left(N_{\rm reco}^j - N_{\rm bg}^j\right)\hbox{,}
\end{equation}
where the index $j$ iterates over bins of observable $X$ at the detector level while the $i$ index labels bins at the parton level; $\Delta X^i$ is the bin width, $\mathcal{B}=0.438$ is the single-lepton branching ratio, $\mathcal{L}$ is the integrated luminosity  and the Bayesian unfolding is symbolized by $\hat{\mathcal{M}}_{ij}^{-1}$.

The integrated cross-section is obtained by integrating the unfolded cross-section over the kinematic bins, and its value is used to compute the normalized differential cross-section $1/\sigma^{\rm full}\cdot{\rm d}\sigma^{\rm full} / {\rm d}X^i$.

To ensure that the results are not biased by the MC generator used for the unfolding procedure, a study is performed in which the particle- and parton-level spectra in simulation are altered by changing the shape of the distributions using continuous functions chosen depending on the observable. The studies confirm that these altered shapes are recovered within statistical uncertainties by the unfolding based on the nominal migration matrices.

\section{Uncertainties} \label{sec:Uncertainties}

This section describes the estimation of systematic uncertainties related to object reconstruction and calibration, MC generator modelling and background estimation. 

To evaluate the impact of each uncertainty after the unfolding, the reconstructed distribution expected from simulation is varied. Corrections based on the nominal \PowHegBox{} signal sample are used to correct for detector effects and the unfolded distribution is compared to the known particle- or parton-level distribution. All detector- and background-related systematic uncertainties have been evaluated using the same generator, while alternative generators have been employed to assess modelling systematic uncertainties (\eg different parton showers). 
In these cases the corrections, derived from the nominal generator, are used to unfold the detector-level spectra of the alternative generator. The relative difference between the unfolded spectra and the corresponding particle- or parton-level spectra of the alternative generator is taken as the uncertainty related to the generator modelling. After the unfolding, each distribution is normalized to unit area.

The covariance matrices for the normalized unfolded spectra due to the statistical and systematic uncertainties are obtained by evaluating the covariance between the kinematic bins using pseudo-experiments. In particular, the correlations due to statistical fluctuations for both data and the signal are evaluated by varying the event counts independently in every bin before unfolding, and then propagating the resulting variations through the unfolding.

\subsection{Object reconstruction and calibration}\label{sec:DetMod}
The jet energy scale uncertainty is derived using a combination of simulations,
test beam data and {\it in situ} measurements~\cite{jes:2013,jer_2,Aad:2012vm}. Additional contributions
from the jet flavour composition, calorimeter response to different jet flavours, and pile-up are taken into
account. Uncertainties in the jet energy resolution are
obtained with an {\it in situ} measurement of the jet response asymmetry 
in dijet events~\cite{jer:2013}.

The efficiency to tag jets containing $b$-hadrons is corrected in
simulation events by applying $b$-tagging scale factors, extracted in 
$\ttbar$ and dijet samples, in order to
account for the residual difference between data and simulation. Scale factors are also applied for jets originating from light quarks that are mis-identified as $b$-jets.
The associated systematic uncertainties are computed by varying the scale factors 
within their uncertainties~\cite{ATLAS-CONF-2014-004,btag, CONF-2012-040}. 

The lepton reconstruction efficiency in simulation is corrected by scale factors 
derived from measurements of these efficiencies in data using a $Z \to \ell^+ \ell^-$ 
enriched control region. 
The lepton trigger and reconstruction efficiency scale factors, energy scale and resolution are varied 
within their uncertainties \cite{atlasElecPerf,muonReso}.

The uncertainty associated with $\Etmiss$ is calculated by propagating the 
energy scale and resolution systematic uncertainties to all jets and leptons in the 
$\Etmiss$ calculation. Additional $\Etmiss$ uncertainties arising from energy deposits not associated with any reconstructed objects are also included \cite{atlasEtmisPerf}.

\subsection{Signal modelling}\label{sec:SigMod}
The uncertainties of the signal modelling
affect the kinematic properties of simulated $\ttbar$ events and reconstruction
efficiencies. 

To assess the uncertainty related to the generator, events simulated with \MCatNLO{}+\Herwig{} are unfolded using the migration matrix and correction factors derived from the \PowHeg{}+\Herwig{} sample. The difference between the unfolded distribution and the known particle- or parton-level distribution of the \MCatNLO{}+\Herwig{} sample is assigned as the relative uncertainty for the fiducial or full phase-space distributions, respectively. This uncertainty is found to be in the range $2$--$5$\%, depending on the variable, increasing up to $10$\% at large \ptt, \mttbar, \ptttbar ~and \absyttbar. The observable that is most affected by these uncertainties is \mttbar{} in the full phase space.

To assess the impact of different parton-shower models, 
unfolded results using events simulated with \PowHeg interfaced 
to \Pythia are compared to events simulated with \PowHeg interfaced 
to \Herwig, using the same procedure described above to evaluate the uncertainty related to the \ttbar generator.
The resulting systematic uncertainties, taken as the symmetrized difference, are found to be  typically at the $1$--$3$\% level.

In order to evaluate the uncertainty related to
the modelling of the ISR/FSR, $\ttbar$ MC samples with modified ISR/FSR modelling are used.
The MC samples used for the evaluation of this uncertainty are generated using the 
\PowHeg generator interfaced to \Pythia,
where the parameters are 
varied as described in Section~\ref{sec:DataSimSamples}.
This uncertainty is found to be in the range $2$--$5$\%, depending on the variable of the $\ttbar$ system 
considered, and reaching the largest values at high \absyt and small \ptttbar.

The impact of the uncertainty related to the PDF is assessed by means of $\ttbar$ samples generated with {\sc MC@NLO} interfaced to {\sc Herwig}. An envelope of spectra is evaluated by reweighting the central prediction of the CT10nlo PDF set, using the full set of 52 eigenvectors at $68$\% CL.
This uncertainty is found to be less than $1$\%.

As a check, the effect of the uncertainty on the top-quark mass was evaluated and found to affect only the efficiency correction by less than 1\%, consistent with what was observed by ATLAS for the analogous measurement with the 7 \TeV{} data~\cite{atlasDiff3}.

\subsection{Background modelling}\label{sec:BkgMod}

Systematics affecting the background are modelled by adding to the signal spectrum the difference of the systematics-varied and nominal backgrounds.

The single-top background is assigned an
uncertainty associated with the theoretical calculations used for its
normalization~\cite{Kidonakis:2011wy,Kidonakis:2010ux,Kidonakis:2010tc}. The overall impact of this systematic uncertainty on the signal is around $0.5$\%.

The systematic uncertainties due to the 
overall normalization and the heavy-flavour fraction of $W$+jets events 
are obtained by varying the data-driven scale factors 
within the statistical uncertainty of the $W$+jets MC sample. The $W$+jets shape uncertainty is 
extracted by varying the renormalization and matching scales in {\sc Alpgen}.
The $W$+jets MC statistical uncertainty is also taken into account. 
The overall impact of this uncertainty is less than $1$\%.

The uncertainty on the background from non-prompt and fake-leptons is evaluated by varying
the definition of loose leptons, changing the selection used to form
the control region and propagating the statistical uncertainty of
parameterizations of the efficiency to pass the tighter lepton requirements for real and fake leptons. 
The combination of all these components also affects the shape of the background.
The overall impact of this systematic uncertainty is less than $1$\%.

A~50\% uncertainty is applied to the normalization of the $Z$+jets background, including the uncertainty on the cross-section and a further 48\% 
due to uncertainties related to the requirement of the presence of at least four jets.
A~40\% uncertainty is applied to the diboson background, including the uncertainty on the cross-section and a further 34\% due to the presence of two additional jets.
The overall impact of these uncertainties is less than $1$\%, and the largest contribution is due to the $Z$+jets background.

\section{Results} \label{sec:Results}

In this section, comparisons between unfolded data distributions and several SM predictions are presented for the different observables discussed in Section~\ref{sec:YieldsAndPlots}. Events are selected by requiring exactly one lepton and at least four jets with at least two of the jets tagged as originating from a~$b$-quark. Normalized differential cross-sections are shown in order to remove systematic uncertainties on the normalization. 

The SM predictions are obtained using different MC generators. The \PowHegBox generator \cite{Frixione:2007vw}, denoted ``PWG'' in the figures, is employed with three different sets of parton shower models, namely \Pythia \cite{Sjostrand:2006za}, \PythiaEight~\cite{pythia8} and \Herwig \cite{HERWIG}. The other NLO generator is \MCatNLO{} \cite{MCATNLO} interfaced with the \Herwig parton shower. Generators at the LO accuracy are represented by \MadGraph~\cite{Alwall:2011uj} interfaced with \Pythia for parton showering, which calculates \ttb{}~matrix elements with up to three additional partons and implements the matrix-element to parton-shower MLM matching scheme~\cite{MLMatching}.

The level of agreement between the measured distributions and simulations with different theoretical predictions is quantified by calculating $\chi^2$ values, employing the full covariance matrices, and inferring $p$-values (probabilities that the $\chi^2$ is larger than or equal to the observed value) from the $\chi^2$ and the number of degrees of freedom (NDF). Uncertainties on the predictions are not included. The normalization constraint used to derive the normalized differential cross-sections lowers by one unit the NDF and the rank of the $N_{\rm b} \times N_{\rm b}$ covariance matrix, where $N_{\rm b}$ is the number of bins of the spectrum under consideration~\cite{multinomial_chi2}. In order to evaluate the $\chi^2$ the following relation is used 

\begin{equation}
\chi^2 = V_{N_{\rm b}-1}^{\rm T} \cdot {\rm Cov}_{N_{\rm b}-1}^{-1} \cdot V_{N_{\rm b}-1} \,,
\end{equation} 

where $V_{N_{\rm b}-1}$ is the vector of differences between data and prediction obtained by discarding one of the $N_{\rm b}$ elements and ${\rm Cov}_{N_{\rm b}-1}$ is the $(N_{\rm b}-1) \times (N_{\rm b}-1)$ sub-matrix derived from the full covariance matrix discarding the corresponding row and column. 
The sub-matrix obtained in this way is invertible and allows the $\chi^2$ to be computed. The $\chi^2$ value does not depend on the choice of the element discarded for the vector $V_{N_{\rm b}-1}$ and the corresponding sub-matrix ${\rm Cov}_{N_{\rm b}-1}$. 

The set of Figures~\ref{fig:results:fiducial:topH:rel}--\ref{fig:results:fiducial:qcd3:rel} presents the normalized \ttbar fiducial phase-space differential cross-sections as a~function of the different observables. In particular, 
Figures~\ref{fig:particle:topH_pt:rel} and \ref{fig:particle:topH_absrap:rel} show the distributions of the hadronic top-quark transverse momentum and the absolute value of the rapidity; Figures~\ref{fig:particle:tt_m:rel}, \ref{fig:particle:tt_pt:rel} and \ref{fig:particle:tt_absrap:rel} present the \ttb{}~system invariant mass, transverse momentum, and absolute value of the rapidity, while the additional observables related to the \ttbar~system and the ratio of the transverse momenta of the hadronically decaying $W$ boson and top quark are shown in Figures~\ref{fig:results:fiducial:qcd1:rel}, \ref{fig:results:fiducial:qcd2:rel} and \ref{fig:results:fiducial:qcd3:rel}.

None of the predictions is able to correctly describe all the distributions, as also witnessed by the $\chi^2$ values and the $p$-values listed in Table \ref{tab:chi2:particle:rel}. In particular, a certain tension between data and all predictions is observed in the case of the hadronic top-quark transverse momentum distribution for values higher than about $400$ \GeV. No electroweak corrections \cite{Kuhn:2005it,Kuhn:2006vh,Bernreuther:2008aw,Kuhn:2006vh,Manohar:2012rs,Kuhn:2013zoa} are included in these predictions, as these have been shown to have a measurable impact only at very high values of the top quark transverse momentum, leading to a slightly softer \ptthad{} spectrum as confirmed by the recent ATLAS measurement of the \ttbar differential distribution of the hadronic top-quark \pt{} for boosted top quarks~\cite{CON-2014-057}. The effect of electroweak corrections alone is not large enough to solve this discrepancy completely \cite{CON-2014-057,ATLAS:ttbar_res_8TeV}. The shape of the \absythad distribution shows only a modest agreement for all the generators, with larger discrepancies observed in the forward region for \PowHeg{}+\Pythia and \PowHeg{}+\PythiaEight.
 
For the \mttbar{} distribution, the \PowHeg{}+\Pythia, \PowHeg{}+\PythiaEight and \PowHeg{}+\Herwig generators are in better agreement with the data. All generators are in good agreement in the \ptttbar{} spectrum except for 
\PowHeg{}+\Herwig in the last bin. This observation suggests that setting \HDampMT~in the \PowHeg samples improves the agreement at high values of the \ttbar~ transverse momentum. 
The data at high values of \ttb{} rapidity is not adequately described by any of the generators considered. The same conclusions hold for the analogous distribution for the absolute spectra, although the overall agreement estimated with the $\chi^2$ values and the $p$-values is better due to the larger uncertainties.

For the variables describing the hard-scattering interaction, the production angle \chittbar{} is well described in the central region. The forward region, described by the tail of this observable and by the tail of the longitudinal boost \boostttbar{}, is not described correctly by any of the generators under consideration. For the variables describing the radiation along the $\ttbar$ pair momentum direction, both \absPoutttbar{} and \DeltaPhittbar{} indicate that the kinematics of top quarks produced in the collinear region (\DeltaPhittbar $\lesssim \pi/2$) are described with fair agreement by all the generators, but the uncertainty is particularly large in this region. The tension observed in the \ptthad{} spectrum is reflected in the tail of the \HTttbar{} distribution. Finally, the ratio of the hadronic $W$ boson and top-quark transverse momenta shows a mis-modelling in the range $1.5$--$3$ for all the generators.

The difficulty in correctly predicting the data in the forward region was further investigated by studying the dependence of the predictions on different PDF sets. The study was performed for the rapidity observables \absythad{}, \absyttbar{} and \boostttbar{}, shown in Figure~\ref{fig:results:fiducial:pdf:rel} and comparing the data with the predictions of \MCatNLO{}+\Herwig for  more recent sets of parton distribution functions. The results exhibit a general improvement in the description of the forward region for the most recent PDF sets (CT14nlo~\cite{CT14:2015}, CJ12mid~\cite{CJ12mid}, MMHT2014nlo~\cite{MMHT2014nlo}, NNPDF 3.0 NLO~\cite{NNPDF3.0nlo}, METAv10LHC~\cite{METAv10LHC}, HERAPDF 2.0 NLO \cite{HERAPDF20}). The improvement with respect to CT10nlo is also clearly shown in Table \ref{tab:chi2:pdf:particle:rel} which lists the $\chi^2$ and corresponding $p$-values for the different sets. The only exception is represented by the \absythad~distribution using HERAPDF 2.0 NLO, for which a disagreement in the forward region is observed.

\begin{figure*}[htbp]
\centering
\subfigure[]{ \includegraphics[width=0.45\textwidth]{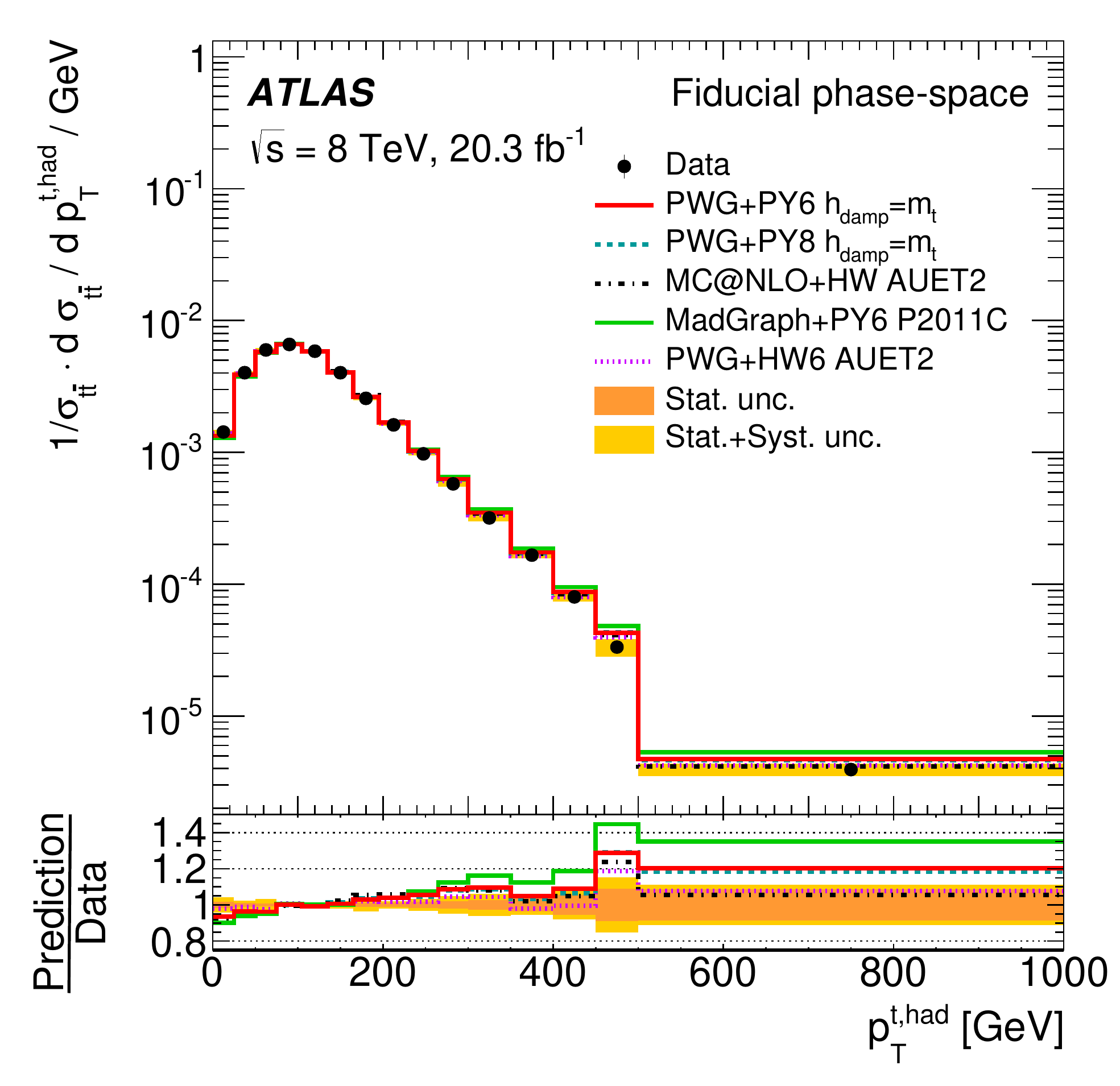}\label{fig:particle:topH_pt:rel}}
\subfigure[]{ \includegraphics[width=0.45\textwidth]{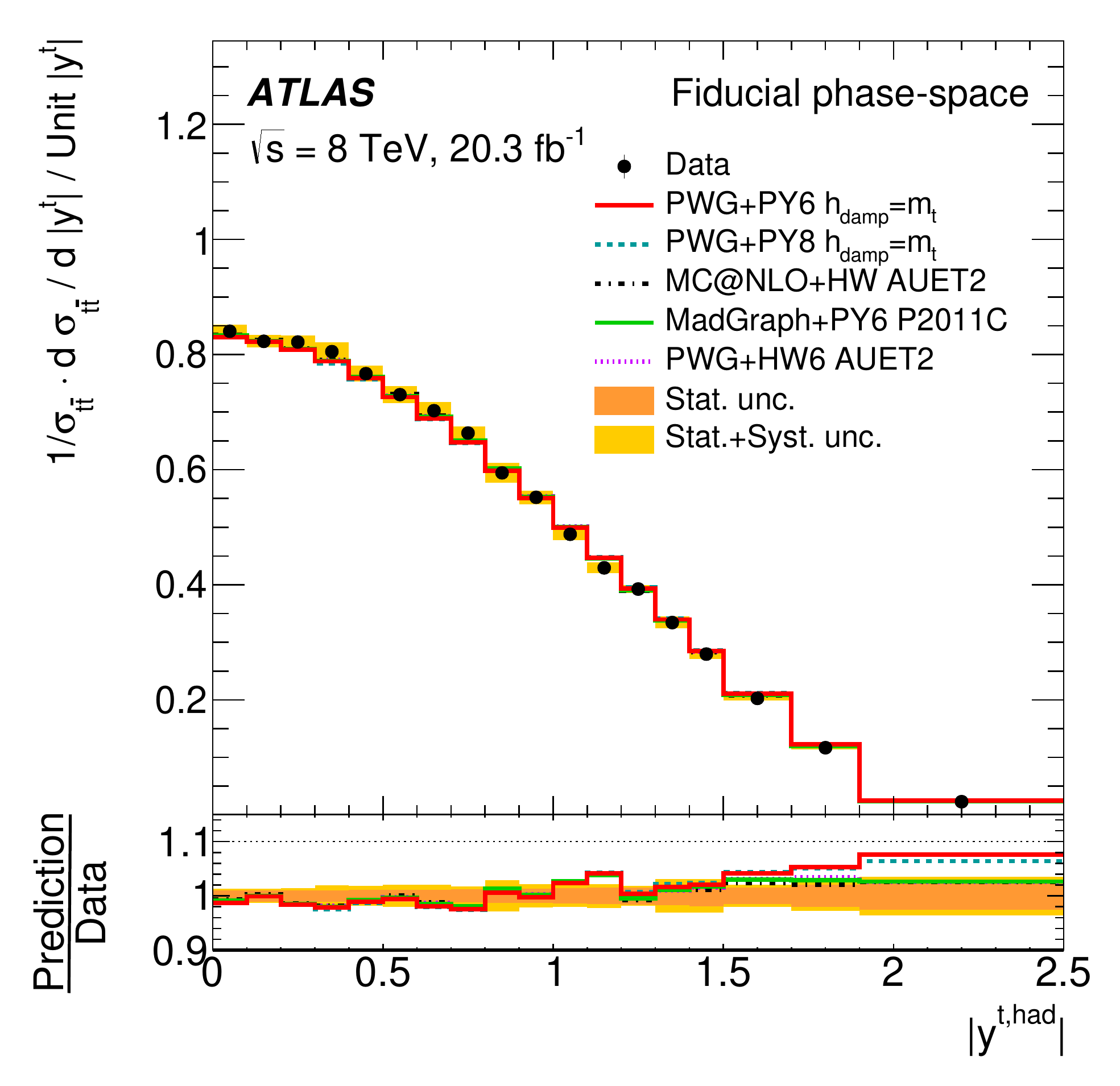}\label{fig:particle:topH_absrap:rel}}
\caption{Fiducial phase-space normalized differential cross-sections as a function of the \subref{fig:particle:topH_pt:rel}~transverse momentum (\ptthad{}) and \subref{fig:particle:topH_absrap:rel}~absolute value of the rapidity (\absythad) of the hadronic top quark. The yellow bands indicate the total uncertainty on the data in each bin. The \PowHeg{}+\Pythia generator with \HDampMT~and the CT10nlo PDF is used as the nominal prediction to correct for detector effects.}
\label{fig:results:fiducial:topH:rel}
\end{figure*}

\begin{figure*}[htbp]
\centering
\subfigure[]{ \includegraphics[width=0.45\textwidth]{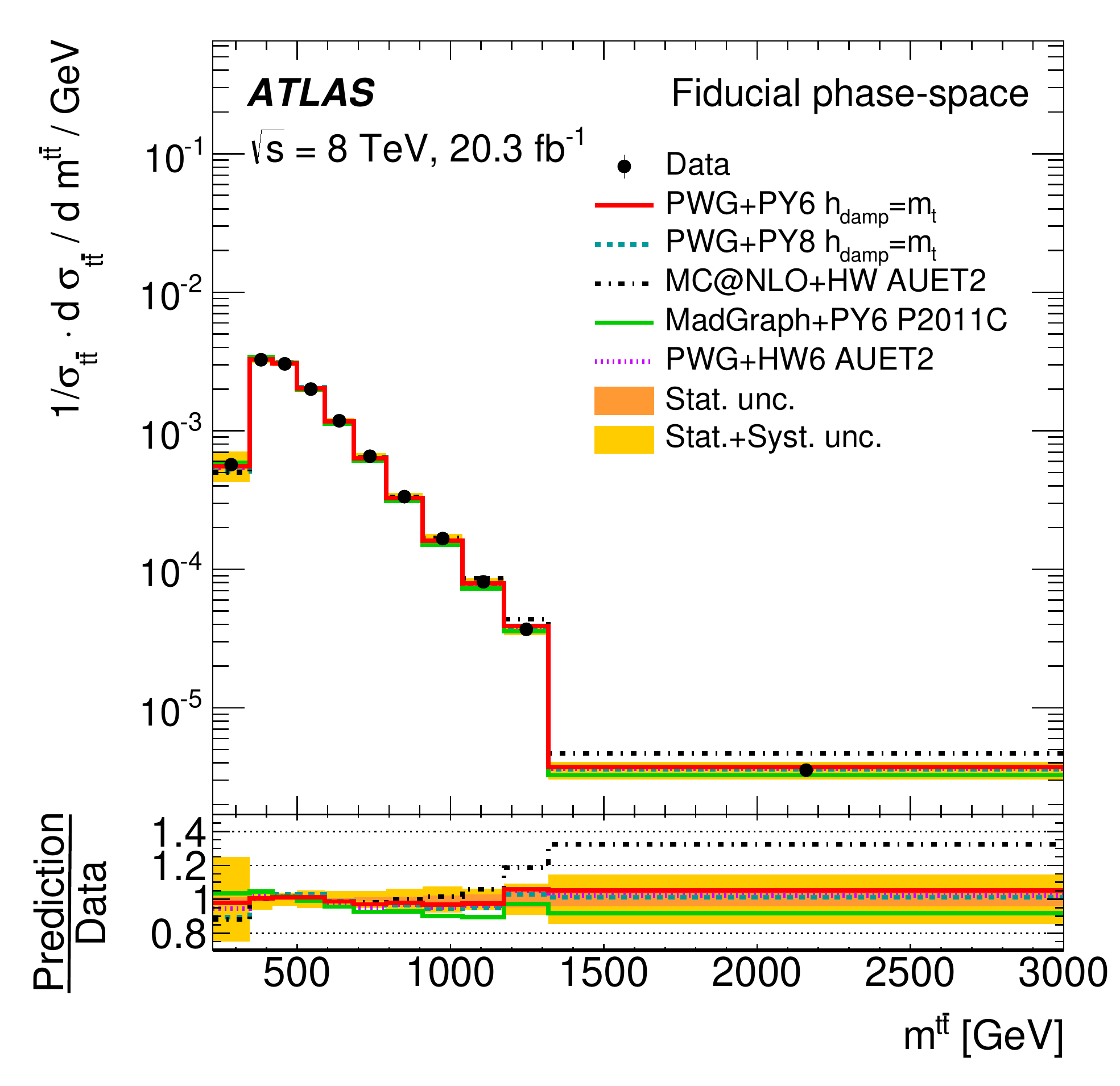}\label{fig:particle:tt_m:rel}}
\subfigure[]{ \includegraphics[width=0.45\textwidth]{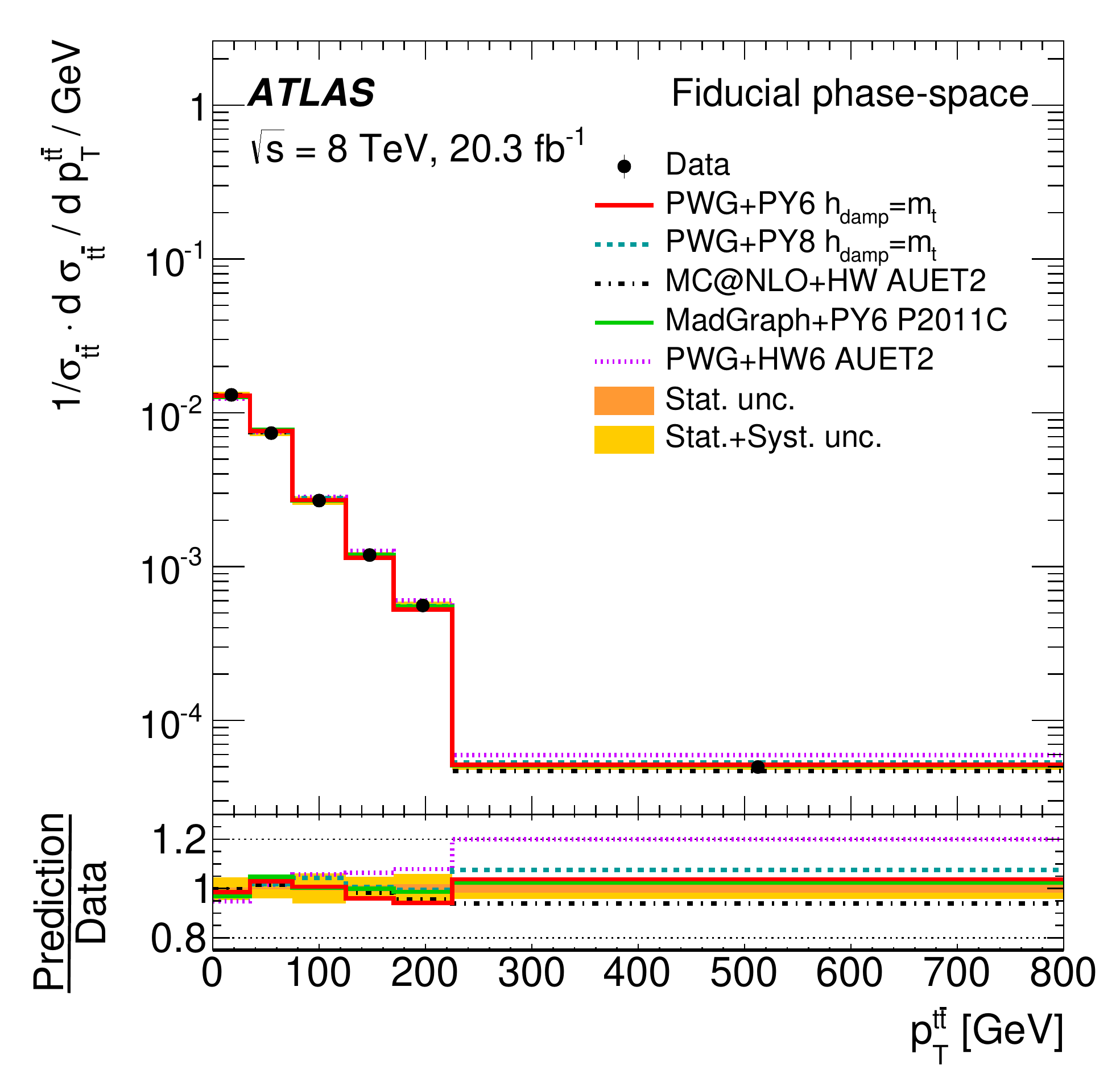}\label{fig:particle:tt_pt:rel}}
\subfigure[]{ \includegraphics[width=0.45\textwidth]{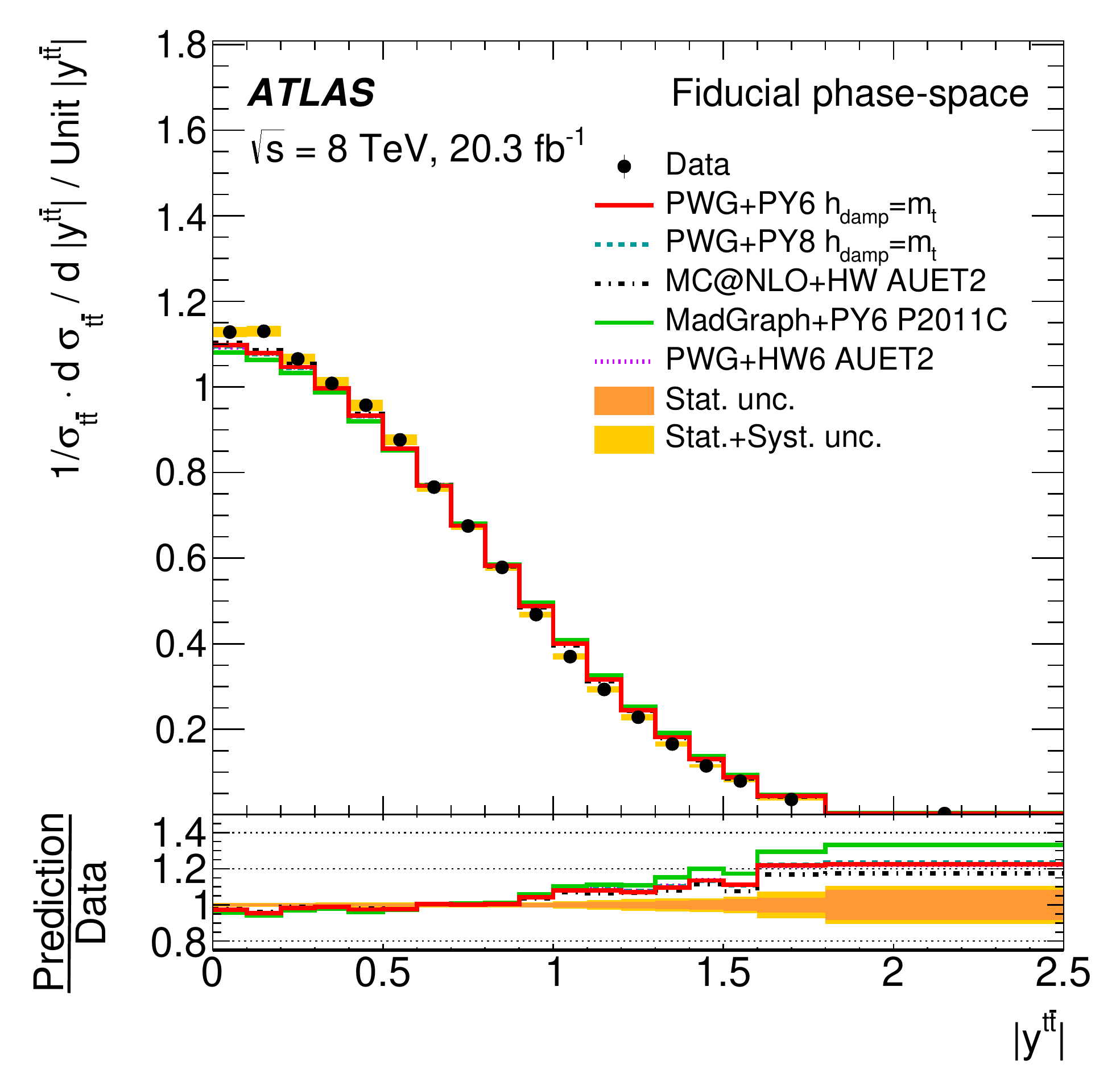}\label{fig:particle:tt_absrap:rel}}
\caption{Fiducial phase-space normalized differential cross-sections as a function of the \subref{fig:particle:tt_m:rel}~invariant mass (\mttbar{}), \subref{fig:particle:tt_pt:rel}~transverse momentum (\ptttbar{}) and \subref{fig:particle:tt_absrap:rel}~absolute value of the rapidity (\absyttbar{}) of the \ttb{}~ system. The yellow bands indicate the total uncertainty on the data in each bin. The \PowHeg{}+\Pythia generator with \HDampMT~and the CT10nlo PDF is used as the nominal prediction to correct for detector effects.}
\label{fig:results:fiducial:tt:rel}
\end{figure*}

\begin{figure*}[htbp]
\centering
\subfigure[]{ \includegraphics[width=0.45\textwidth]{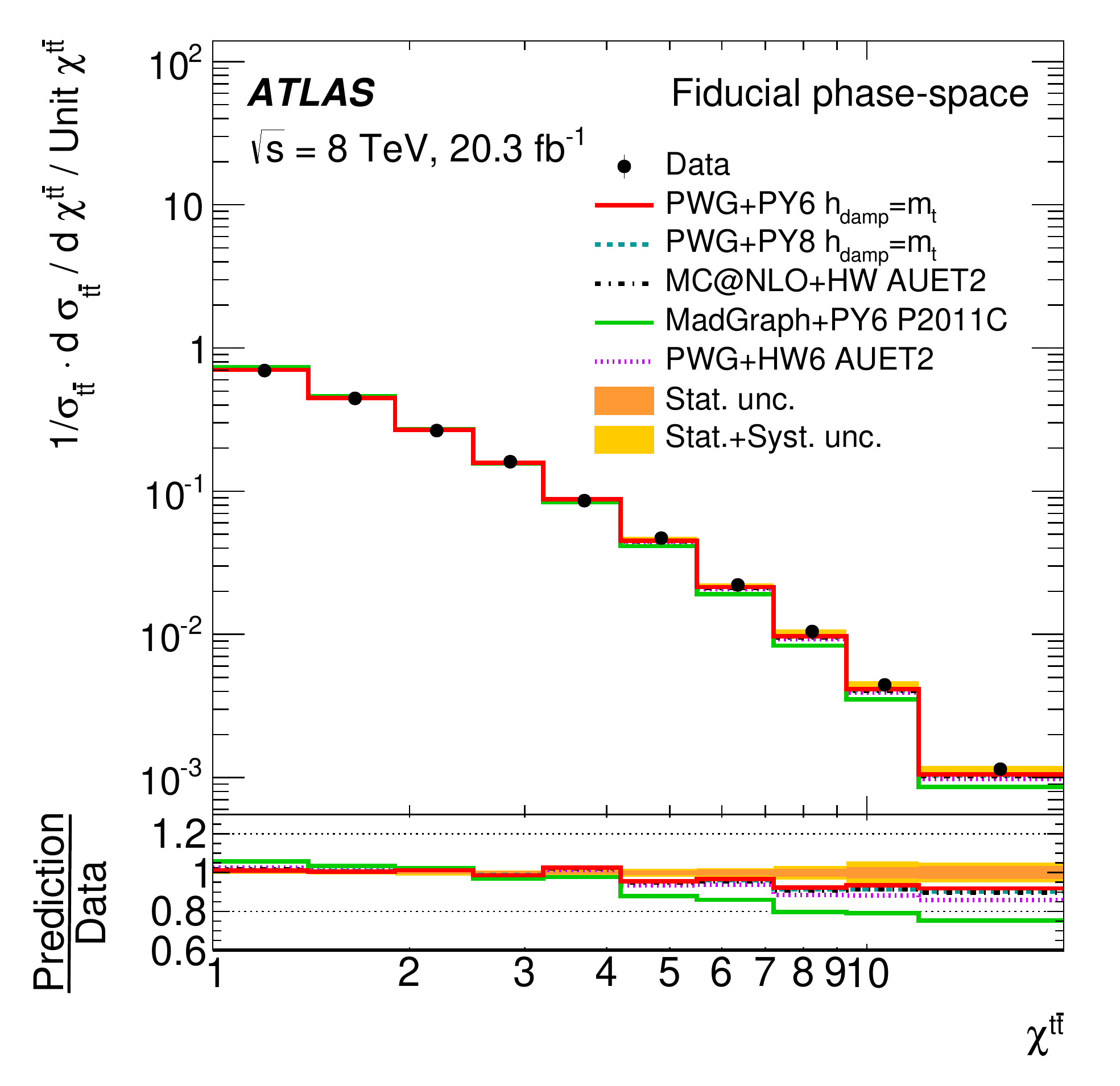}\label{fig:particle:Chi_ttbar:rel}}
\subfigure[]{ \includegraphics[width=0.45\textwidth]{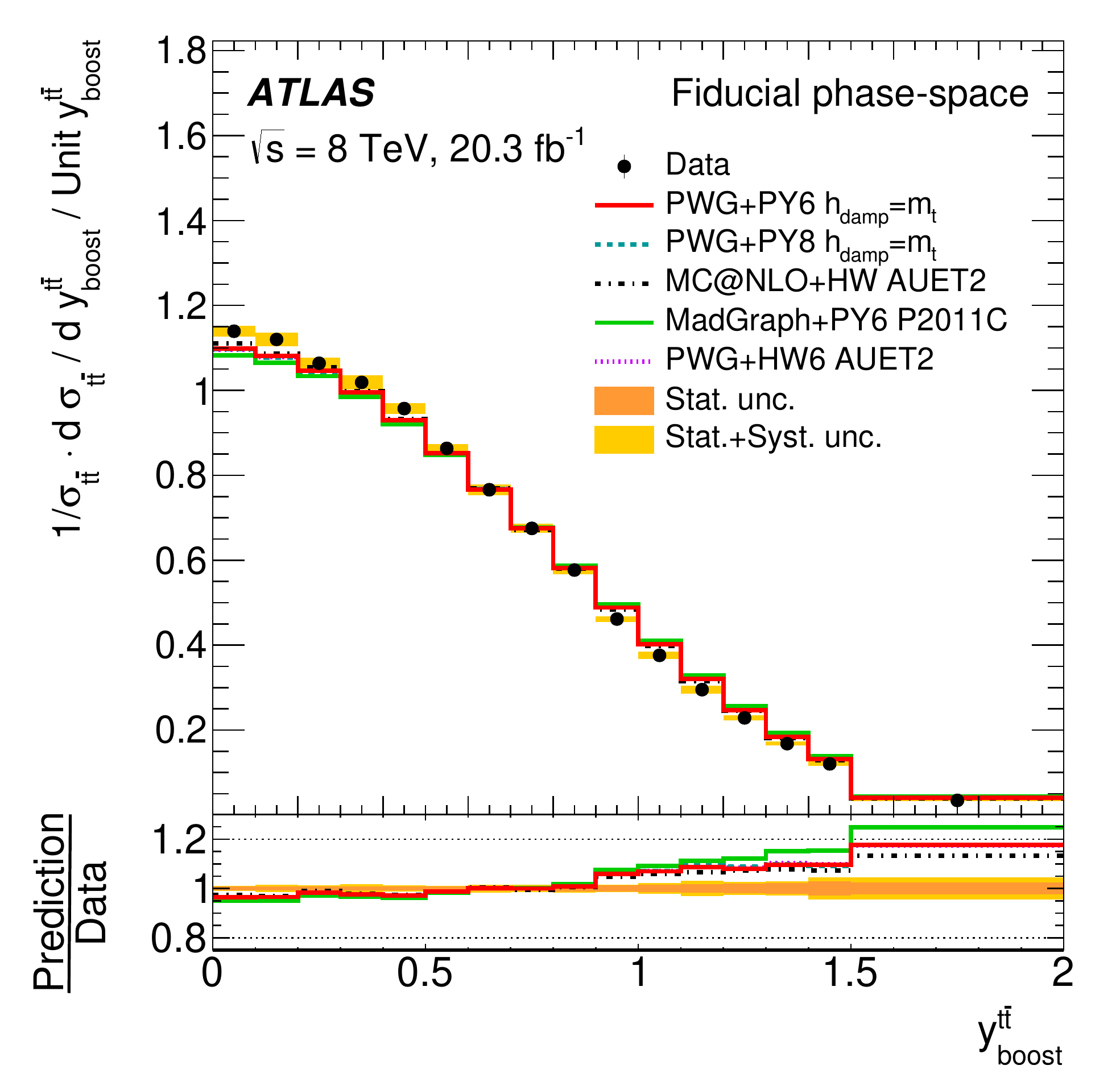}\label{fig:particle:Yboost:rel}}
\caption{Fiducial phase-space normalized differential cross-sections as a function of the \ttb{}~ \subref{fig:particle:Chi_ttbar:rel}~production angle (\chittbar{}) and \subref{fig:particle:Yboost:rel}~longitudinal boost (\boostttbar{}). The yellow bands indicate the total uncertainty on the data in each bin. The \PowHeg{}+\Pythia generator with \HDampMT~and the CT10nlo PDF is used as the nominal prediction to correct for detector effects.}
\label{fig:results:fiducial:qcd1:rel}
\end{figure*}

\begin{figure*}[htbp]
\centering
\subfigure[]{ \includegraphics[width=0.45\textwidth]{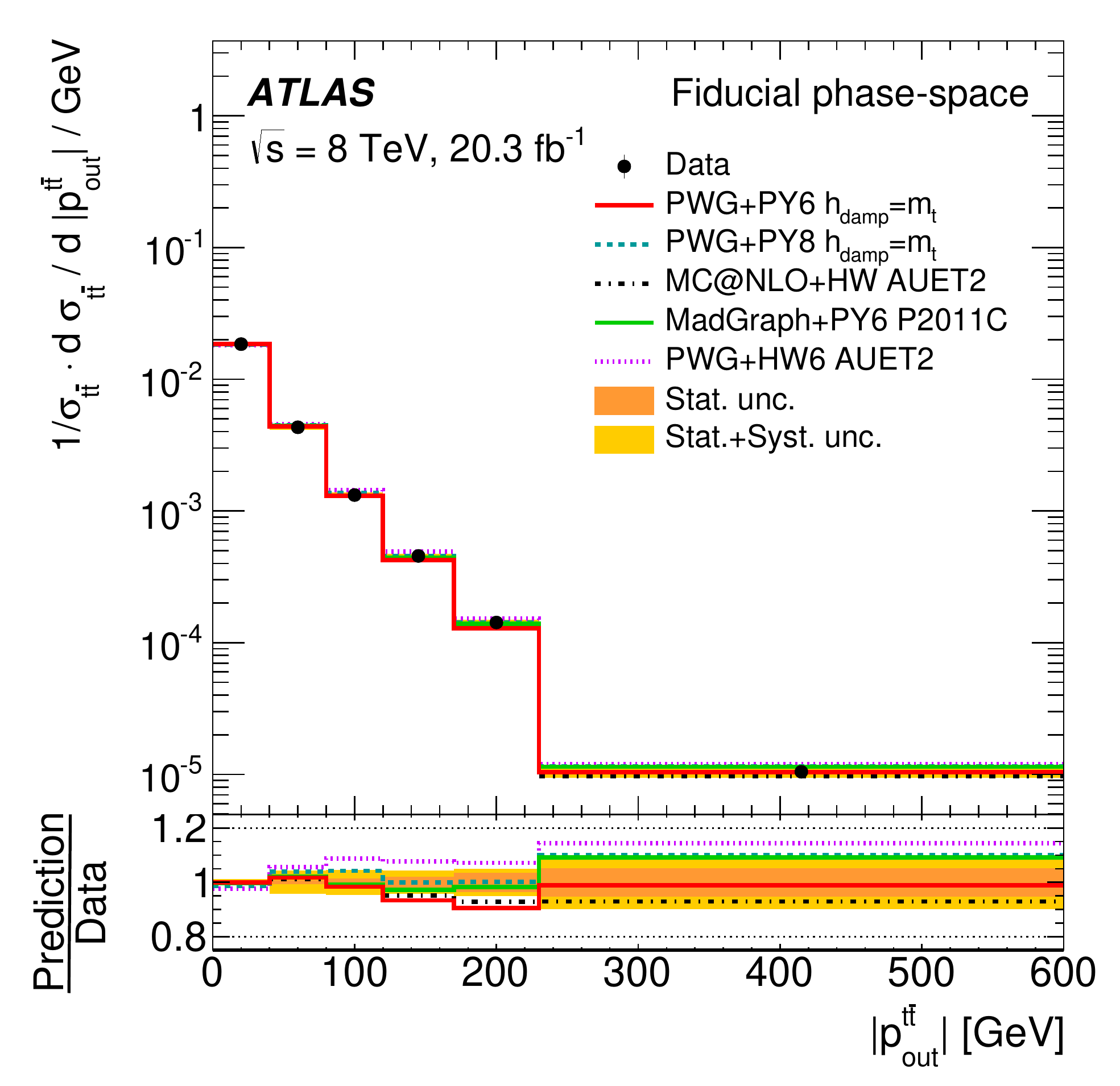}\label{fig:particle:Pout:rel}}
\subfigure[]{ \includegraphics[width=0.45\textwidth]{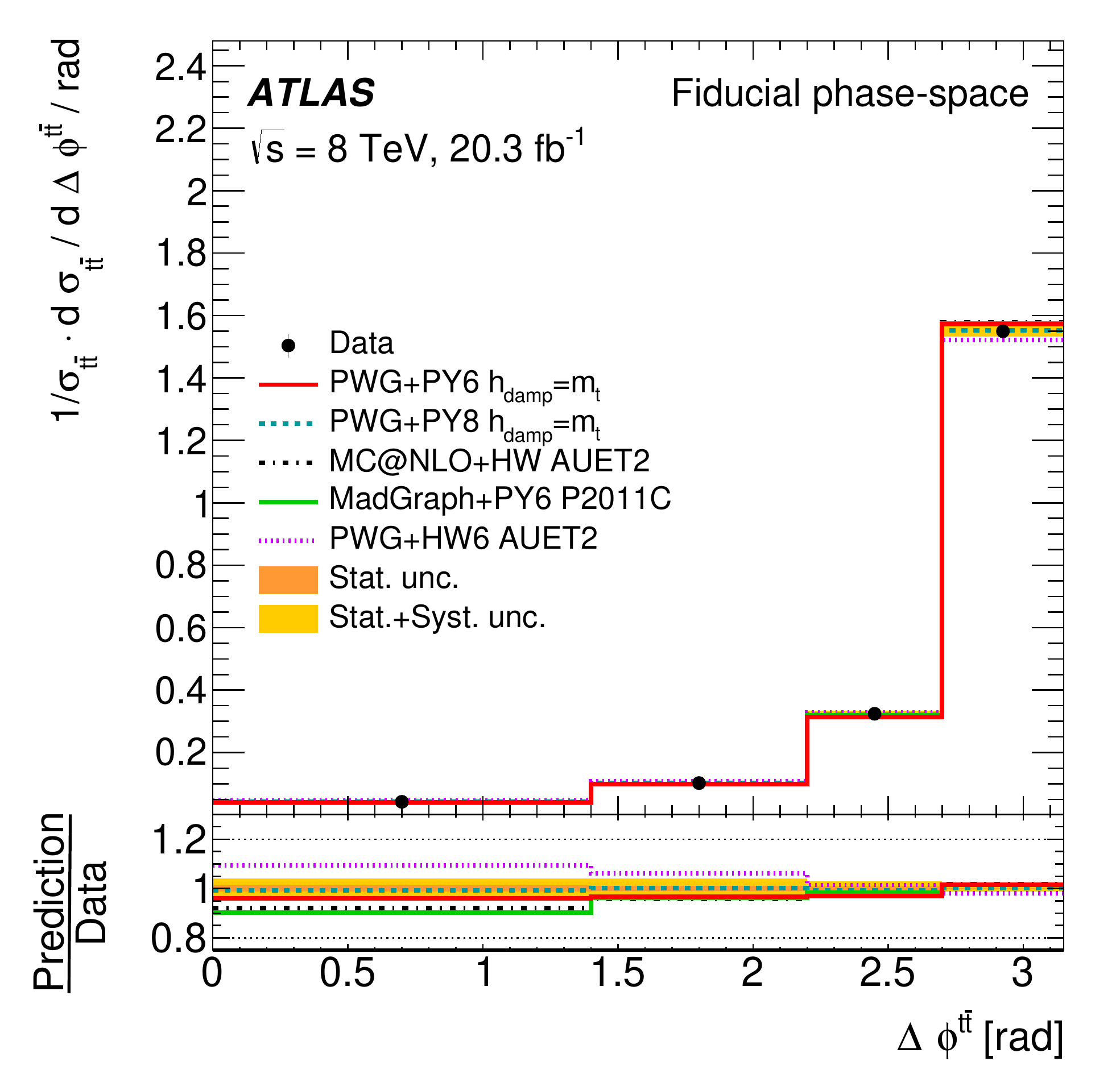}\label{fig:particle:tt_dPhi:rel}}
\caption{Fiducial phase-space normalized differential cross-sections as a function of the \ttb{}~ \subref{fig:particle:Pout:rel}~out-of-plane momentum (\absPoutttbar{}) and \subref{fig:particle:tt_dPhi:rel}~azimuthal angle (\DeltaPhittbar). The yellow bands indicate the total uncertainty on the data in each bin. The \PowHeg{}+\Pythia generator with \HDampMT~and the CT10nlo PDF is used as the nominal prediction to correct for detector effects.}
\label{fig:results:fiducial:qcd2:rel}
\end{figure*}

\begin{figure*}[htbp]
\centering
\subfigure[]{ \includegraphics[width=0.45\textwidth]{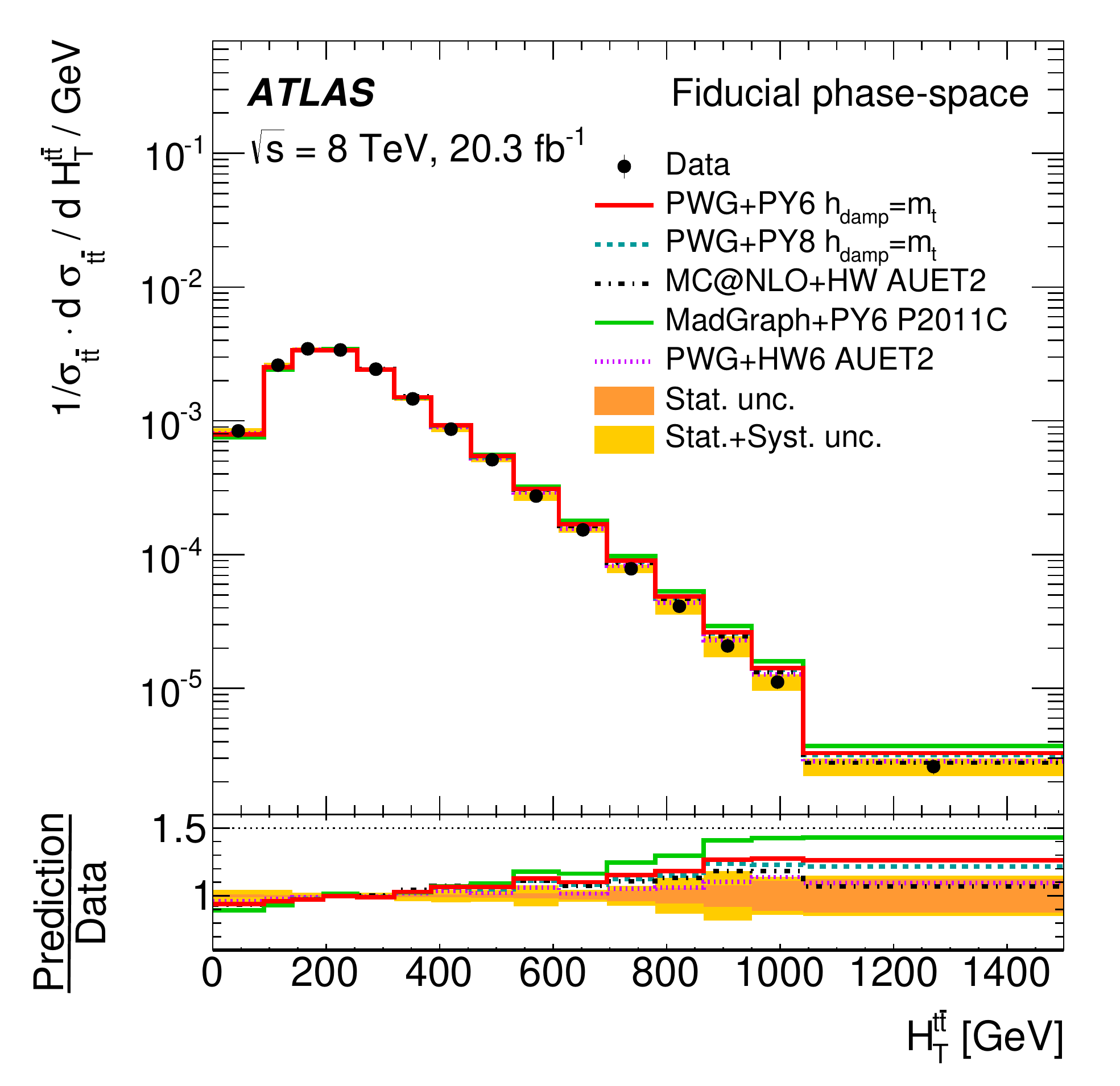}\label{fig:particle:HT_ttbar:rel}}
\subfigure[]{ \includegraphics[width=0.45\textwidth]{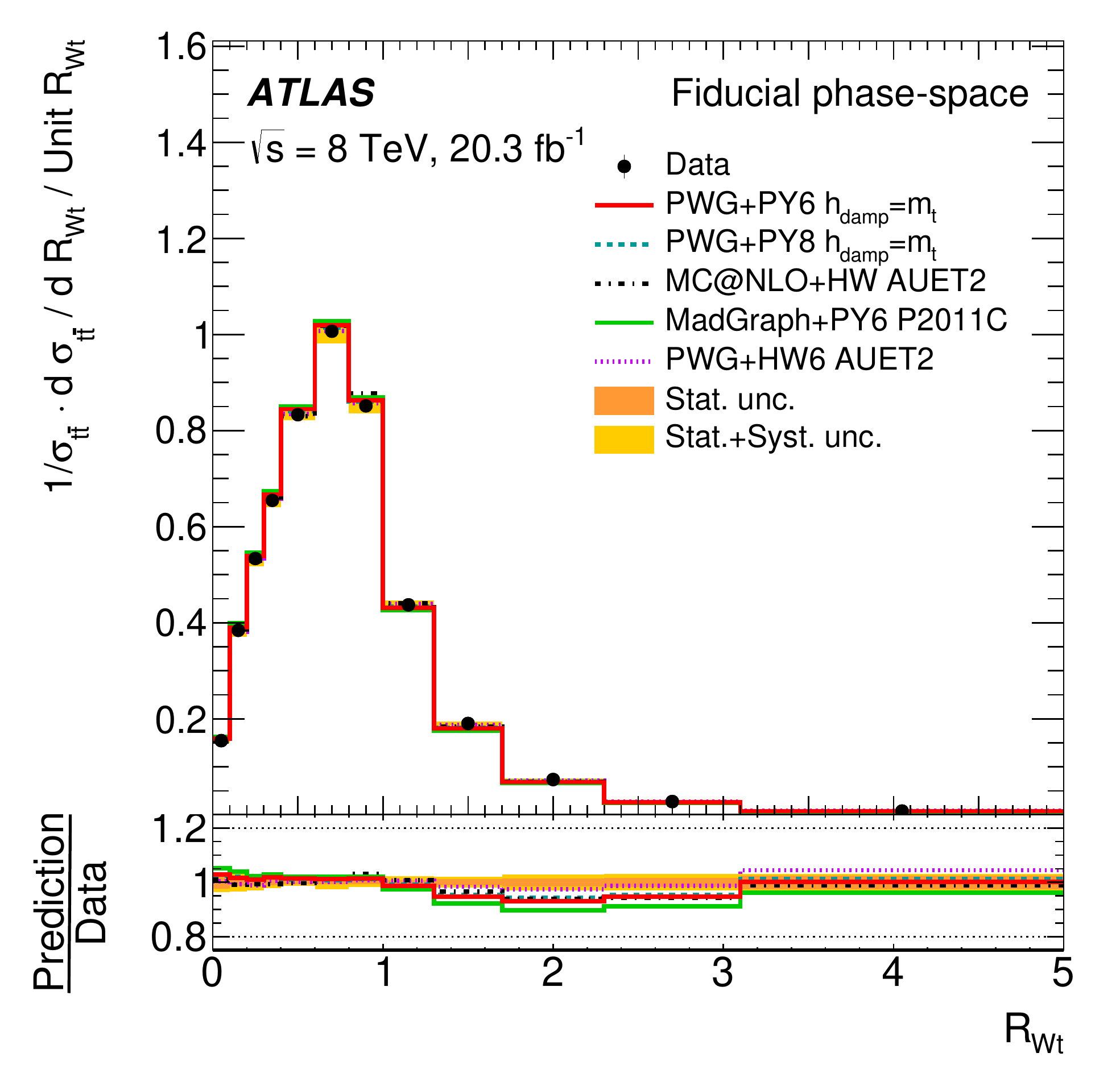}\label{fig:particle:R_Wt_had:rel}}
\caption{Fiducial phase-space normalized differential cross-sections as a function of the \subref{fig:particle:HT_ttbar:rel}~ scalar sum of the transverse momenta of the hadronic and leptonic top quarks (\HTttbar) and \subref{fig:particle:R_Wt_had:rel}~ the ratio of the hadronic $W$ and the hadronic top transverse momenta (\RWtttbar). The yellow bands indicate the total uncertainty on the data in each bin. The \PowHeg{}+\Pythia generator with \HDampMT~and the CT10nlo PDF is used as the nominal prediction to correct for detector effects.}
\label{fig:results:fiducial:qcd3:rel}
\end{figure*}

\begin{figure*}[htbp]
\centering
\subfigure[]{ \includegraphics[width=0.45\textwidth]{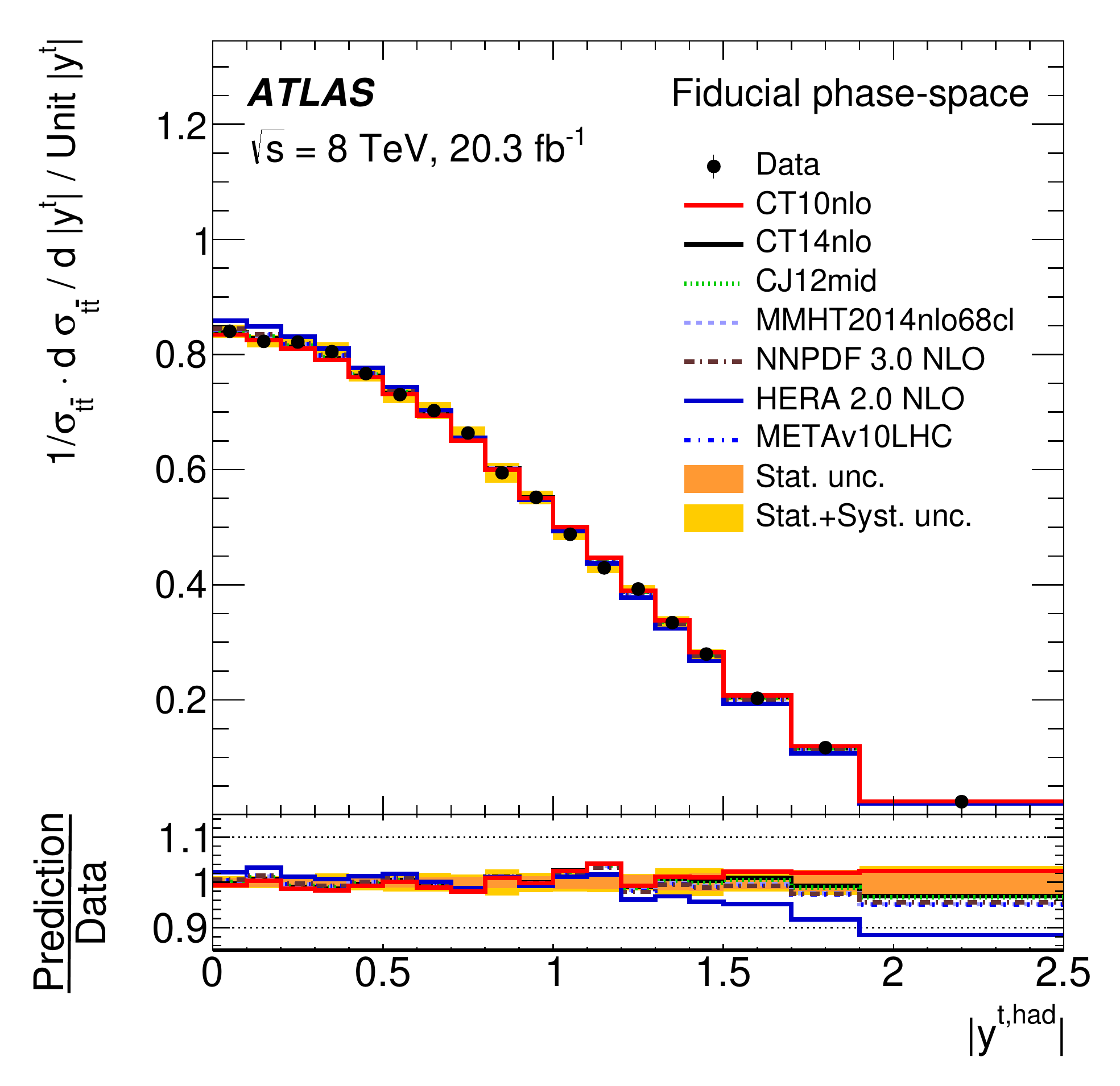}\label{fig:pdf:particle:topH_absrap:rel}}
\subfigure[]{ \includegraphics[width=0.45\textwidth]{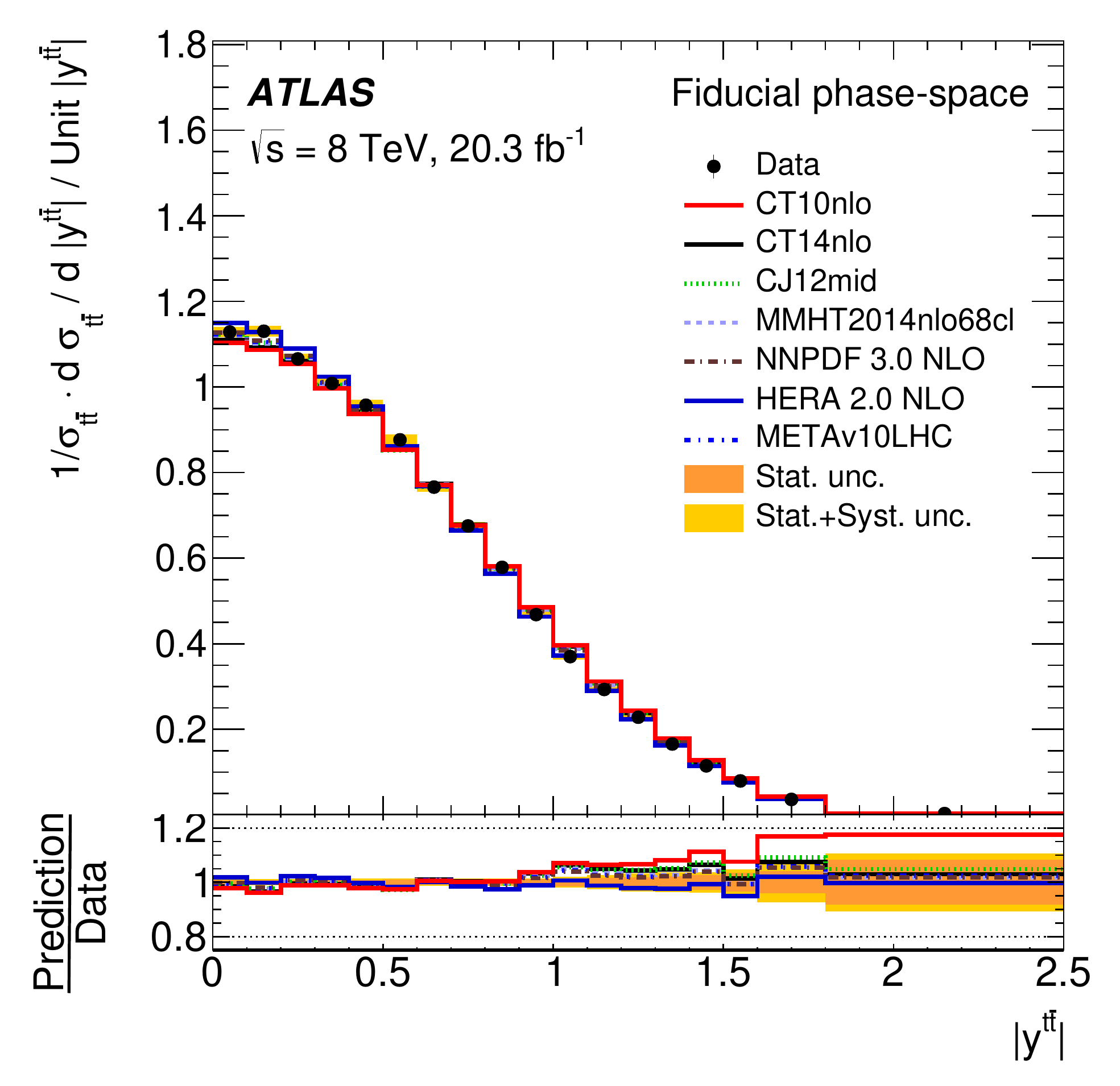}\label{fig:pdf:particle:tt_absrap:rel}}
\subfigure[]{ \includegraphics[width=0.45\textwidth]{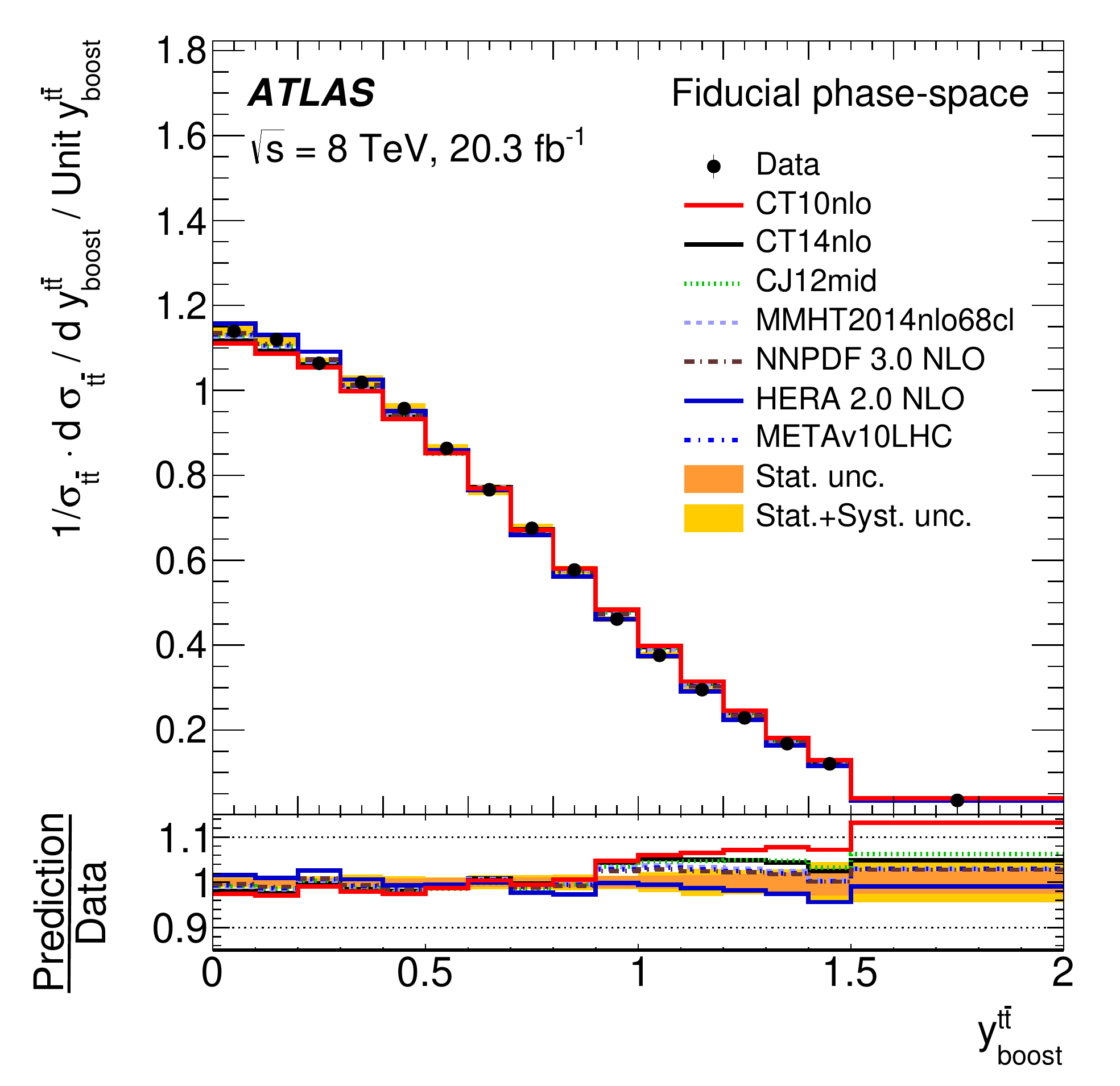}\label{fig:pdf:particle:Yboost:rel}}\caption{Fiducial phase-space normalized differential cross-sections as a function of the \subref{fig:pdf:particle:topH_absrap:rel}~absolute value of the rapidity of the hadronic top quark (\absythad), \subref{fig:pdf:particle:tt_absrap:rel}~absolute value of the rapidity (\absyttbar{}) of the \ttb{}~ system and \subref{fig:pdf:particle:Yboost:rel}~longitudinal boost (\boostttbar{}). The yellow bands indicate the total uncertainty on the data in each bin. The \MCatNLO{}+\Herwig generator is reweighted using the new PDF sets to produce the different predictions. The \PowHeg{}+\Pythia generator with \HDampMT~and the CT10nlo PDF is used as the nominal prediction to correct for detector effects.}
\label{fig:results:fiducial:pdf:rel}
\end{figure*}

\clearpage

The set of Figures~\ref{fig:results:full:topH:rel}--\ref{fig:results:full:qcd2:rel} presents the normalized \ttbar~full phase-space differential cross-sections as a function of the different observables. In particular, Figures~\ref{fig:parton:topH_pt:rel} and \ref{fig:parton:topH_absrap:rel} show the top-quark transverse momentum and the absolute value of the rapidity; Figures~\ref{fig:parton:tt_m:rel}, \ref{fig:parton:tt_pt:rel} and \ref{fig:parton:tt_absrap:rel} present the \ttb{}~system invariant mass, transverse momentum and absolute value of the rapidity while the additional observables related to the \ttbar system are shown in Figures~\ref{fig:results:full:qcd1:rel} and \ref{fig:results:full:qcd2:rel}. 
Regarding the comparison between data and predictions, the general picture, already outlined for the fiducial phase-space measurements, is still valid 
even though the uncertainties are much larger due to the full phase-space extrapolation. In particular, the predictions for the top-quark \pt{} and \HTttbar{}~ tend to be in a better agreement with the data than what is observed in the fiducial phase-space. 
The $\chi^2$ and corresponding $p$-values for the different observables and predictions are shown in Table~\ref{tab:chi2:parton:rel}.

In Figures~\ref{fig:results:theory:topH:rel}--\ref{fig:results:NNLO:tt:rel} the normalized \ttbar full phase-space differential cross-section as a function of \ptt, \absyt, \mttbar{} and \absyttbar{} are compared with theoretical higher-order QCD calculations.

The measurements are compared to four calculations that offer beyond--NLO accuracy:
\begin{itemize}
\item an approximate next-to-next-to-leading-order (\aNNLO) calculation based on QCD threshold expansions beyond the leading logarithmic approximation \cite{DiffTop} using the CT14nnlo PDF \cite{CT14:2015};
\item an approximate next-to-next-to-next-to-leading-order (\aNNNLO) calculation based on the resummation of soft-gluon contributions in the double-differential cross section at next-to-next-to-leading-logarithm (\NNLL) accuracy in the moment-space approach in perturbative QCD \cite{Kidonakis:aNNNLO} using the MSTW2008nnlo PDF \cite{MSTW};
\item an approximate \NLO+\NNLL{} calculation \cite{nnloMtt} using the MSTW2008nnlo PDF \cite{MSTW}.
\item a full NNLO calculation \cite{stripper} using the MSTW2008nnlo PDF \cite{MSTW}. The \NNLO{} prediction does not cover the highest bins in \ptt{} and \mttbar. \end{itemize}

These predictions have been interpolated in order to match the binning of the presented measurements. Table \ref{tab:chi2:theory:topH:parton:rel} shows the $\chi^2$ and $p$-values for these higher-order QCD calculations. 

Figures \ref{fig:results:theory:topH:rel} and \ref{fig:results:NNLO:topH:rel} show a comparison of the \ptt{} and \absyt{} distributions to the \aNNLO{} and \aNNNLO{}, and to the \NNLO{} calculations respectively. The \aNNNLO{} calculation is seen to improve the agreement compared to the \PowHeg{}+\Pythia{} generator in \absyt, but not in \ptt. The \aNNLO{} prediction produces a \ptt{} distribution that is softer than the data at high transverse momentum and does not improve the description of \absyt. The \NNLO{} calculation is in good agreement with both the \ptt{} and \absyt{} distributions, in particular the disagreement seen at high \ptt{} for the \NLO{} generators is resolved by the \NNLO{} calculation.

The measurement of the invariant mass and transverse momentum of the \ttbar{} system is compared to the \NLO+\NNLL{} prediction in Figure \ref{fig:results:theory:tt:rel}. The \NLO+\NNLL{} calculation shows a good agreement in the \mttbar{} spectrum and a very large discrepancy for high values of the \ttbar{} transverse momentum.
Figure \ref{fig:results:NNLO:tt:rel} shows a comparison of the \NNLO{} calculation to the \mttbar{} and \absyttbar{} measurements. For the rapidity of the \ttbar{} system, the \NNLO{} calculation improves the agreement slightly compared to the \PowHeg+\Pythia{} prediction, but some shape difference can be seen between data and prediction.

\begin{figure*}[htbp]
\centering
\subfigure[]{ \includegraphics[width=0.45\textwidth]{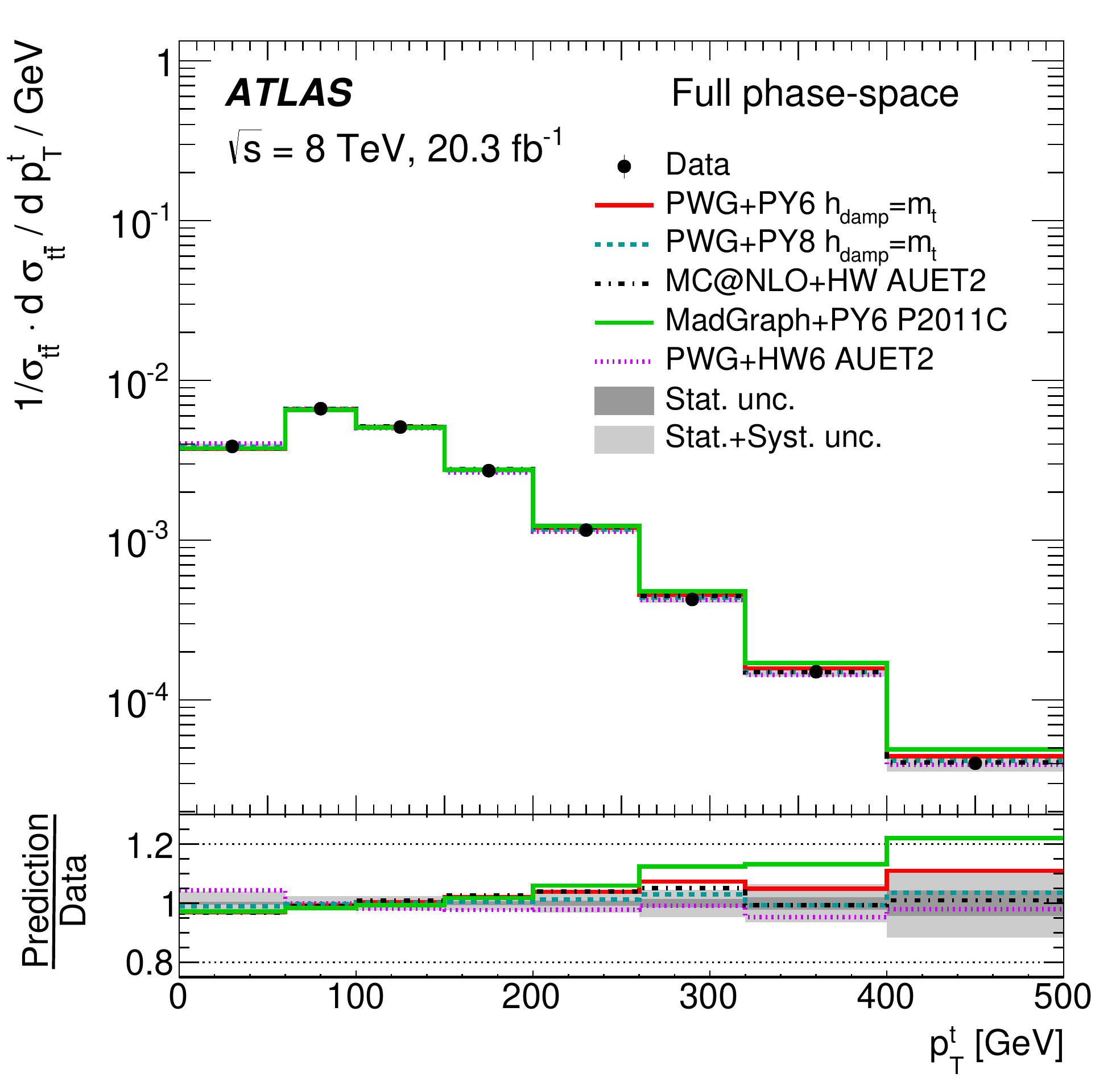}\label{fig:parton:topH_pt:rel}}
\subfigure[]{ \includegraphics[width=0.45\textwidth]{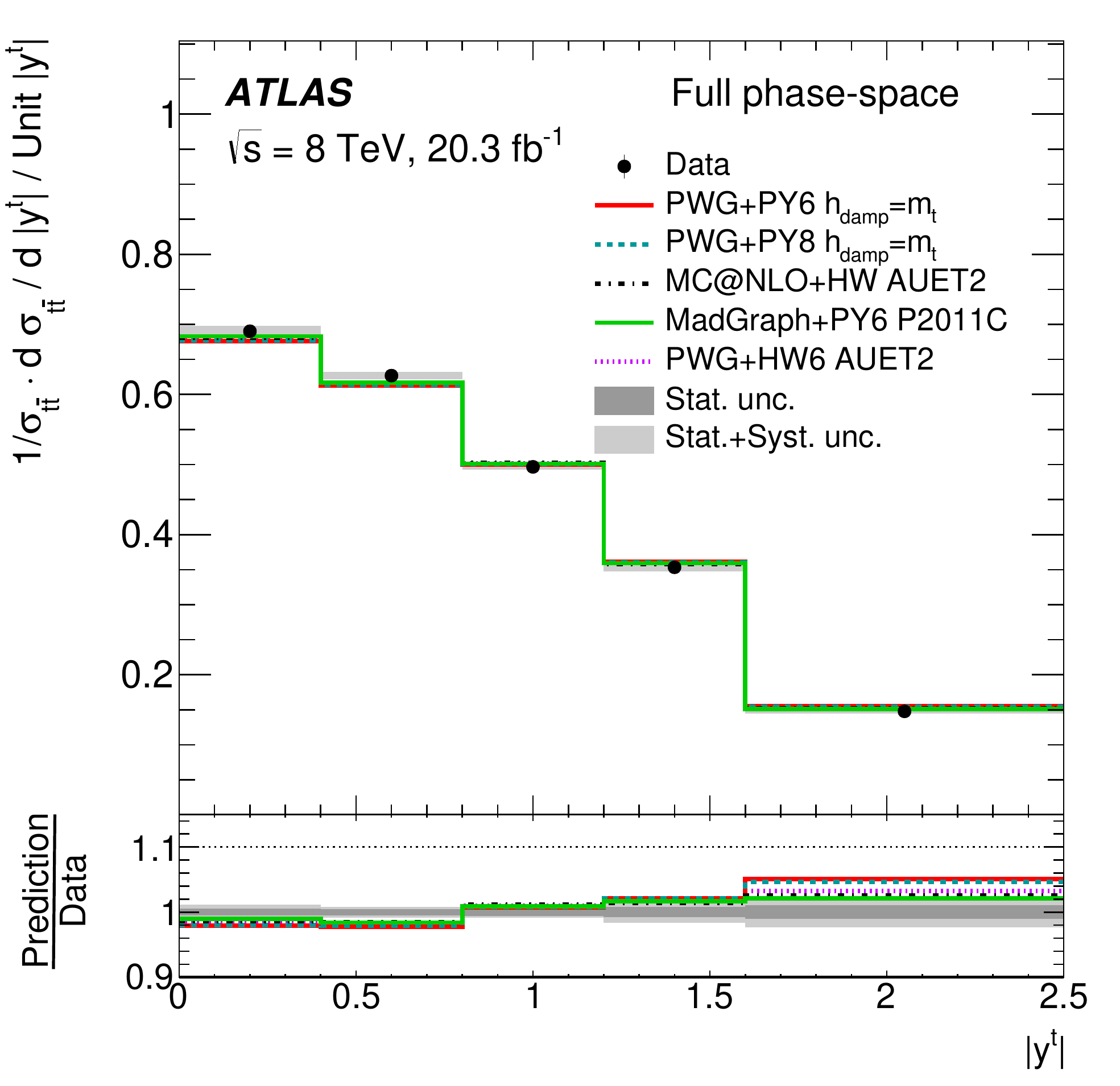}\label{fig:parton:topH_absrap:rel}}
\caption{Full phase-space normalized differential cross-sections as a function of the \subref{fig:parton:topH_pt:rel}~transverse momentum (\ptt{}) and \subref{fig:parton:topH_absrap:rel}~the absolute value of the rapidity (\absyt{}) of the top quark. The grey bands indicate the total uncertainty on the data in each bin. The \PowHeg{}+\Pythia generator with \HDampMT~and the CT10nlo PDF is used as the nominal prediction to correct for detector effects.}
\label{fig:results:full:topH:rel}
\end{figure*}

\begin{figure*}[htbp]
\centering
\subfigure[]{ \includegraphics[width=0.45\textwidth]{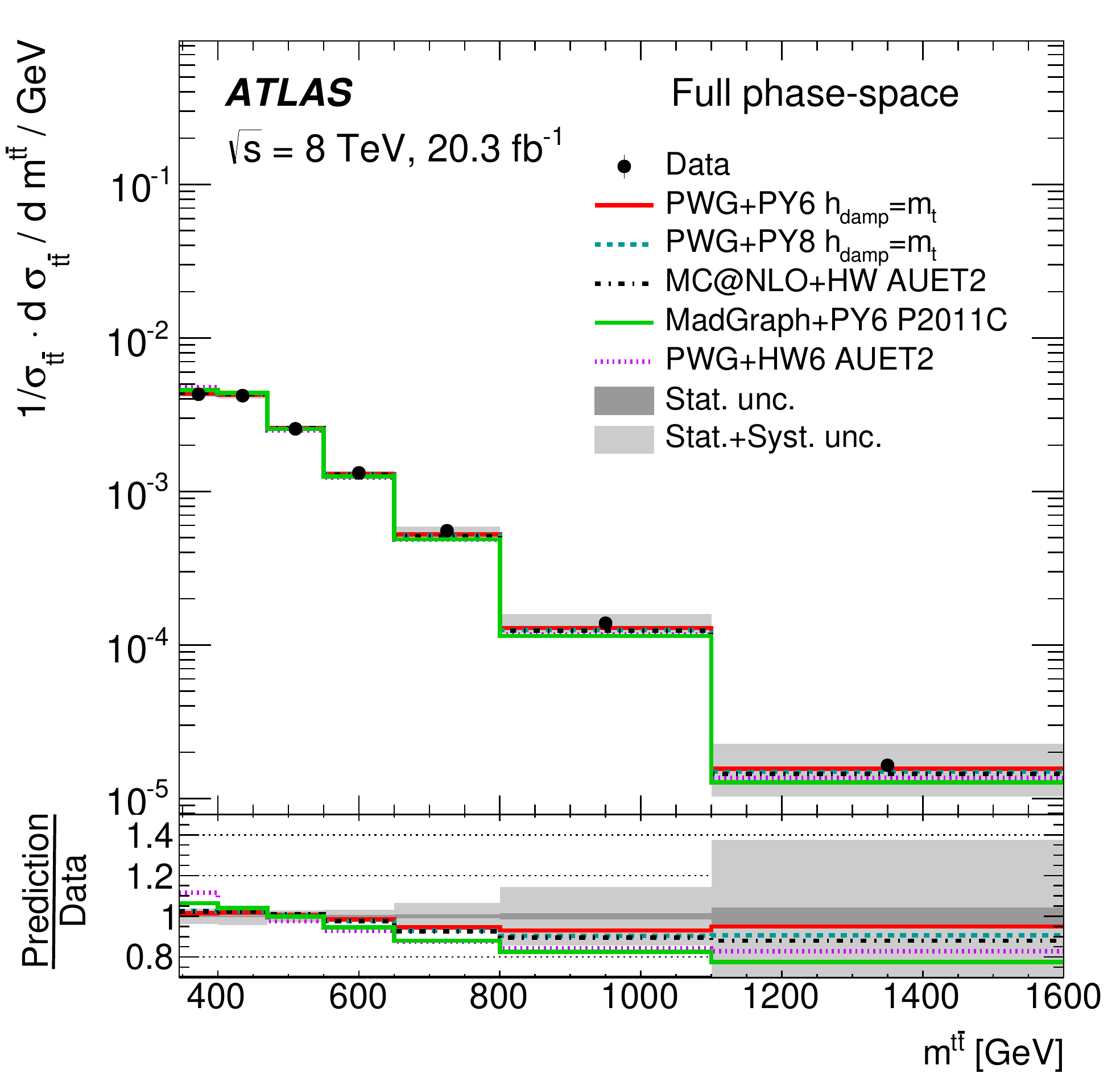}\label{fig:parton:tt_m:rel}}
\subfigure[]{ \includegraphics[width=0.45\textwidth]{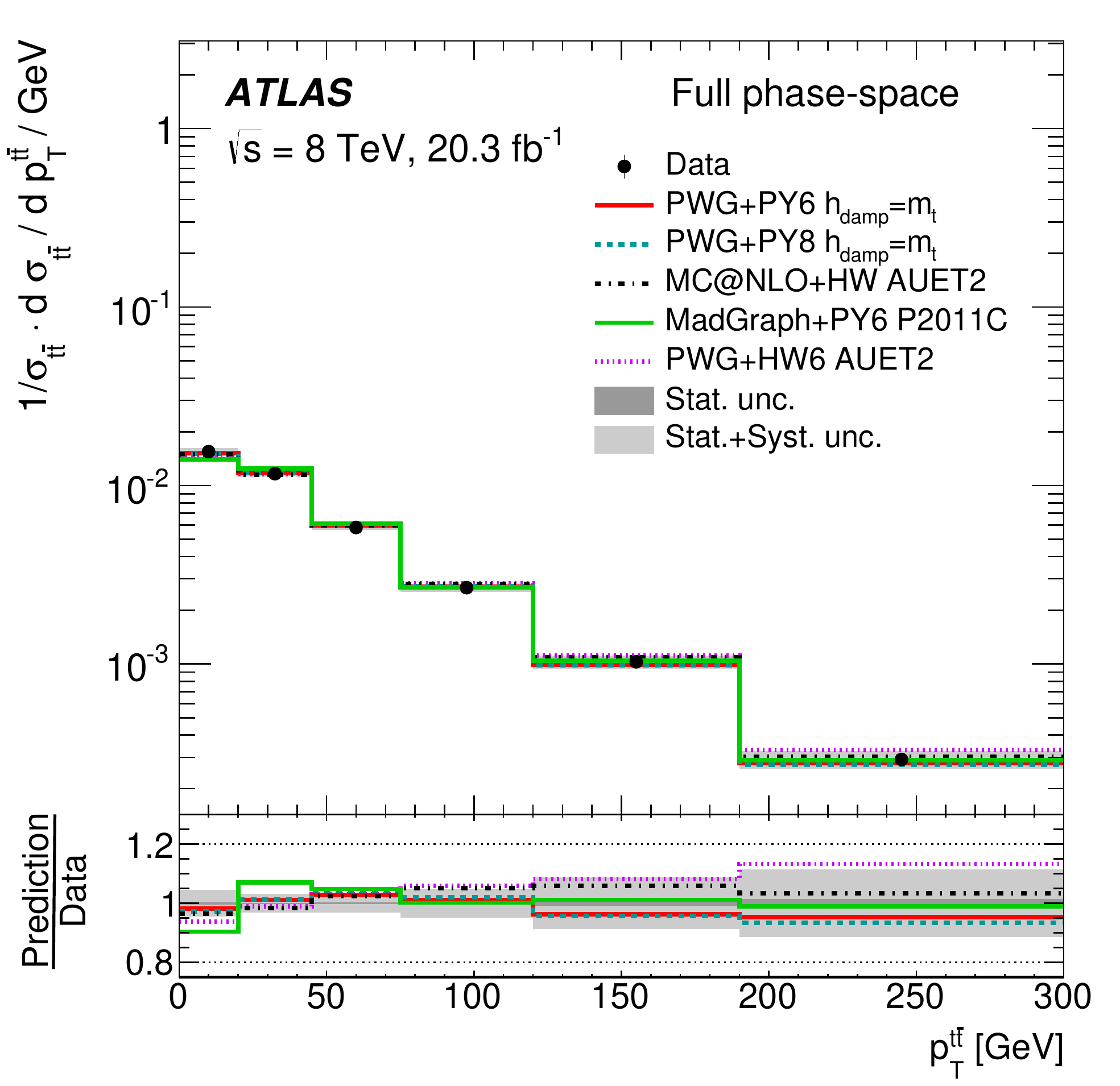}\label{fig:parton:tt_pt:rel}}
\subfigure[]{ \includegraphics[width=0.45\textwidth]{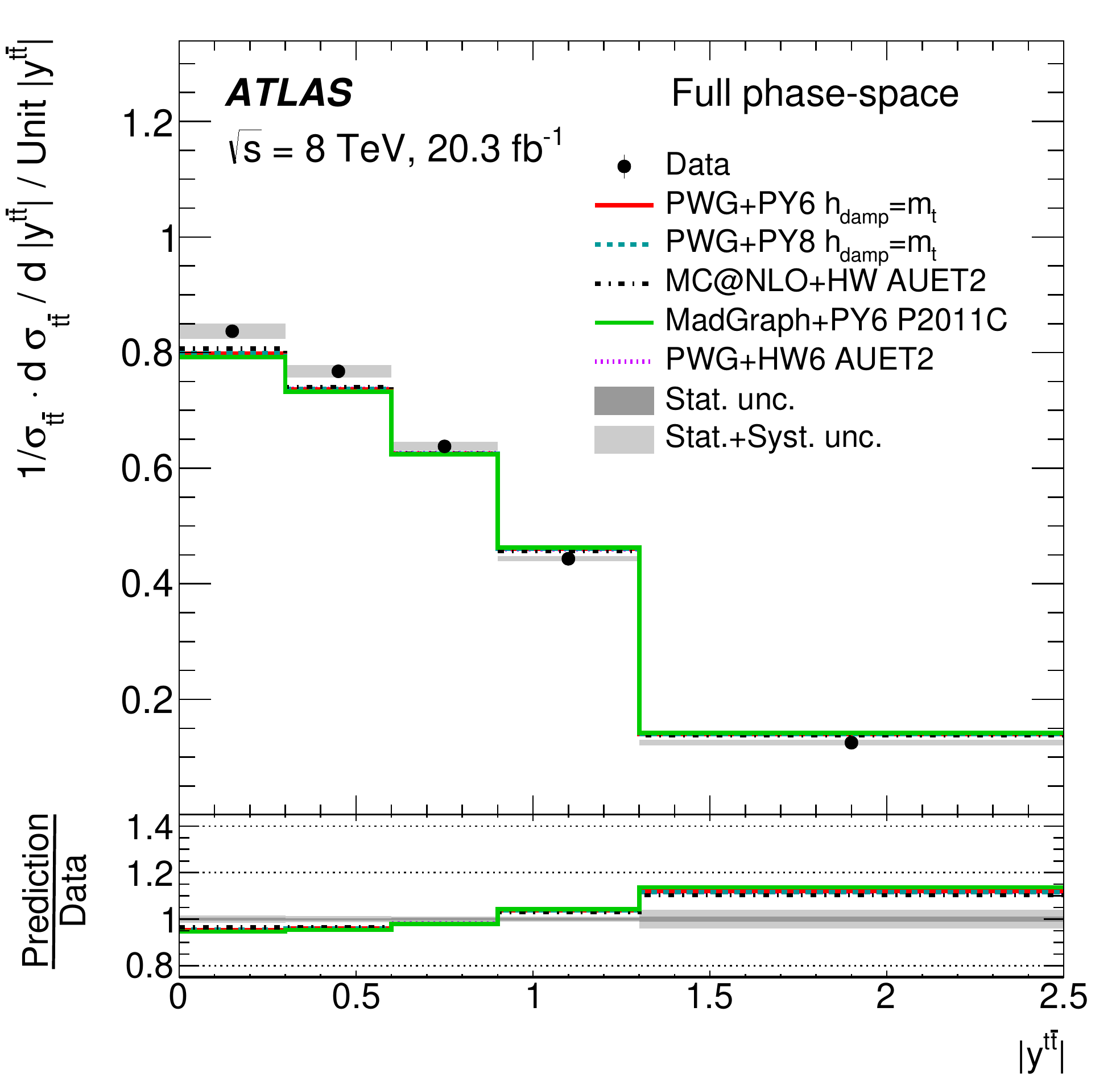}\label{fig:parton:tt_absrap:rel}}
\caption{Full phase-space normalized differential cross-sections as a function of the~ \subref{fig:parton:tt_m:rel}~invariant mass (\mttbar{}), \subref{fig:parton:tt_pt:rel}~transverse momentum (\ptttbar{}) and \subref{fig:parton:tt_absrap:rel}~absolute value of the rapidity (\absyttbar{}) of the \ttb{}~system. The grey bands indicate the total uncertainty on the data in each bin. The \PowHeg{}+\Pythia generator with \HDampMT~and the CT10nlo PDF is used as the nominal prediction to correct for detector effects.}
\label{fig:results:full:tt:rel}
\end{figure*}

\begin{figure*}[htbp]
\centering
\subfigure[]{ \includegraphics[width=0.45\textwidth]{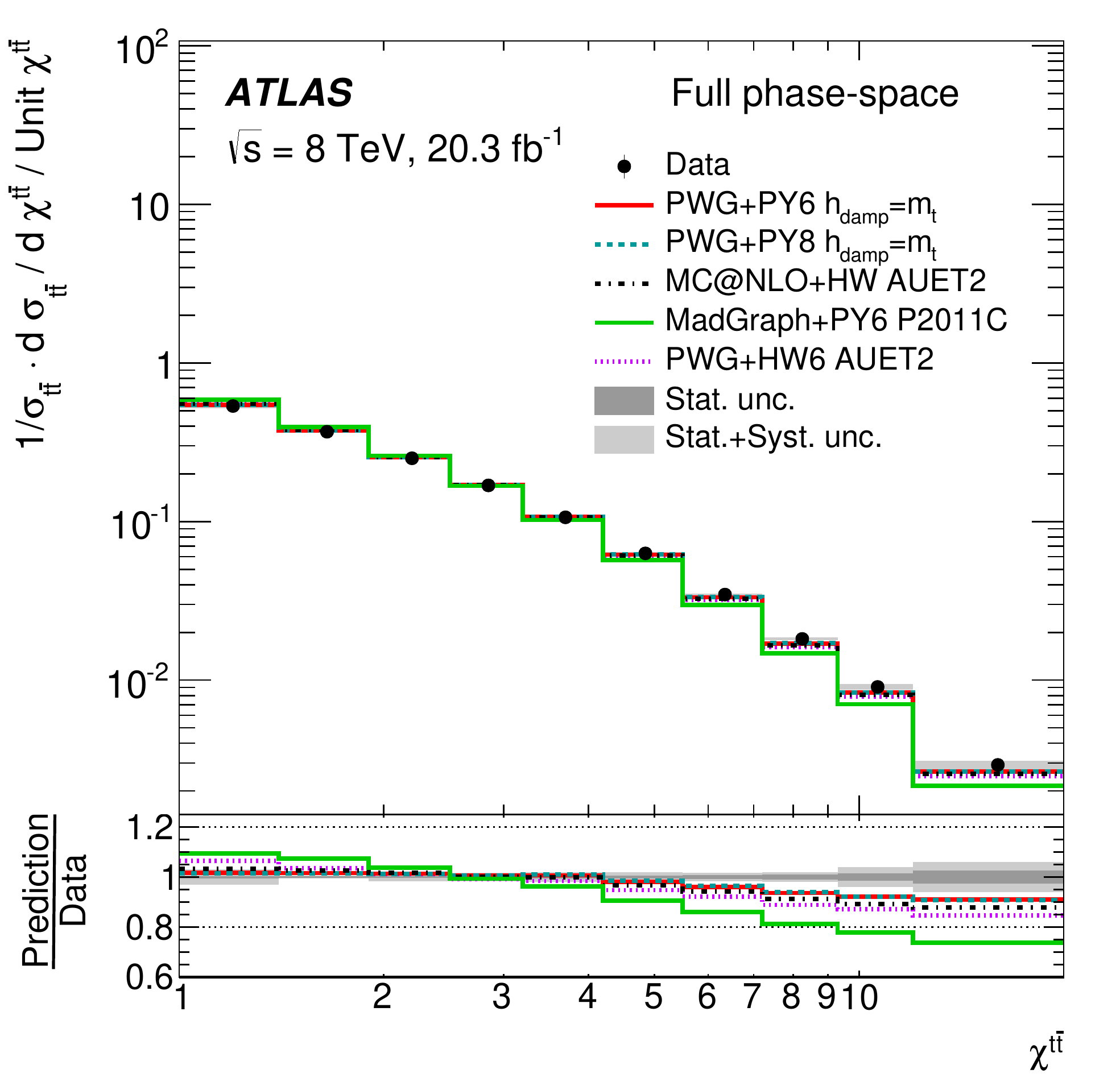}\label{fig:parton:Chi_ttbar:rel}}
\subfigure[]{ \includegraphics[width=0.45\textwidth]{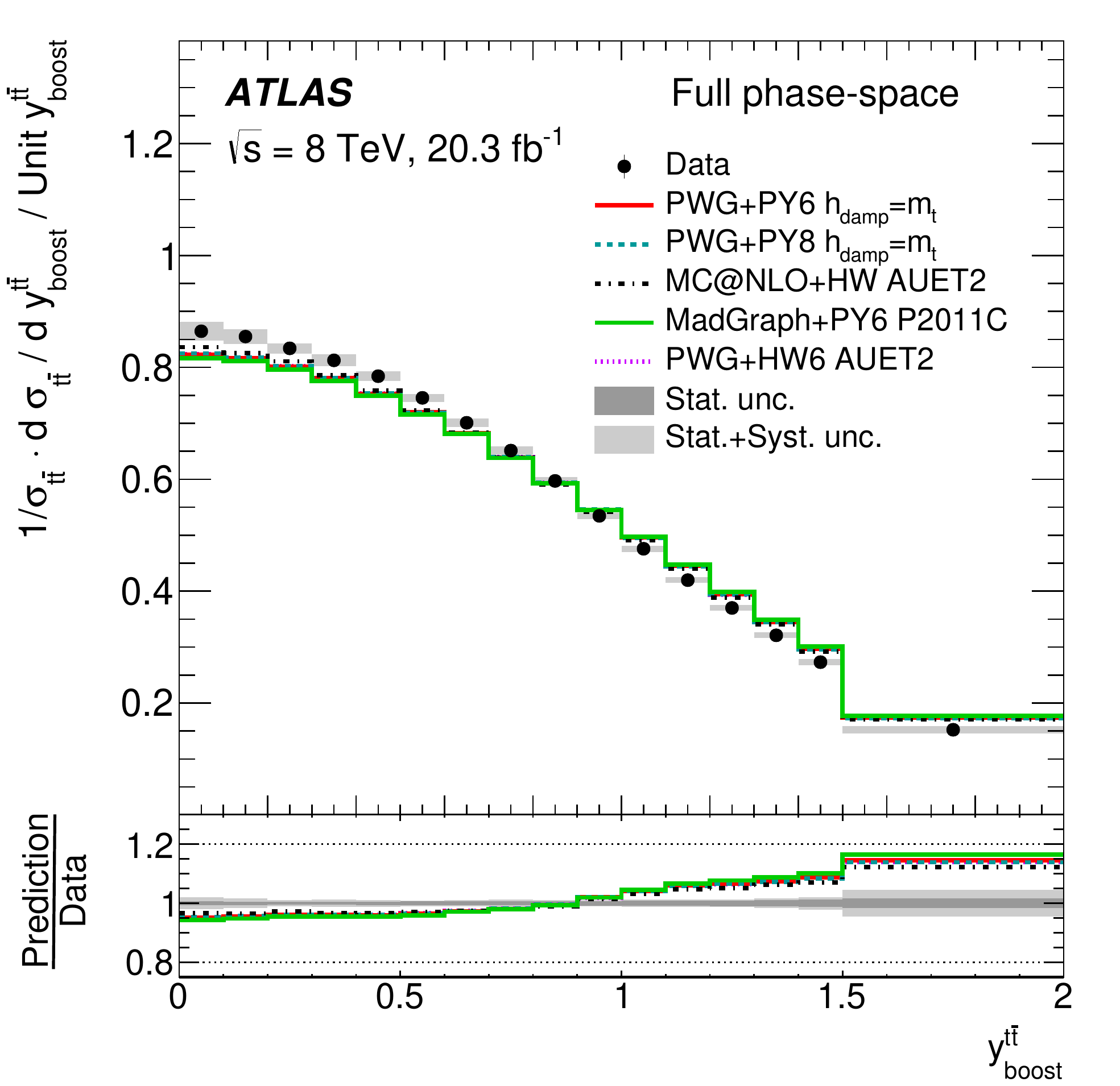}\label{fig:parton:Yboost:rel}}
\caption{Full phase-space normalized differential cross-sections as a function of the~ \subref{fig:parton:Chi_ttbar:rel}~production angle (\chittbar{}) and \subref{fig:parton:Yboost:rel}~longitudinal boost (\boostttbar{}) of the \ttb{}~system. The grey bands indicate the total uncertainty on the data in each bin. The \PowHeg{}+\Pythia generator with \HDampMT~and the CT10nlo PDF is used as the nominal prediction to correct for detector effects.}
\label{fig:results:full:qcd1:rel}
\end{figure*}

\begin{figure*}[htbp]
\centering
\subfigure[]{ \includegraphics[width=0.45\textwidth]{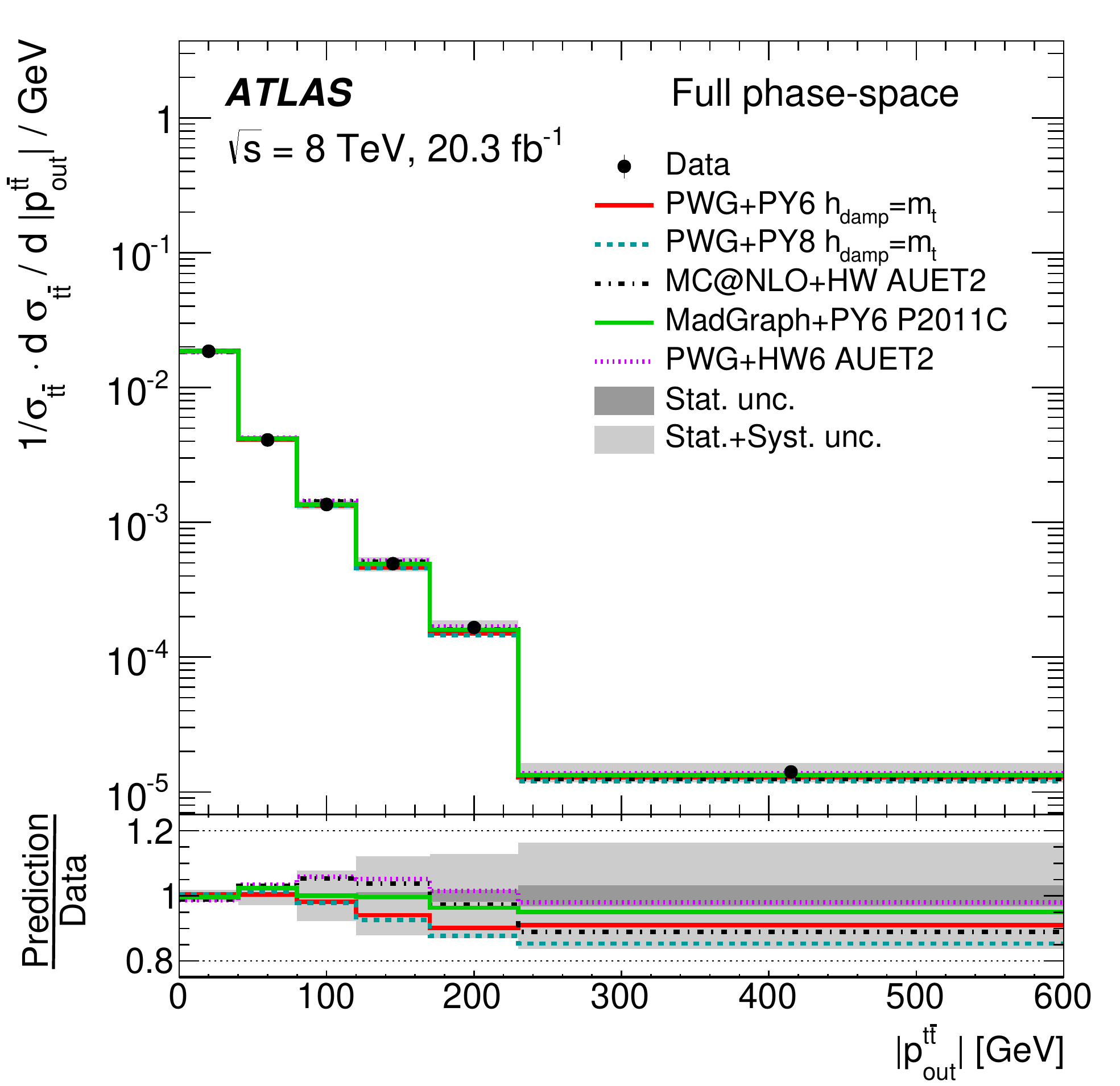}\label{fig:parton:Pout:rel}}
\subfigure[]{ \includegraphics[width=0.45\textwidth]{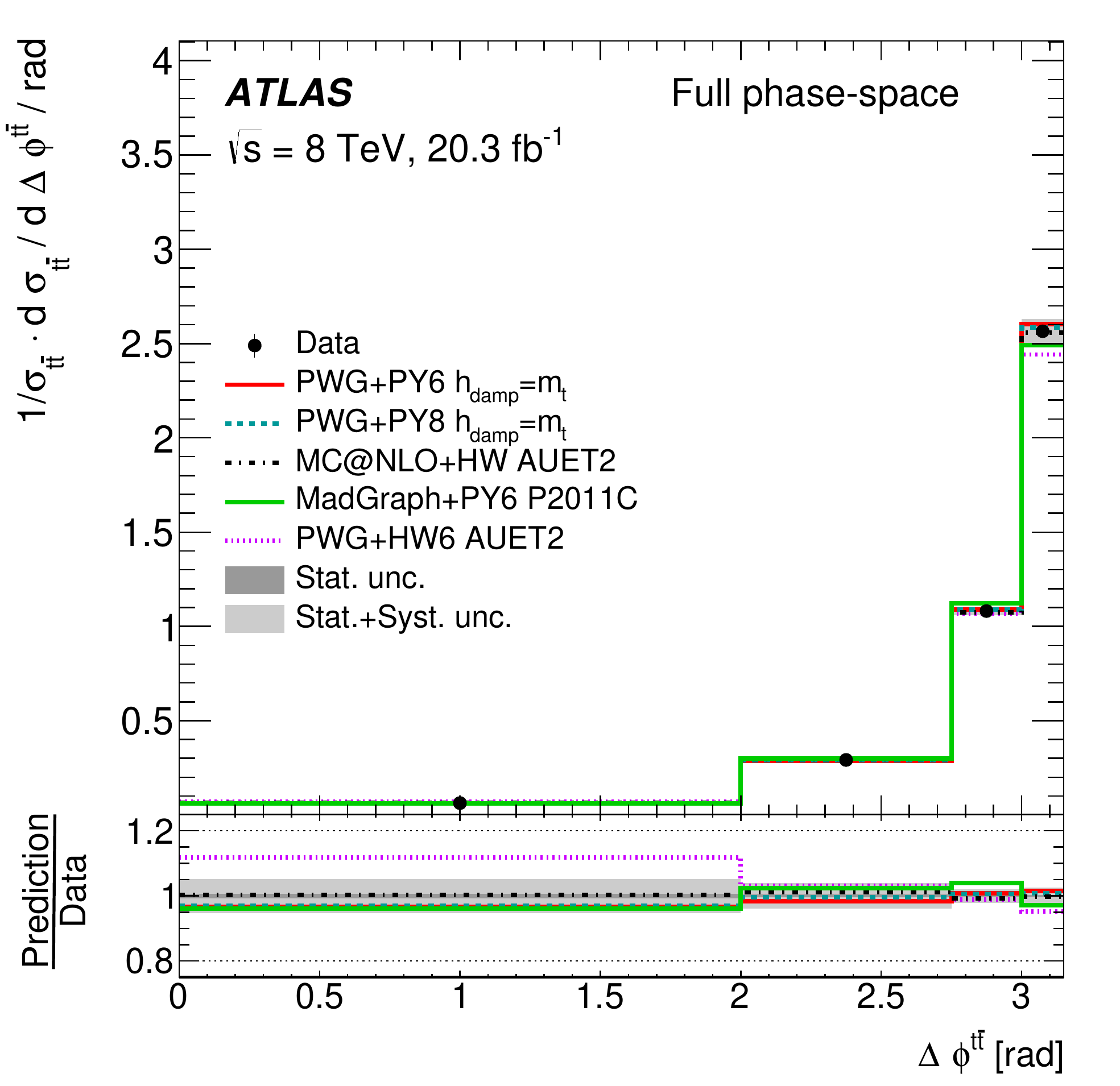}\label{fig:parton:tt_dPhi:rel}}
\subfigure[]{ \includegraphics[width=0.45\textwidth]{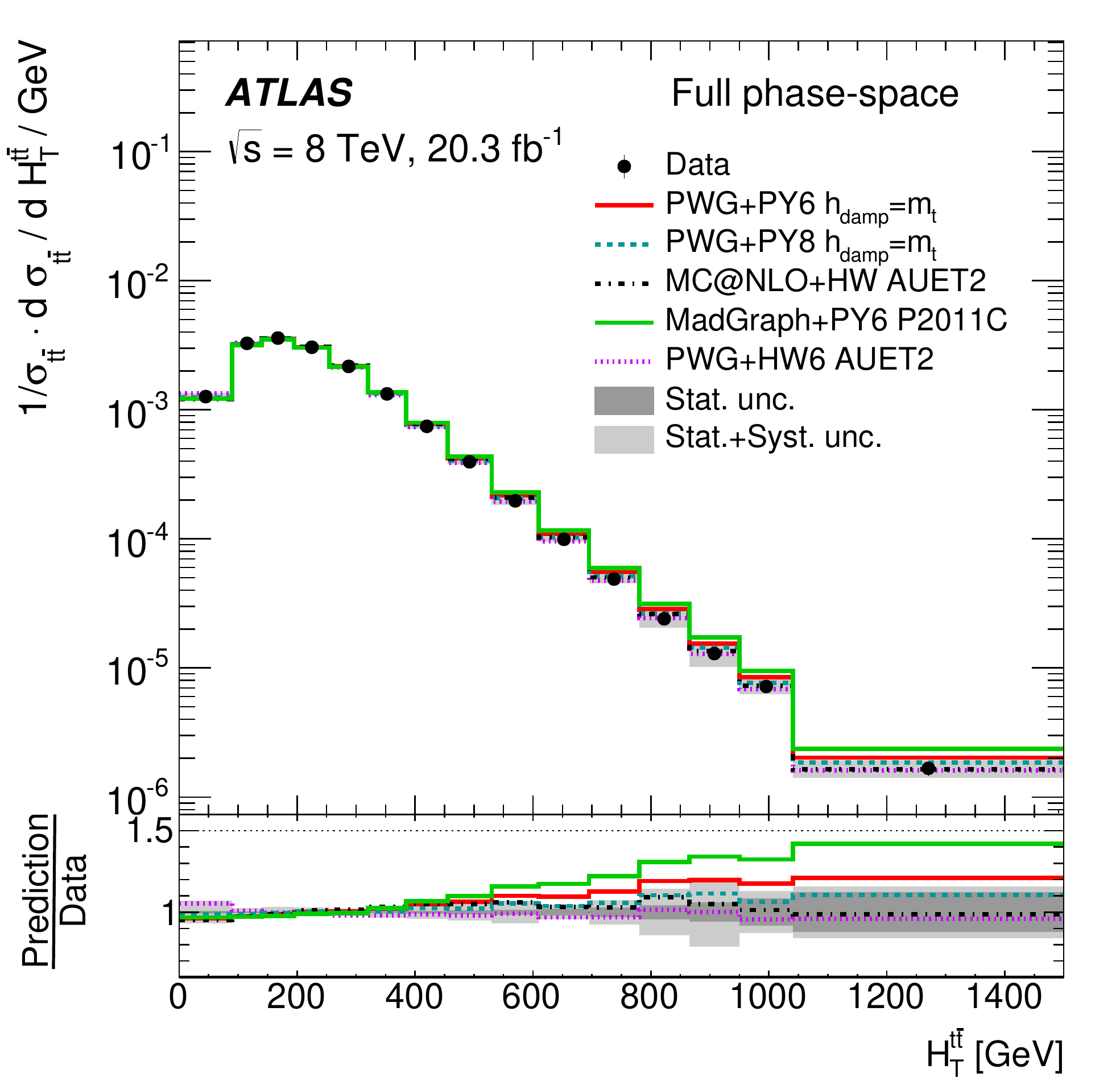}\label{fig:parton:HT_ttbar:rel}}
\caption{Full phase-space normalized differential cross-sections as a function of the~ \subref{fig:parton:Pout:rel}~out-of-plane momentum (\absPoutttbar{}),  \subref{fig:parton:tt_dPhi:rel}~azimuthal angle (\DeltaPhittbar), and \subref{fig:parton:HT_ttbar:rel}~ scalar sum of the transverse momenta of the hadronic and leptonic top quarks (\HTttbar)) of the \ttb{}~system. The grey bands indicate the total uncertainty on the data in each bin. The \PowHeg{}+\Pythia generator with \HDampMT~and the CT10nlo PDF is used as the nominal prediction to correct for detector effects.}
\label{fig:results:full:qcd2:rel}
\end{figure*}

\begin{figure*}[htbp]
\centering
\subfigure[]{ \includegraphics[width=0.45\textwidth]{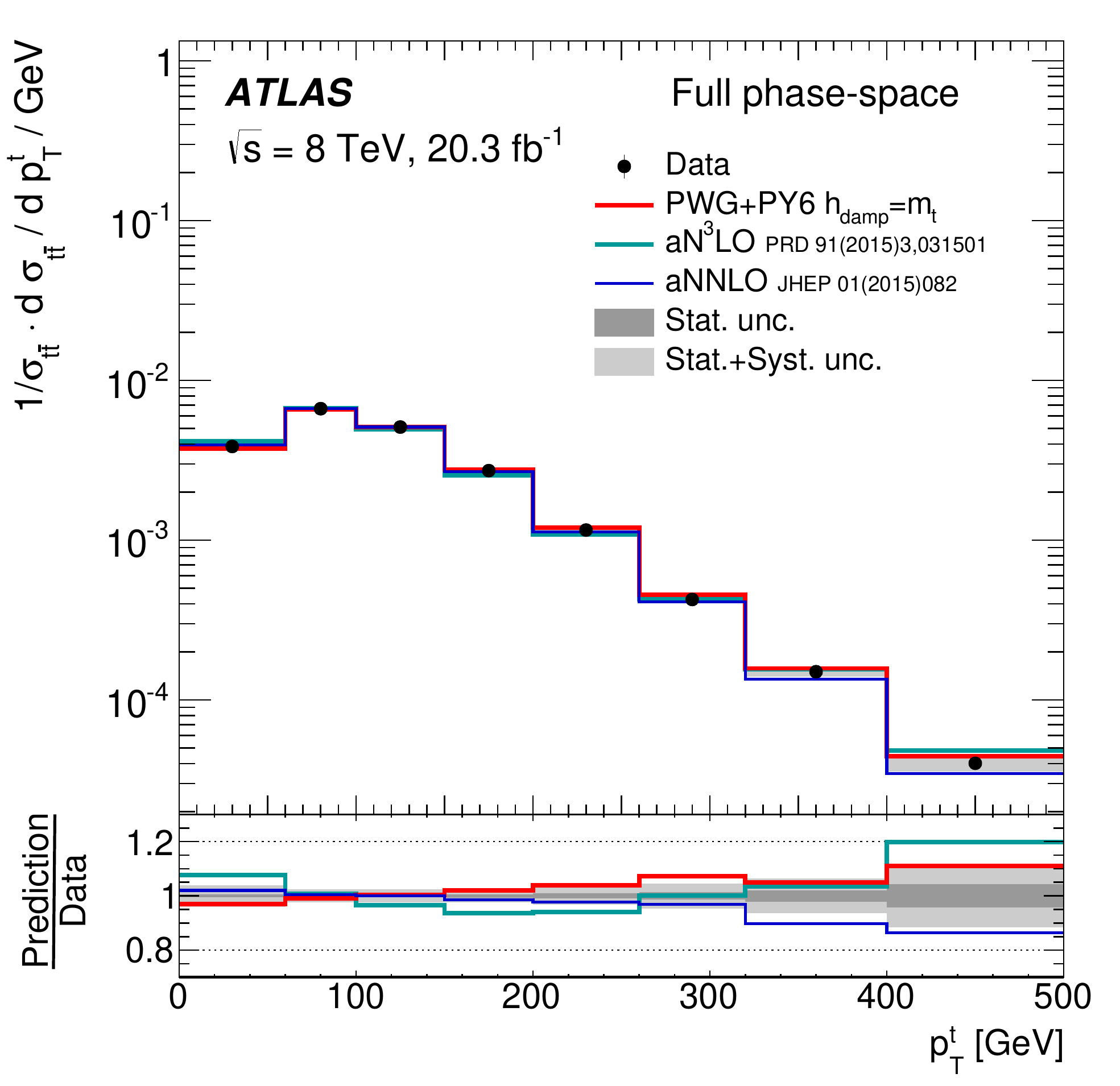}\label{fig:theory:topH_pt:rel}}
\subfigure[]{ \includegraphics[width=0.45\textwidth]{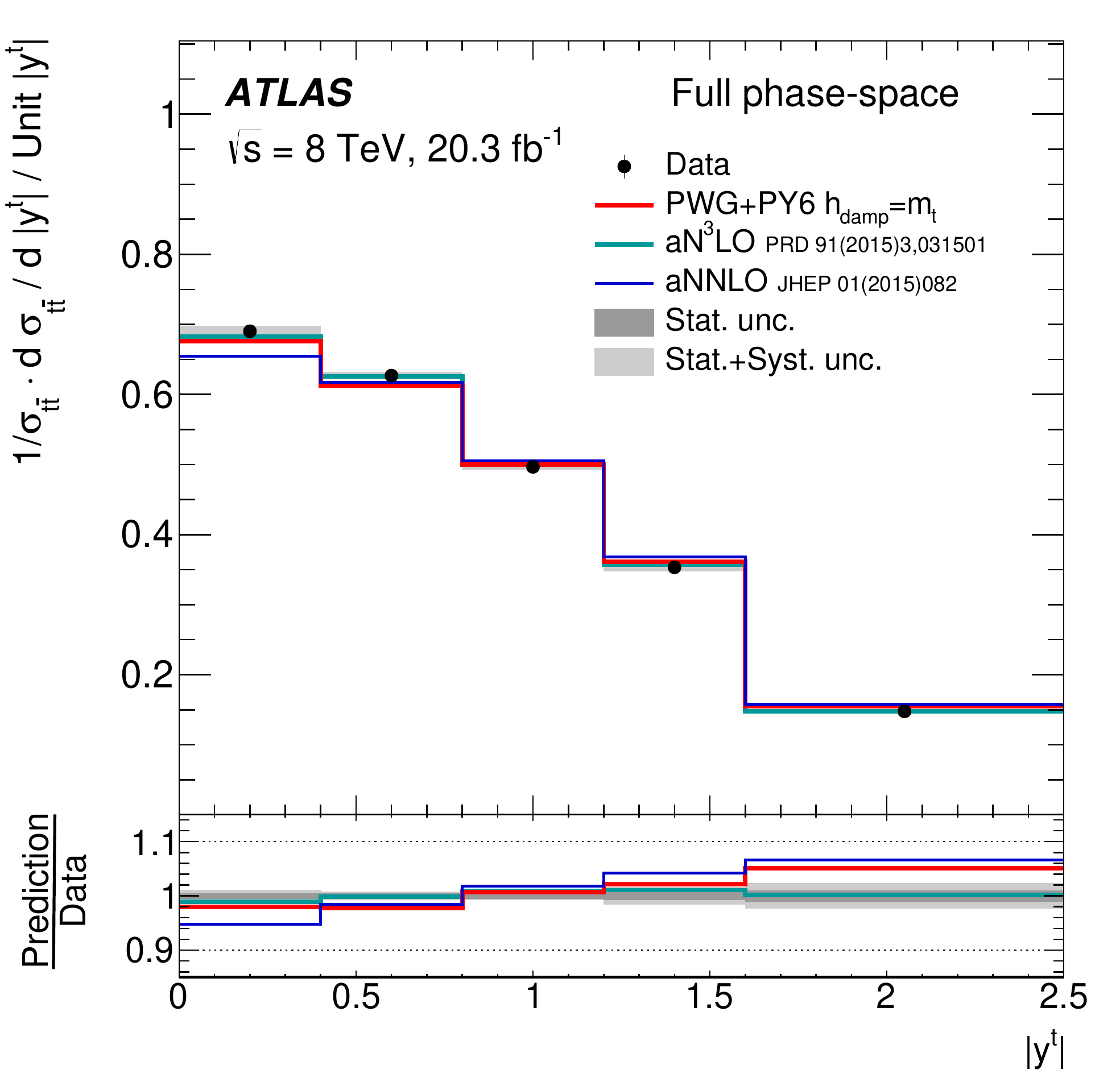}\label{fig:theory:topH_absrap:rel}}
\caption{Full phase-space normalized differential cross-section as a function of the \subref{fig:theory:topH_pt:rel}~transverse momentum (\ptt{}) and \subref{fig:theory:topH_absrap:rel}~absolute value of the rapidity of the top quark (\absyt) compared to higher-order theoretical calculations. The grey band indicates the total uncertainty on the data in each bin. The \PowHeg{}+\Pythia generator with \HDampMT~and the CT10nlo PDF is used as the nominal prediction to correct for detector effects.}
\label{fig:results:theory:topH:rel}
\end{figure*}

\begin{figure*}[htbp]
\centering
\subfigure[]{ \includegraphics[width=0.45\textwidth]{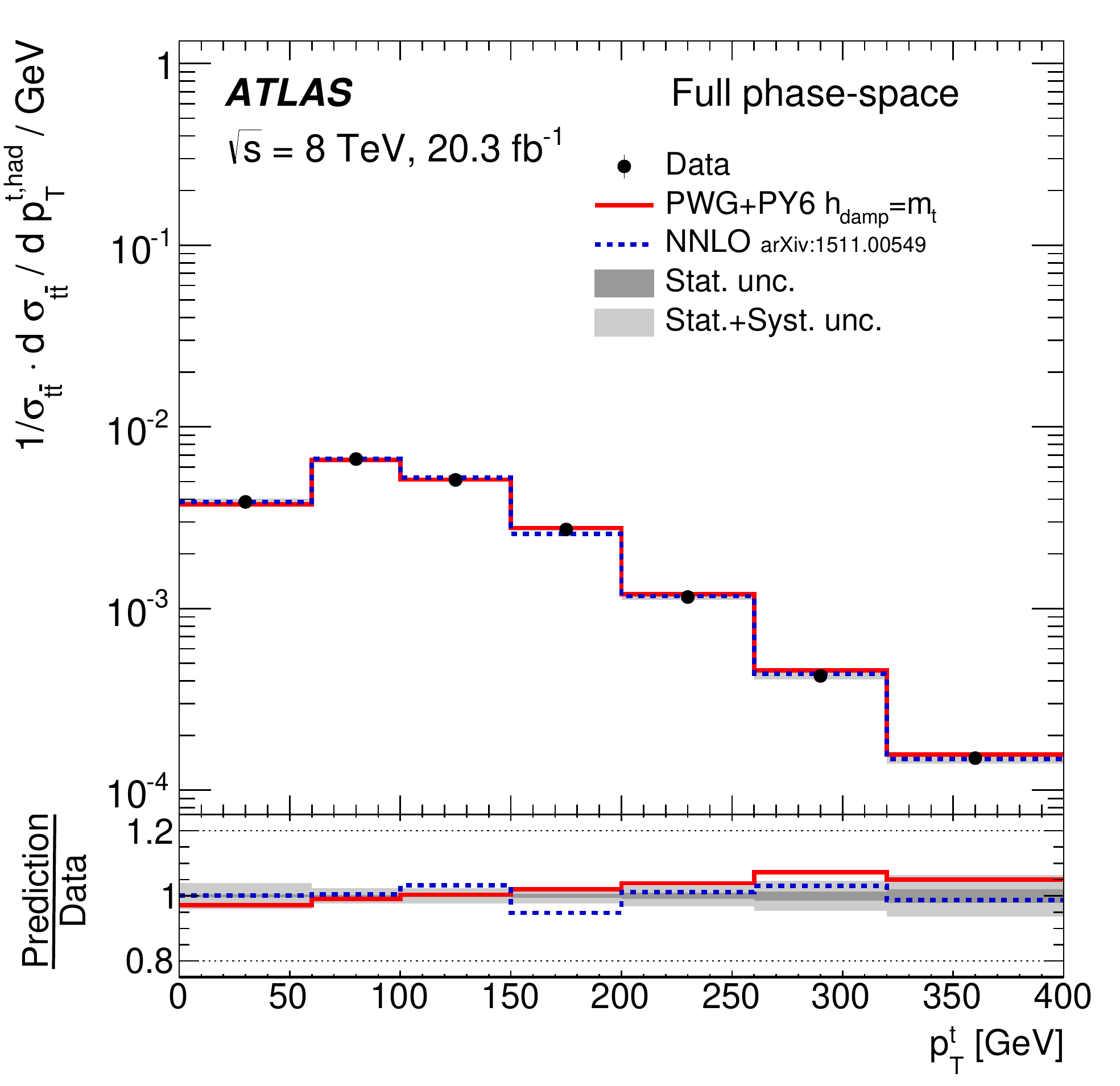}\label{fig:NNLO:topH_pt:rel}}
\subfigure[]{ \includegraphics[width=0.45\textwidth]{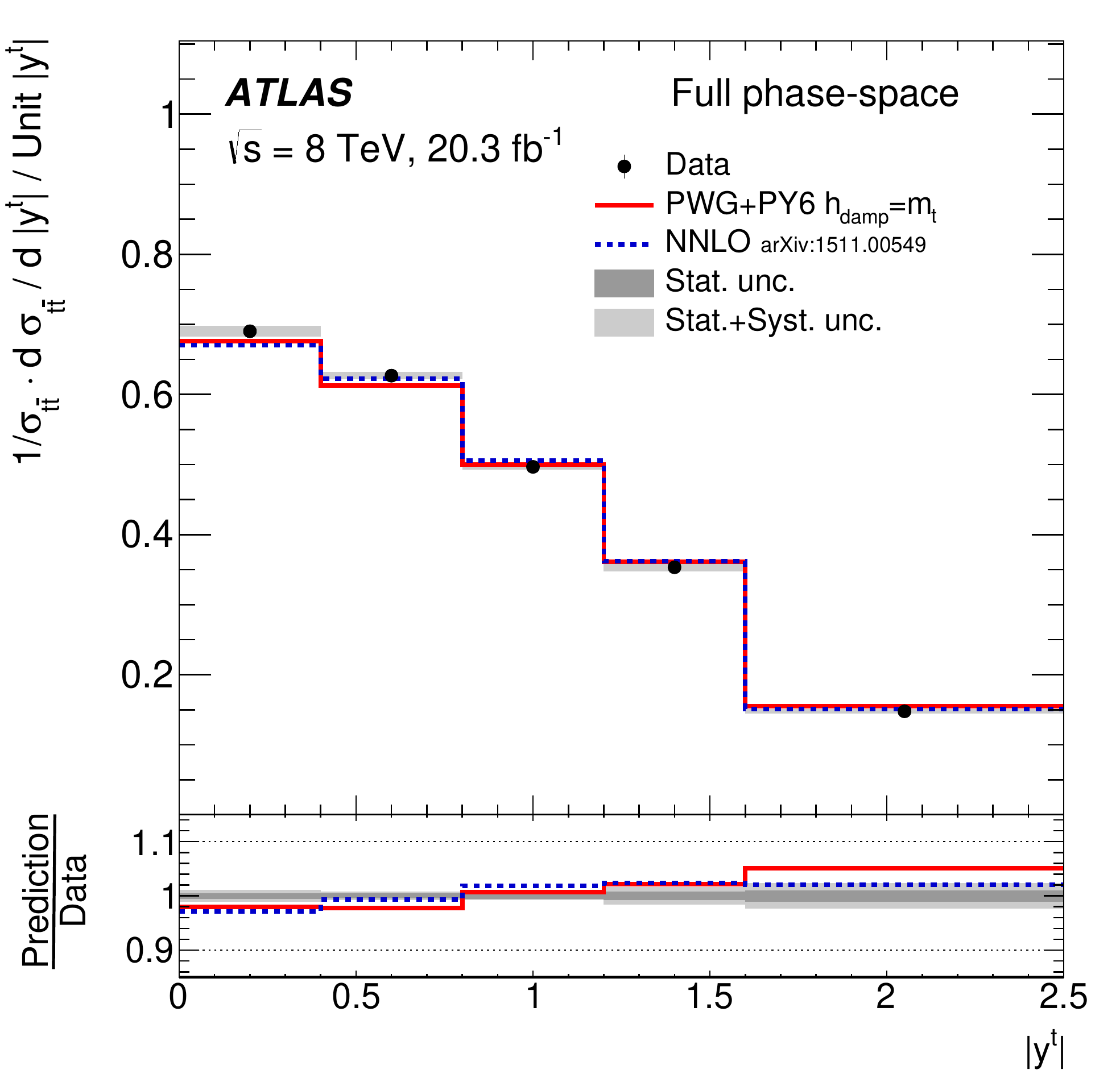}\label{fig:NNLO:topH_absrap:rel}}
\caption{Full phase-space normalized differential cross-section as a function of the \subref{fig:NNLO:topH_pt:rel}~transverse momentum (\ptt{}) and \subref{fig:NNLO:topH_absrap:rel}~absolute value of the rapidity of the top quark (\absyt) compared to NNLO theoretical calculations \cite{stripper} using the MSTW2008nnlo PDF set. The grey band indicates the total uncertainty on the data in each bin. The \PowHeg{}+\Pythia generator with \HDampMT~and the CT10nlo PDF is used as the nominal prediction to correct for detector effects.}
\label{fig:results:NNLO:topH:rel}
\end{figure*}

\begin{figure*}[htbp]
\centering
\subfigure[]{ \includegraphics[width=0.45\textwidth]{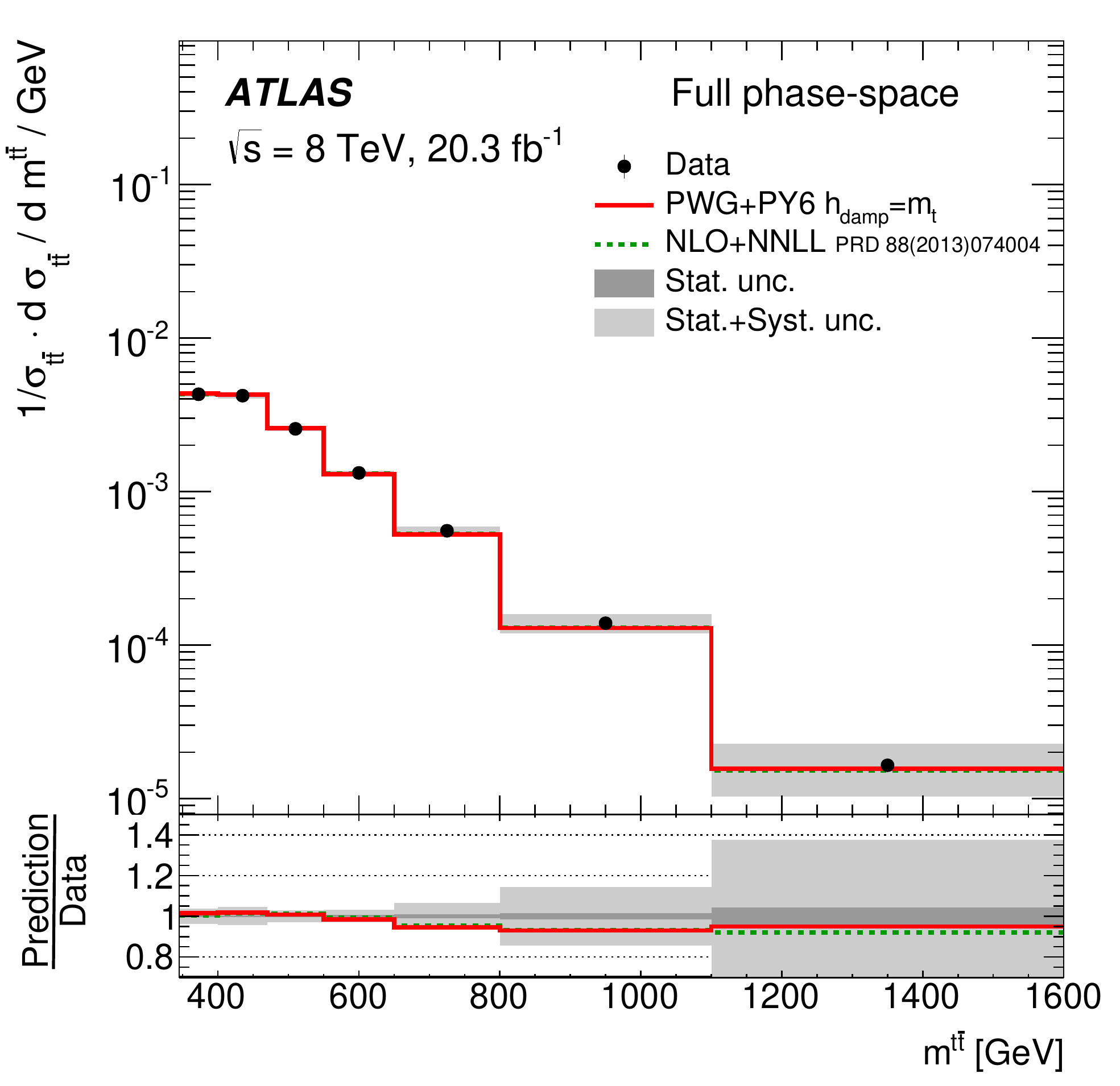}\label{fig:theory:parton:tt_m:rel}}
\subfigure[]{ \includegraphics[width=0.45\textwidth]{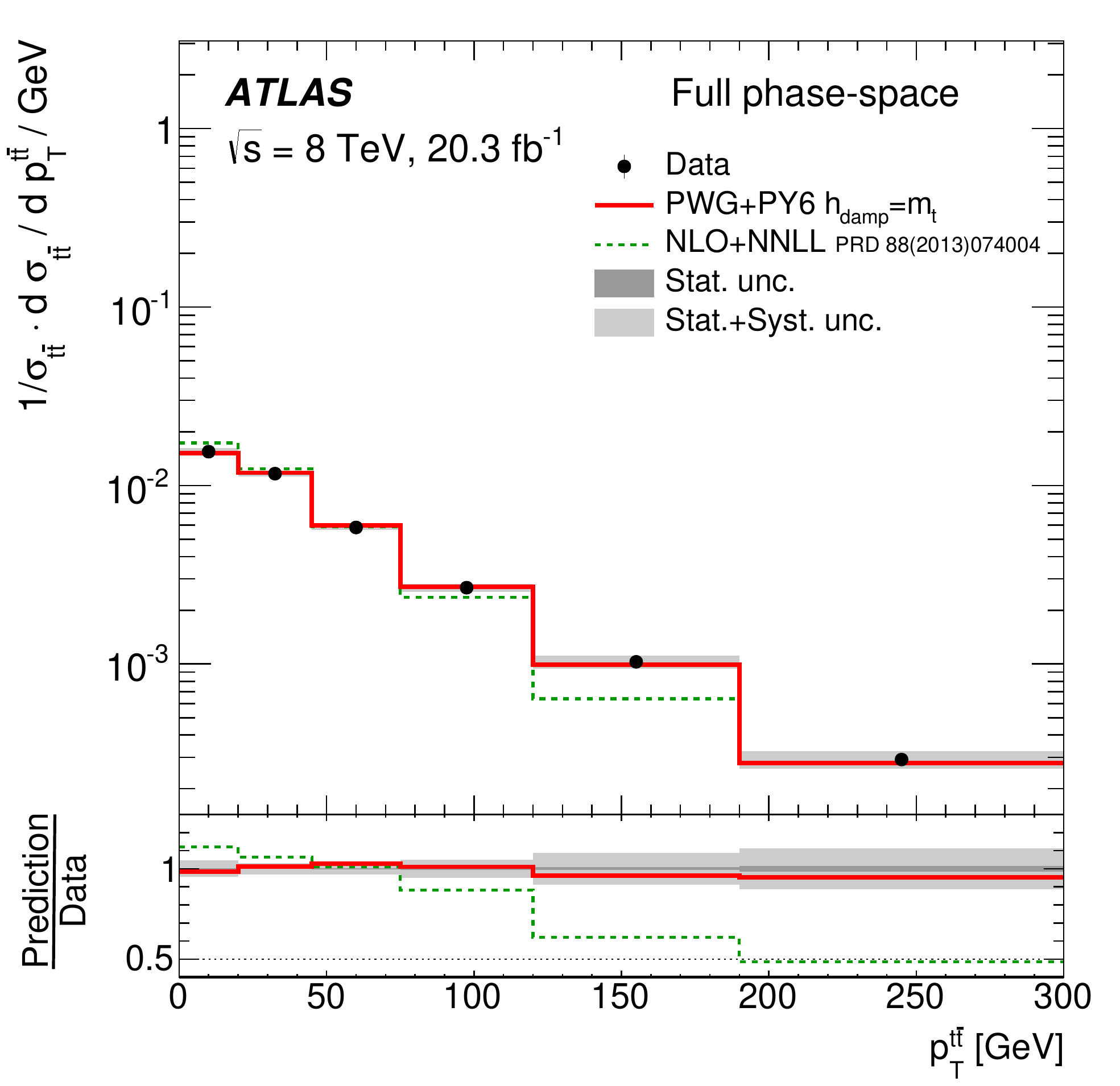}\label{fig:theory:parton:tt_pt:rel}}
\caption{Full phase-space normalized differential cross-section as a function of the \subref{fig:theory:parton:tt_m:rel}~invariant mass (\mttbar{}~) and  \subref{fig:theory:parton:tt_pt:rel}~transverse momentum (\ptttbar) of the \ttbar~system compared to higher-order theoretical calculations. The grey band indicates the total uncertainty on the data in each bin. The \PowHeg{}+\Pythia generator with \HDampMT~and the CT10nlo PDF is used as the nominal prediction to correct for detector effects.}
\label{fig:results:theory:tt:rel}
\end{figure*}

\begin{figure*}[htbp]
\centering
\subfigure[]{ \includegraphics[width=0.45\textwidth]{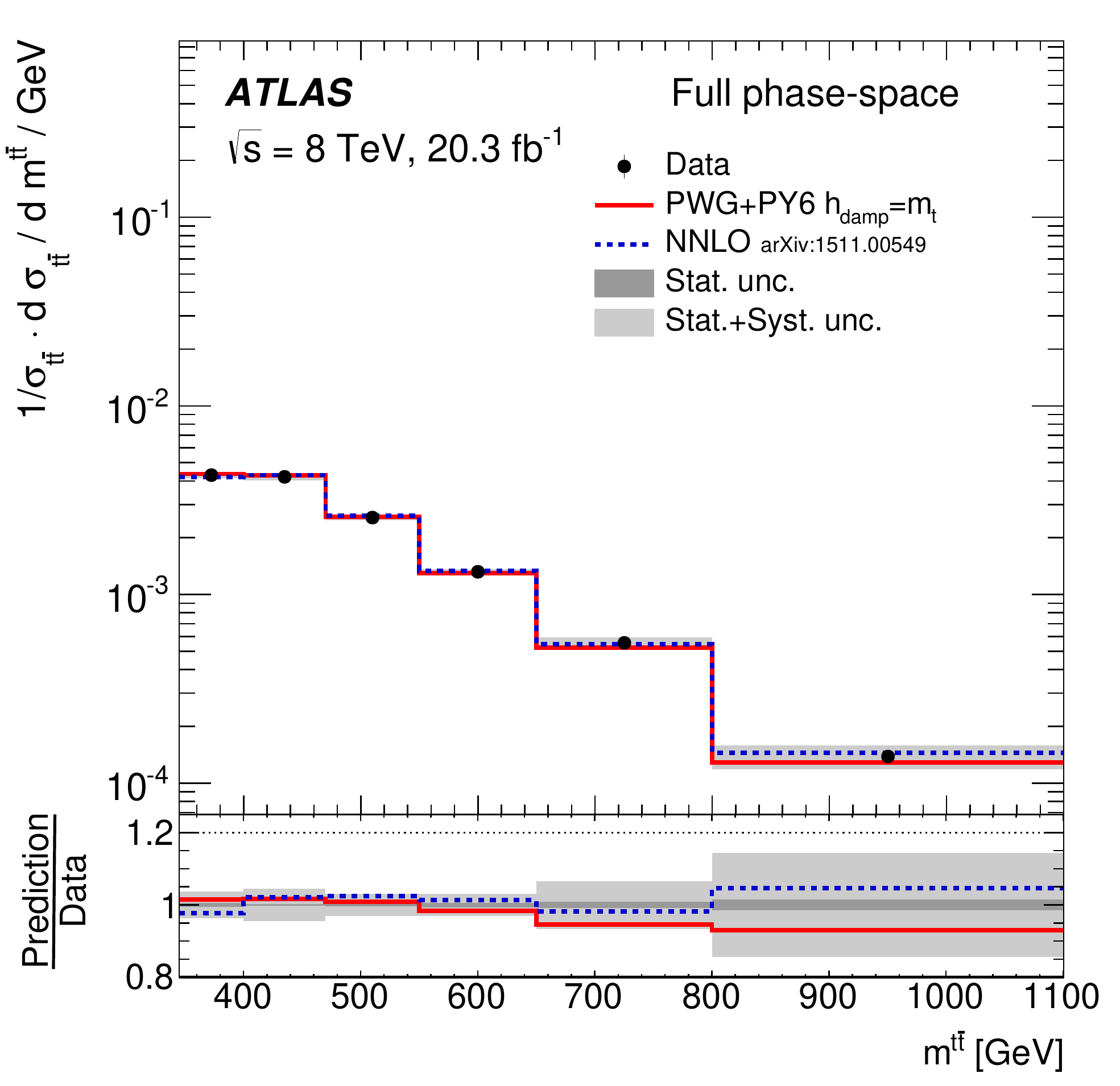}\label{fig:NNLO:parton:tt_m:rel}}
\subfigure[]{ \includegraphics[width=0.45\textwidth]{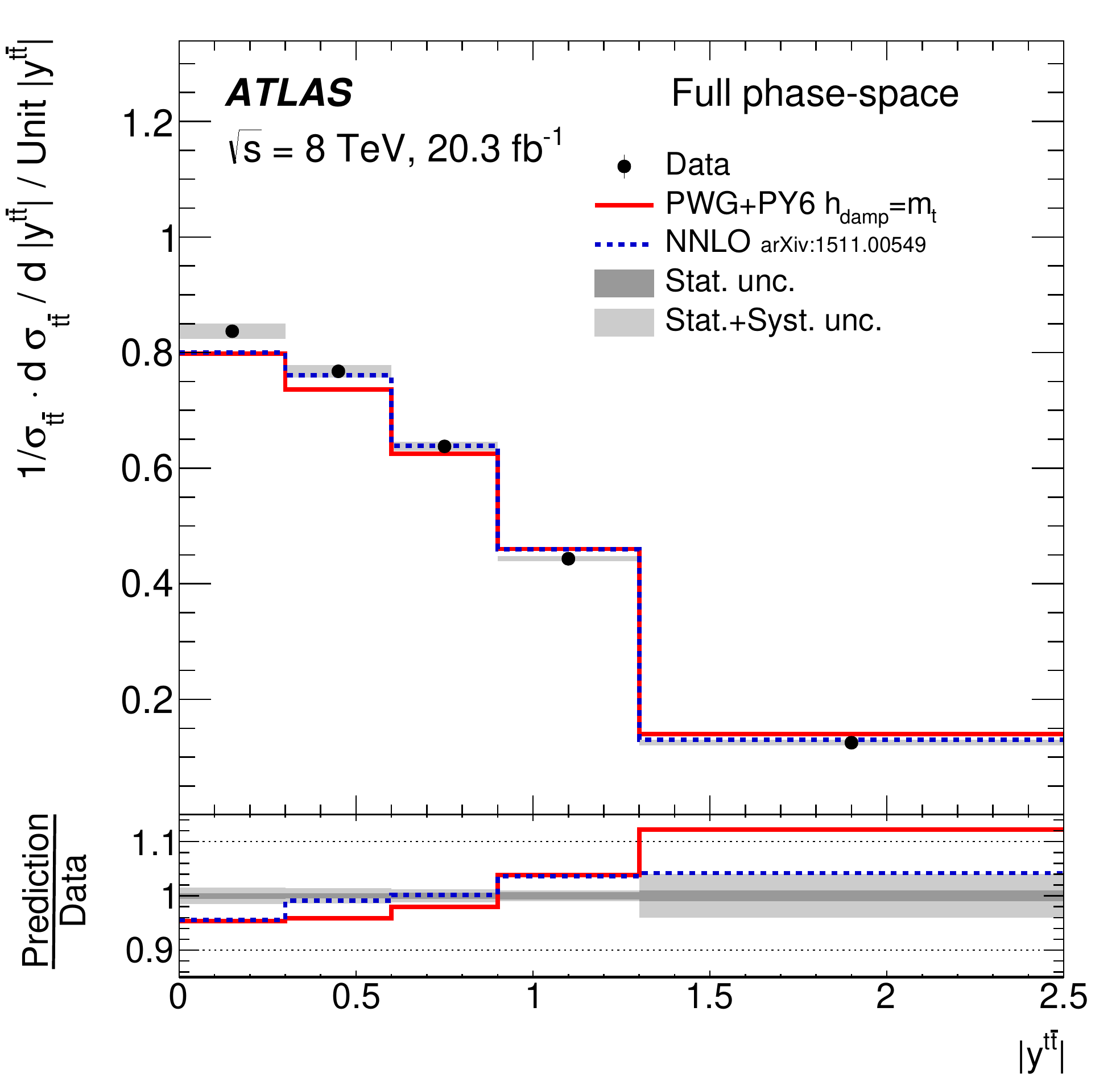}\label{fig:NNLO:parton:tt_absrap:rel}}
\caption{Full phase-space normalized differential cross-section as a function of the \subref{fig:NNLO:parton:tt_m:rel}~invariant mass (\mttbar{}~) and \subref{fig:NNLO:parton:tt_absrap:rel} absolute value of the rapidity (\absyttbar) of the \ttbar~system compared to NNLO theoretical calculations \cite{stripper} using the MSTW2008nnlo PDF set. The grey band indicates the total uncertainty on the data in each bin. The \PowHeg{}+\Pythia generator with \HDampMT~and the CT10nlo PDF is used as the nominal prediction to correct for detector effects.}
\label{fig:results:NNLO:tt:rel}
\end{figure*}

\clearpage

\begin{table*} [p]
\footnotesize
\centering
\noindent\makebox[\textwidth]{
\begin{tabular}{c | r @{/} l c  | r @{/} l c  | r @{/} l c  | r @{/} l c  | r @{/} l c }
\hline
Variable  & \multicolumn{3}{|c|}{PWG+PY8} & \multicolumn{3}{|c|}{MC@NLO+HW} & \multicolumn{3}{|c|}{PWG+PY6} & \multicolumn{3}{|c|}{PWG+HW6} & \multicolumn{3}{|c}{MadGraph+PY6} \\
  & \multicolumn{3}{|c|}{CT10 $h_{\rm damp}=m_{t}$} & \multicolumn{3}{|c|}{CT10 AUET2} & \multicolumn{3}{|c|}{CT10 $h_{\rm damp}=m_{t}$} & \multicolumn{3}{|c|}{CT10 $h_{\rm damp}=\infty$} & \multicolumn{3}{|c}{P2011C} \\
& \multicolumn{2}{|c}{$\chi^{2}$/NDF} &  ~$p$-value  & \multicolumn{2}{|c}{$\chi^{2}$/NDF} &  ~$p$-value  & \multicolumn{2}{|c}{$\chi^{2}$/NDF} &  ~$p$-value  & \multicolumn{2}{|c}{$\chi^{2}$/NDF} &  ~$p$-value  & \multicolumn{2}{|c}{$\chi^{2}$/NDF} &  ~$p$-value  \\
\hline
\hline
$ p_{T}^{t,{\rm had}}$ & {\ } 9.5 & 14 & 0.80  &  {\ } 13 & 14 & 0.56  &  {\ } 11 & 14 & 0.68  &  {\ } 4.8 & 14 & 0.99  &  {\ } 41 & 14 & <0.01 \\
$              R_{Wt}$ & {\ } 16 & 11 & 0.14  &  {\ } 14 & 11 & 0.23  &  {\ } 21 & 11 & 0.03  &  {\ } 5.6 & 11 & 0.90  &  {\ } 48 & 11 & <0.01 \\
$     \chi^{t\bar{t}}$ & {\ } 18 & 9 & 0.04  &  {\ } 24 & 9 & <0.01  &  {\ } 17 & 9 & 0.04  &  {\ } 34 & 9 & <0.01  &  {\ } 130 & 9 & <0.01 \\
$      |y^{t\bar{t}}|$ & {\ } 35 & 17 & <0.01  &  {\ } 25 & 17 & 0.10  &  {\ } 31 & 17 & 0.02  &  {\ } 33 & 17 & 0.01  &  {\ } 58 & 17 & <0.01 \\
$        m^{t\bar{t}}$ & {\ } 17 & 10 & 0.08  &  {\ } 33 & 10 & <0.01  &  {\ } 11 & 10 & 0.38  &  {\ } 16 & 10 & 0.11  &  {\ } 18 & 10 & 0.05 \\
$y_{boost}^{t\bar{t}}$ & {\ } 39 & 15 & <0.01  &  {\ } 25 & 15 & 0.06  &  {\ } 35 & 15 & <0.01  &  {\ } 38 & 15 & <0.01  &  {\ } 65 & 15 & <0.01 \\
$|p_{out}^{t\bar{t}}|$ & {\ } 3.4 & 5 & 0.63  &  {\ } 3.1 & 5 & 0.69  &  {\ } 7.7 & 5 & 0.18  &  {\ } 5.6 & 5 & 0.35  &  {\ } 5.9 & 5 & 0.31 \\
$   |y^{t,{\rm had}}|$ & {\ } 19 & 17 & 0.33  &  {\ } 13 & 17 & 0.75  &  {\ } 17 & 17 & 0.47  &  {\ } 14 & 17 & 0.69  &  {\ } 13 & 17 & 0.74 \\
$    p_{T}^{t\bar{t}}$ & {\ } 4.2 & 5 & 0.52  &  {\ } 4.0 & 5 & 0.54  &  {\ } 8.7 & 5 & 0.12  &  {\ } 14 & 5 & 0.01  &  {\ } 4.6 & 5 & 0.47 \\
$    H_{T}^{t\bar{t}}$ & {\ } 16 & 14 & 0.34  &  {\ } 13 & 14 & 0.55  &  {\ } 18 & 14 & 0.20  &  {\ } 9.5 & 14 & 0.80  &  {\ } 50 & 14 & <0.01 \\
$\Delta \phi^{t\bar{t}}$ & {\ } 0.3 & 3 & 0.96  &  {\ } 3.7 & 3 & 0.29  &  {\ } 1.2 & 3 & 0.74  &  {\ } 5.4 & 3 & 0.14  &  {\ } 6.0 & 3 & 0.11 \\
\hline
\end{tabular}}

\caption{\small{Comparison between the measured fiducial phase-space normalized differential cross-sections and the predictions from several MC generators. For each variable and prediction a $\chi^2$ and a $p$-value are calculated using the covariance matrix of each measured spectrum. 
The number of degrees of freedom (NDF) is equal to $N_{\rm b}-1$ where $N_{\rm b}$ is the number of bins in the distribution.}}
\label{tab:chi2:particle:rel}
\end{table*}
\begin{table*} [p]
\footnotesize
\centering
\noindent\makebox[\textwidth]{
\begin{tabular}{c | r @{/} l c  | r @{/} l c  | r @{/} l c  | r @{/} l c  | r @{/} l c }
\hline
Variable  & \multicolumn{3}{|c|}{PWG+PY8} & \multicolumn{3}{|c|}{MC@NLO+HW} & \multicolumn{3}{|c|}{PWG+PY6} & \multicolumn{3}{|c|}{PWG+HW6} & \multicolumn{3}{|c}{MadGraph+PY6} \\
  & \multicolumn{3}{|c|}{CT10 $h_{\rm damp}=m_{t}$} & \multicolumn{3}{|c|}{CT10 AUET2} & \multicolumn{3}{|c|}{CT10 $h_{\rm damp}=m_{t}$} & \multicolumn{3}{|c|}{CT10 $h_{\rm damp}=\infty$} & \multicolumn{3}{|c}{MadGraph+PY6 P2011C} \\
& \multicolumn{2}{|c}{$\chi^{2}$/NDF} &  ~$p$-value  & \multicolumn{2}{|c}{$\chi^{2}$/NDF} &  ~$p$-value  & \multicolumn{2}{|c}{$\chi^{2}$/NDF} &  ~$p$-value  & \multicolumn{2}{|c}{$\chi^{2}$/NDF} &  ~$p$-value  & \multicolumn{2}{|c}{$\chi^{2}$/NDF} &  ~$p$-value  \\
\hline
\hline
$           p_{\rm T}^{t}$ & {\ } 0.7 & 7 & 1.00  &  {\ } 5.1 & 7 & 0.65  &  {\ } 5.8 & 7 & 0.56  &  {\ } 3.8 & 7 & 0.80  &  {\ } 16 & 7 & 0.03 \\
$     \chi^{t\bar{t}}$ & {\ } 29 & 9 & <0.01  &  {\ } 69 & 9 & <0.01  &  {\ } 32 & 9 & <0.01  &  {\ } 120 & 9 & <0.01  &  {\ } 400 & 9 & <0.01 \\
$      |y^{t\bar{t}}|$ & {\ } 34 & 4 & <0.01  &  {\ } 24 & 4 & <0.01  &  {\ } 35 & 4 & <0.01  &  {\ } 33 & 4 & <0.01  &  {\ } 44 & 4 & <0.01 \\
$        m^{t\bar{t}}$ & {\ } 3.6 & 6 & 0.73  &  {\ } 3.8 & 6 & 0.71  &  {\ } 1.9 & 6 & 0.93  &  {\ } 22 & 6 & <0.01  &  {\ } 13 & 6 & 0.04 \\
$y_{\rm boost}^{t\bar{t}}$ & {\ } 140 & 15 & <0.01  &  {\ } 93 & 15 & <0.01  &  {\ } 140 & 15 & <0.01  &  {\ } 140 & 15 & <0.01  &  {\ } 180 & 15 & <0.01 \\
$|p_{\rm out}^{t\bar{t}}|$ & {\ } 1.8 & 5 & 0.88  &  {\ } 1.9 & 5 & 0.86  &  {\ } 1.1 & 5 & 0.96  &  {\ } 2.5 & 5 & 0.78  &  {\ } 0.8 & 5 & 0.98 \\
$             |y^{t}|$ & {\ } 2.3 & 4 & 0.69  &  {\ } 1.5 & 4 & 0.83  &  {\ } 2.5 & 4 & 0.65  &  {\ } 1.8 & 4 & 0.77  &  {\ } 1.2 & 4 & 0.87 \\
$    p_{\rm T}^{t\bar{t}}$ & {\ } 2.7 & 5 & 0.75  &  {\ } 2.8 & 5 & 0.72  &  {\ } 1.2 & 5 & 0.94  &  {\ } 5.0 & 5 & 0.41  &  {\ } 11 & 5 & 0.05 \\
$    H_{\rm T}^{t\bar{t}}$ & {\ } 3.2 & 14 & 1.00  &  {\ } 7.3 & 14 & 0.92  &  {\ } 16 & 14 & 0.29  &  {\ } 3.2 & 14 & 1.00  &  {\ } 44 & 14 & <0.01 \\
$\Delta \phi^{t\bar{t}}$ & {\ } 0.5 & 3 & 0.93  &  {\ } 0.2 & 3 & 0.97  &  {\ } 0.8 & 3 & 0.85  &  {\ } 6.2 & 3 & 0.10  &  {\ } 4.3 & 3 & 0.23 \\
\hline
\end{tabular}}

\caption{\small{Comparison between the measured full phase-space normalized differential cross-sections and the predictions from several MC generators. For each variable and prediction a $\chi^2$ and a $p$-value are calculated using the covariance matrix of each measured spectrum. 
The number of degrees of freedom (NDF) is equal to $N_{\rm b}-1$ where $N_{\rm b}$ is the number of bins in the distribution.}}
\label{tab:chi2:parton:rel}
\end{table*}

\begin{landscape}
\begin{table*} [p]
\footnotesize
\centering
\noindent\makebox[\textwidth]{
\begin{tabular}{c | r @{/} l c  | r @{/} l c  | r @{/} l c  | r @{/} l c  | r @{/} l c  | r @{/} l c  | r @{/} l c }
\hline
Variable  & \multicolumn{3}{|c|}{CT14nlo} & \multicolumn{3}{|c|}{CJ12mid} & \multicolumn{3}{|c|}{MMHT2014nlo68cl} & \multicolumn{3}{|c|}{NNPDF30nlo} & \multicolumn{3}{|c|}{CT10nlo} & \multicolumn{3}{|c|}{METAv10LHC} & \multicolumn{3}{|c}{HERA20NLO} \\
& \multicolumn{2}{|c}{$\chi^{2}$/NDF} &  ~$p$-value  & \multicolumn{2}{|c}{$\chi^{2}$/NDF} &  ~$p$-value  & \multicolumn{2}{|c}{$\chi^{2}$/NDF} &  ~$p$-value  & \multicolumn{2}{|c}{$\chi^{2}$/NDF} &  ~$p$-value  & \multicolumn{2}{|c}{$\chi^{2}$/NDF} &  ~$p$-value  & \multicolumn{2}{|c}{$\chi^{2}$/NDF} &  ~$p$-value  & \multicolumn{2}{|c}{$\chi^{2}$/NDF} &  ~$p$-value  \\
\hline
\hline
$      |y^{t\bar{t}}|$ & {\ } 24 & 17 & 0.14  &  {\ } 18 & 17 & 0.36  &  {\ } 16 & 17 & 0.51  &  {\ } 14 & 17 & 0.70  &  {\ } 25 & 17 & 0.10  &  {\ } 14 & 17 & 0.64  &  {\ } 24 & 17 & 0.12 \\
$   |y^{t,{\rm had}}|$ & {\ } 15 & 17 & 0.60  &  {\ } 13 & 17 & 0.71  &  {\ } 14 & 17 & 0.66  &  {\ } 12 & 17 & 0.79  &  {\ } 13 & 17 & 0.75  &  {\ } 13 & 17 & 0.71  &  {\ } 26 & 17 & 0.08 \\
$y_{boost}^{t\bar{t}}$ & {\ } 21 & 15 & 0.15  &  {\ } 18 & 15 & 0.29  &  {\ } 12 & 15 & 0.68  &  {\ } 8.8 & 15 & 0.89  &  {\ } 25 & 15 & 0.06  &  {\ } 10 & 15 & 0.84  &  {\ } 17 & 15 & 0.32 \\
\hline
\end{tabular}}

\caption{\small{Comparison between the measured fiducial phase-space normalized differential cross-sections and the predictions from new PDF sets using the \MCatNLO{}+\Herwig generator. For each variable and prediction a $\chi^2$ and a $p$-value are calculated using the covariance matrix of each measured spectrum. 
The number of degrees of freedom (NDF) is equal to $N_{\rm b}-1$ where $N_{\rm b}$ is the number of bins in the distribution.}}
\label{tab:chi2:pdf:particle:rel}
\end{table*}

\begin{table*} [p]
\footnotesize
\centering
\noindent\makebox[\textwidth]{
\begin{tabular}{c | r @{/} l c  | r @{/} l c }
\hline
Variable  & \multicolumn{3}{|c|}{\aNNNLO} & \multicolumn{3}{|c}{aNNLO} \\
& \multicolumn{2}{|c}{$\chi^{2}$/NDF} &  ~$p$-value  & \multicolumn{2}{|c}{$\chi^{2}$/NDF} &  ~$p$-value  \\
\hline
\hline
$           p_{T}^{t}$ & {\ } 18 & 7 & 0.01  &  {\ } 4.0 & 7 & 0.78 \\
$             |y^{t}|$ & {\ } 0.6 & 4 & 0.96  &  {\ } 9.2 & 4 & 0.06 \\
\hline
\end{tabular}}

\caption{\small{Comparison between the measured full phase-space normalized differential cross-sections and higher-order QCD calculations. For each variable and prediction a $\chi^2$ and a $p$-value are calculated using the covariance matrix of each measured spectrum. 
The number of degrees of freedom (NDF) is equal to $N_{\rm b}-1$ where $N_{\rm b}$ is the number of bins in the distribution.}}
\label{tab:chi2:theory:topH:parton:rel}
\end{table*}
\end{landscape}

\clearpage
\section{Conclusions} \label{sec:Conclusion}

Kinematic distributions of the top quarks in \ttbar{} events, selected in the lepton+jets channel, are measured in the fiducial and full phase space using data from 8 \TeV{} proton--proton collisions collected by the ATLAS detector at the Large Hadron Collider, corresponding to an integrated luminosity of 20.3 fb$^{-1}$. Normalized differential cross-sections are measured as a function of the hadronic top-quark transverse momentum and rapidity, and as a function of the mass, transverse momentum, and rapidity of the \ttbar{} system. In addition, a new set of observables describing the hard-scattering interaction (\chittbar, \boostttbar) and sensitive to the emission of radiation along with the \ttb{} pair (\deltaPhittbar, \absPoutttbar, \HTttbar, \RWtttbar) are presented.

The measurements presented here exhibit, for most distributions and in large part of the phase space, a precision of the order of 5\% or better and an overall agreement with the Monte Carlo predictions of the order of 10\%.

The \absyttbar and \boostttbar distributions are not well modelled by any generator under consideration in the fiducial phase space, however the agreement improves when new parton distribution functions are used with the \MCatNLO{}+\Herwig generator. 

All the generators under consideration consistently predict a ratio of the hadronic $W$ boson and top-quark transverse momenta (\RWtttbar) with a mis-modelling of up to 10\% in the range $1.5$--$3$.

The tail of the \ptthad{} distribution is harder in all predictions than what is observed in data, an effect previously observed in measurements by ATLAS and CMS. The agreement improves when using the \Herwig parton shower with respect to \Pythia. The tension observed for \PowHeg{}+\Pythia{}, \PowHeg{}+\PythiaEight{}~ and \MadGraph{}+\Pythia{} in the \ptt{} spectrum is reflected in the tail of the \HTttbar{} distribution. 

Similarly, both \aNNNLO{} and aNNLO predictions have a poor agreement in the \ptt{} spectrum in the full phase space. However, the full \NNLO{} calculation, which has just become available, is in good agreement with the \ptt{} distribution, indicating the disagreement seen with the generators and other calculations is due to missing higher-order terms. The \NNLO{} calculation also shows good agreement in the \absyt{} and \mttbar{} distributions.

\section*{Acknowledgements}

We thank CERN for the very successful operation of the LHC, as well as the support staff from our institutions without whom ATLAS could not be operated efficiently.

We acknowledge the support of ANPCyT, Argentina; YerPhI, Armenia; ARC,
Australia; BMWF and FWF, Austria; ANAS, Azerbaijan; SSTC, Belarus; CNPq and FAPESP,
Brazil; NSERC, NRC and CFI, Canada; CERN; CONICYT, Chile; CAS, MOST and NSFC,
China; COLCIENCIAS, Colombia; MSMT CR, MPO CR and VSC CR, Czech Republic;
DNRF, DNSRC and Lundbeck Foundation, Denmark; EPLANET, ERC and NSRF, European Union;
IN2P3-CNRS, CEA-DSM/IRFU, France; GNSF, Georgia; BMBF, DFG, HGF, MPG and AvH
Foundation, Germany; GSRT and NSRF, Greece; ISF, MINERVA, GIF, DIP and Benoziyo Center,
Israel; INFN, Italy; MEXT and JSPS, Japan; CNRST, Morocco; FOM and NWO,
Netherlands; BRF and RCN, Norway; MNiSW, Poland; GRICES and FCT, Portugal; MERYS
(MECTS), Romania; MES of Russia and ROSATOM, Russian Federation; JINR; MSTD,
Serbia; MSSR, Slovakia; ARRS and MIZ\v{S}, Slovenia; DST/NRF, South Africa;
MICINN, Spain; SRC and Wallenberg Foundation, Sweden; SER, SNSF and Cantons of
Bern and Geneva, Switzerland; NSC, Taiwan; TAEK, Turkey; STFC, the Royal
Society and Leverhulme Trust, United Kingdom; DOE and NSF, United States of
America.

The crucial computing support from all WLCG partners is acknowledged
gratefully, in particular from CERN and the ATLAS Tier-1 facilities at
TRIUMF (Canada), NDGF (Denmark, Norway, Sweden), CC-IN2P3 (France),
KIT/GridKA (Germany), INFN-CNAF (Italy), NL-T1 (Netherlands), PIC (Spain),
ASGC (Taiwan), RAL (UK) and BNL (USA) and in the Tier-2 facilities
worldwide.

\appendix

\clearpage
\label{app:References}
\printbibliography

\newpage 
\begin{flushleft}
{\Large The ATLAS Collaboration}

\bigskip

G.~Aad$^{\rm 85}$,
B.~Abbott$^{\rm 113}$,
J.~Abdallah$^{\rm 151}$,
O.~Abdinov$^{\rm 11}$,
R.~Aben$^{\rm 107}$,
M.~Abolins$^{\rm 90}$,
O.S.~AbouZeid$^{\rm 158}$,
H.~Abramowicz$^{\rm 153}$,
H.~Abreu$^{\rm 152}$,
R.~Abreu$^{\rm 116}$
% ...
\bigskip
\\
% ...
$^{11}$ Institute of Physics, Azerbaijan Academy of Sciences, Baku, Azerbaijan\\
% ...
$^{85}$ CPPM, Aix-Marseille Universit{\'e} and CNRS/IN2P3, Marseille, France\\
% ...
$^{90}$ Department of Physics and Astronomy, Michigan State University, East Lansing MI, United States of America\\
% ...
$^{107}$ Nikhef National Institute for Subatomic Physics and University of Amsterdam, Amsterdam, Netherlands\\
% ...
$^{113}$ Homer L. Dodge Department of Physics and Astronomy, University of Oklahoma, Norman OK, United States of America\\
% ...
$^{116}$ Center for High Energy Physics, University of Oregon, Eugene OR, United States of America\\
% ...
$^{151}$ Institute of Physics, Academia Sinica, Taipei, Taiwan\\
$^{152}$ Department of Physics, Technion: Israel Institute of Technology, Haifa, Israel\\
$^{153}$ Raymond and Beverly Sackler School of Physics and Astronomy, Tel Aviv University, Tel Aviv, Israel\\
% ...
$^{158}$ Department of Physics, University of Toronto, Toronto ON, Canada\\
% ...
\end{flushleft}


\clearpage

\end{document}